\documentclass[dvips]{acta}
\usepackage{supertabular,lscape,epsfig}
\usepackage{amssymb}
\usepackage{amsmath}
\usepackage{booktabs}
\usepackage{rotating}
\usepackage{graphicx}
\usepackage{subcaption}

\SetPages{0}{0}

\SetVol{72}{2022}

\begin{document}

\begin{Titlepage}
\Title{A Method of Improving Standard Stellar Luminosities with Multiband Standard Bolometric Corrections}
\Author{Bak{\i}\c{s}, V. and Eker, Z.}{Department of Space Sciences and Technologies, Faculty of Sciences, Akdeniz University, Antalya, TR.\\
e-mail:volkanbakis@akdeniz.edu.tr}

\Received{April, 2022}
\end{Titlepage}

\Abstract{Standard luminosity ($L$) of 406 main-sequence stars with the most accurate astrophysical parameters are predicted from their absolute magnitudes and bolometric corrections at Johnson $B,V$, and Gaia EDR3 $G$, $G_{BP}$, $G_{RP}$ filters. Required multiband $BC$ and $BC-T_{eff}$ relations are obtained first from the parameters of 209 DDEB (Double-lined Detached Eclipsing Binaries) with main-sequence components and Gaia EDR3 parallaxes. A simplified SED is formulated to give filter dependent component light contributions and interstellar dimming, which are essential in computing $BC$ of a component virtually at any filter. The mean standard $L$ of a star is calculated from the mean $M_{Bol}$ which is a mathematical average of independent $M_{Bol}$ values predicted at different filters, while the uncertainty of $L$ is the uncertainty propagated from the uncertainty of the mean $M_{Bol}$. The mean standard $L$ of the sample stars are compared to the corresponding $L$ values according to the Stefan-Boltzmann law. A very high correlation ($R^2>0.999$) is found. Comparing histogram distributions of errors shows that uncertainties associated with the mean standard $L$ (peak at $\sim2.5$ per cent) are much smaller than the uncertainties of $L$ (peak at $\sim8$ per cent) by the Stefan-Boltzmann law. Increasing the number of filters used in predicting the mean $M_{Bol}$ increases the accuracy of the standard stellar luminosity. Extinction law, color-color relations and color excess - color excess relations for Gaia passbands are demonstrated for main-sequence stars for the first time.}{stars: fundamental parameters, (stars:) binaries: eclipsing, Sun: general}

\section{Introduction}
\label{sec:intro}

An unobserved missing fraction of a stellar luminosity ($L$) by a photometric observation is named bolometric correction ($BC$) if it is expressed in magnitude units. As if fixing a defect or restoring the missing part; adding this fraction ($BC$) to the apparent ($V$) or absolute ($M_V$) magnitudes of a star, which are known to be naturally limited in a certain wavelength range, one obtains the apparent ($m_{Bol}$) or absolute ($M_{Bol}$) bolometric magnitudes representing the total $L$ of the star.  Although $BC$ is useful, ingeniously invented, and the ready-to-use quantity to get those quantities from an apparent magnitude ($V$) and a parallax, major difficulty confronted by the earlier astronomers (Kuiper 1938; Mc Donald \& Underhill 1952; Popper 1959; Wildey 1963; Smak 1966; Johnson 1966; Weidemann \& Bues 1967; Heintze 1973) is that a pre-required $BC$ must be determined first from observations but there is neither a telescope nor a detector to measure a bolometric magnitude while $L$ is already known to be un-observable. 

Therefore, a relation between an apparent magnitude (e.g. $V$) and observable part of the luminosity (e.g. $L_V$) recognised in all of the photometric passbands with pre-established filter transmissions operated in the Vega system of magnitudes, where Vega is a common reference, cannot be established between a bolometric magnitude and $L$ using direct observations.

This difficulty, however, was overcome by assuming arbitrary zero points for both the bolometric magnitudes and $BC$ scale. The arbitrariness attributed to $BC$ and bolometric magnitude scales was then caused publications of many different $BC$ tables; some containing all negative (Kuiper 1938; Popper 1959; Wildey 1963; Cox 2000; Pecaut \& Mamajec 2013) $BC$ values as if intentionally opposing the rest containing a limited number of positive $BC$ (Code et al. 1976; Johnson 1964, 1966; Flower 1977, 1996; Bessel et al. 1998; Sung et al. 2013; Cassagrande \& VandenBerg 2018; Eker et al. 2020). Incorrect usage of tabulated $BC$ values was discussed by Torres (2010). The biggest of the problems, however, is that a star could be found to have several $BC$ from several tables, implying that several different $M_{Bol}$ representing a single $L$.

The problems of arbitrariness attributed to the $BC$ scale were studied recently by Eker et al. (2021a), who introduced the concept of standard $BC$. The standardisation was necessary to avoid problems caused by arbitrariness of the $BC$ scale and for unifying the $BC$ and $M_{Bol}$ values, which is the easiest way of assuring consistent $L$ of a star if it is predicted from astrometric (parallax) and photometric observations. 

Accuracy of the classical methods of computing a stellar luminosity (1- a direct method from radii ($R$) and effective temperatures ($T_{eff}$); 2- a method using a mass-luminosity relation ($MLR$); 3- a method requiring a bolometric correction) is later studied by Eker et al. (2021b), who introduced the concept of standard stellar $L$.  If $L$ of a star is calculated from one of its absolute magnitudes ($M_\xi$, where $\xi$ indicates a filter in a photometric system) and corresponding standard $BC$, it is called standard $L$ while $L$ according to the Stefan-Boltzmann law is standard by definition.

The methods (2) and (3) are indirect because a pre-determined $MLR$ is required for method (2) and a pre-determined $BC-T_{eff}$ relation is necessary for method (3).  In the absence of these pre-determined relations, both are not operable.

Eker et al. (2021b) claimed the indirect methods are less accurate than the direct method providing a stellar $L$ with a typical accuracy of 8.2 - 12.2 per cent, which could be as high as a few per cent; e.g. primary of V505 Per has $L= 2.688  L_\odot$ and its uncertainty ($\Delta L/L$) is 2.53 per cent implied by very small relative uncertainties of its radius $\Delta R / R = 1.09$ per cent and effective temperature $\Delta T_{eff} / T_{eff}= 0.32$ per cent (Tomasella et al. 2008). Only if a unique $BC$ directly determined from an observed SED with very high spectral resolution is used in the third method, then the relative uncertainty of predicted $L$ could be improved up to a one per cent or more level. Otherwise, using a standard $BC$ predicted from the standard $BC-T_{eff}$ relation, method 3 cannot provide an accuracy better than the direct method.

However, using a unique $BC$ in method 3 is just a speculation, that is, it is impractical nowadays as expressed by Eker et al. (2021b). Therefore, the primary aim of this study is not to speculate, but to investigate how to improve the standard stellar luminosities  obtained by the third method realistically using multiband standard $BC-T_{eff}$ relations. To achieve this aim, a new method is introduced for estimating relative light contributions of binary components from a simplified SED operable virtually at any photometric passband. Then, the classical method of Eker et al. (2020), which requires an apparent magnitude of a binary system, the light ratio of components, a reliable parallax and an interstellar extinction, is used for predicting the multiband $BC-T_{eff}$ relations for the Gaia $G$, $G_{BP}$ and $G_{RP}$, and Johnson $B$, $V$ passbands. Data and input parameters are described in \S2. The new method is explained in \S3. Calibrations of multiband $BC-T_{eff}$ relations are described in \S4. How to improve standard $L$ of a star is discussed in \S5. Discussions are found in \S6. Conclusions are in \S7.

\section{Data}
\label{sec:data}

Having essentially the same purpose, to obtain most reliable empirical $BC$ from the most reliable stellar parameters first and then to calibrate the most reliable $BC$-$T_{eff}$ relations, this study and Eker et al. (2020) rely upon the same original data set of DDEB (Double-Lined Detached Eclipsing Binaries) published by Eker et al. (2018). The 509 main-sequence stars with most reliable masses ($M$) and radii ($R$) accurate within 15 per cent and with published effective temperatures ($T_{eff}$) as the components of DDEB having metallicities 0.008 $\leq$ Z $\leq$ 0.040  in the solar neighbourhood of the Galactic disk were originally used by Eker et al. (2018) for calibrating interrelated mass-luminosity (MLR), mass-radius (MRR) and mass-effective temperature (MTR) relations.  Later, Eker et al. (2020) combined this data set with the data set of Graczyk et al. (2019), who studied the global zero-point shift between the photometric fluxes of 81 detached eclipsing binaries and Gaia DR2 trigonometric parallaxes (Gaia Collaboration et al. 2018) in order to increase the number of available systems with component light ratios in Johnson $B$ and $V$-bands, essential for computing the apparent magnitude of a component from the total brightness of a system; a step required before computing $BC$ of a component. 

This combined data set of Eker et al. (2020) contained 290 DDEB having at least one component on the main sequence. Aiming to calibrate a main-sequence $BC_V$ - $T_{eff}$ relation, Eker et al. (2020) could find only 400 main-sequence stars (194 binaries, 8 primaries, and 4 secondaries, that is, 206 systems) from the available 580 component stars (290 DDEB). 290 minus 206, that is, 84 systems were lost because of a missing systemic apparent brightness ($V_{tot}$), or a missing light ratio ($l_2$ / $l_1$) in the V band, or a missing parallax, or a missing interstellar dimming ($A_V$). if any one of those parameters is missing for a system, $BC$ of its components cannot be calculated, that is, this system is not eligible for calculating $BC$. A percentage of a third light is also needed if it is detected in the light curve of a system. 206 DDEB has 412 components. After subtracting the non-main-sequence stars, the number of stars with a computed $BC_V$ reduced to 400, from which a main-sequence  $BC_V$ - $T_{eff}$ relation is calibrated by Eker et al. (2020).  

The original data set containing 290 systems investigated by Eker et al. (2020) is considered for this study for two purposes: First of all, we wanted to test whether or not the new method used in this study provides reliable component light contributions. For this, 206 systems with $V$ and $B$-passband light ratios published by Eker et al. (2020) are ideal for testing by comparing to the light ratios found in this study. Nevertheless, the new method also has its own limitations, to be discussed in the next section. That is, 206 systems will naturally be reduced further. To compensate for such a further loss and to be able to use its applicability to the systems even without standard $V$ light curve solutions, the larger data set (290 system) was chosen as the main sample for this study.

Here in this study, nearly the same number (406) of main-sequence stars are found as the components of 209 DDEB (197 system, 9 primaries and 3 secondaries) eligible to calculate their $BC$ values and then to continue calibrating $BC$-$T_{eff}$ relations for the Gaia photometric passbands $G$, $G_{BP}$ and $G_{RP}$. Among this sub sample 312 main-sequence stars (152 binaries, 5 primaries and 3 secondaries) are found as a sub sub sample containing stars common between the present study and Eker et al. (2020), which are sufficient for testing and verifying the validity of the component contributions predicted by the new method involving SED.

The basic physical parameters of the components and total brightness in $B$, $V$ and Gaia passbands of DDEBs used in this study are listed in Table~1, where the columns are self-explanatory to indicate order, name of the system, the component (primary or secondary) are in the first three columns. Note that the primaries and the secondaries identified as non-main-sequence stars are indicated by p* or s*. From the fourth to seventh columns, masses and radii with their errors are given. In columns 9 and 10, temperatures and errors are given. References for physical parameters are in columns 8 and 11. The total brightness of the systems in $B$ and $V$-bands with their errors are in columns 12-15 with their references in the 16th column. In the rest of the columns (17-22), the total brightness of the systems in the Gaia passbands which are available in the Gaia EDR3 are given.

In Table~2, in addition to the first three columns which are the same as in Table~1, the fourth and fifth columns give cross references where Xref(1) (column 4) indicates the order number in Eker et al. (2018) and Xref(2) (column 5) indicates the order number in Eker et al. (2020). Every system in this study has at least one Xref number in columns 4 and 5. Only five binaries from the list of Graczyk et al. (2019) do not have Xref in column 4. Systems without Xref numbers in column 5 are the ones excluded by Eker et al. (2020) either because of missing $V$-band light ratio or without interstellar extinction etc. In this study, we preferred to display component contributions (columns 6 and 7) rather than light ratios which were preferred by Eker et al. (2020). Thus, the sum of the contributions of the primary and the secondary is one. 

The advantage here is that the component contributions of all systems in Table\,2 are predictable by the new method. There are two possibilities for this: from the reddened SED, and the other from the unreddened SED. The unreddened light contributions are listed in Table\,2 (columns  8, 9, 10, 11, 12) for $B$, $V$, $G$, $G_{BP}$ and $G_{RP}$ passbands, respectively. Since reddened component contributions are the same in four digits after the decimal, to save space, they are not shown in Table\,2.

At last, component apparent magnitudes in accord with the contributions are given under columns 13, 15, 17, 19 and 21 for the same bands in the same order. Rather than propagating observational random errors, Eker et al. (2020) had no option but to assume the same uncertainty for the component magnitudes as the systemic brightness because the light ratio ($L_2$/$L_1$) of components predicted from light curve solutions are usually published without an uncertainty. Using the same method here, the same steps are followed in this study. That is, the uncertainty of the component magnitudes were assumed to have the same uncertainty as the uncertainty of the systemic brightness.  
 
One may notice that many systems in Table 2 do not have $B$-band apparent magnitudes (Column 13) in contrast to apparent magnitudes at Gaia passbands, which are almost complete excluding only $\beta$ Aur and $\alpha$ CrB (both are too bright for reliable measurement), while ten systems are missing in the $V$-band. Those are the systems even the new method does not help to compute their $BC$ values because of missing total brightness.
 
Including VV Pyx, there are eleven more systems (marked in Table 3) which do not produce reliable SED model with EDR3 parallaxes. For them, DR2 parallaxes produce better SED model. VV Pyx, $\beta$ Aur and $\alpha$ CrB are three systems we found better SED models with Hipparcos parallaxes. 

\begin{figure}
	\includegraphics[width=\columnwidth]{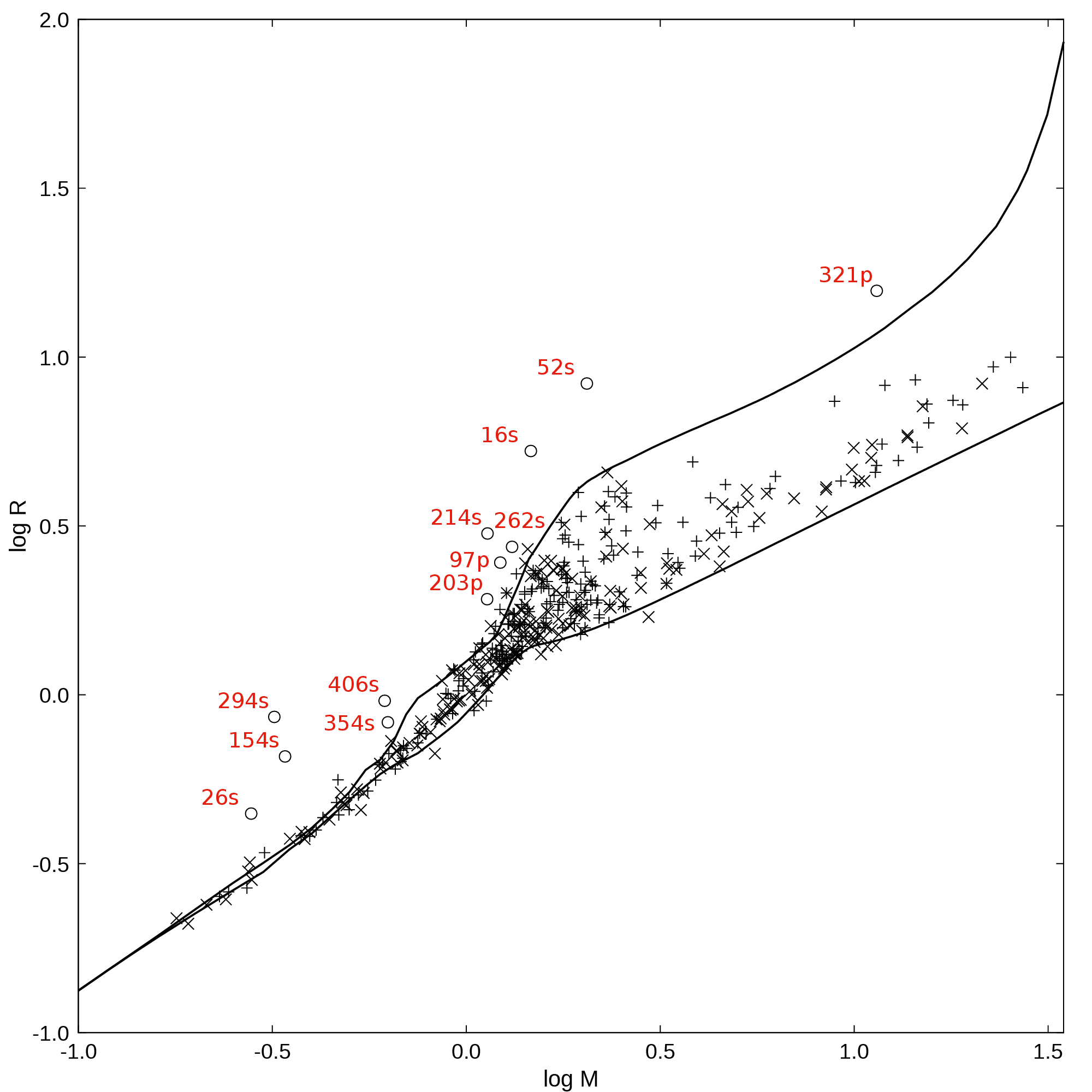}
    \caption{Primary (+) and secondary (x) components on the main-sequence and components discarded (empty circles). Numbers are in the order in Table~1, p and s for primary and secondary. Solid lines show ZAMS and TAMS limits for solar metallicity (Z=0.014).}
    \label{fig:logM_logR} 
\end{figure}

Before going to the next section, where the new method is described, we must certify the 406 stars considered in this study are all in the main-sequence stage of evolution, which is very obvious in Fig.~1. The three primaries and nine secondaries identified as non-main-sequence stars appear located outside of the main-sequence limits, which are shown by continuous lines, according to ZAMS (Zero Age Main Sequence) and TAMS (Terminal Age Main Sequence) limits for the solar abundance $Z = 0.014$ from PARSEC evolutionary models (Bressan et al. 2012). 

Metal abundance distribution of DDEB at the solar neighbourhood within the disk of the Milky Way is already discussed by Eker et al. (2018), where the peak at the solar metallicity (Z=0.014-0.017) together with the lower (Z=0.008) and the upper (Z=0.040) limits were indicated. To save space and to be clearer at identifications of non-main-sequence stars in Fig.~1, the ZAMS and TAMS curves for Z=0.008, Z=0.040 and discussions therein are not repeated in this study. However, because the present sub-sample is mainly selected from the sample of Eker et al. (2018), metal abundances of the main-sequences stars used in this study are also expected to have a similar metallicity distribution within the limits 0.008 $\leq$ Z $\leq$ 0.040.

\section{A New method involving SED to obtain light ratio of components}

When computing $BC$ of stars, unlike previous studies (Code et al. 1976; Cayrel et al. 1997; Girardi et al. 2008; Andrae et al. 2018; Chen et al. 2019) utilizing the well-known relation directly with $S_\lambda(V)$ (sensitivity function of the $V$ magnitude system), $f_\lambda$, (monochromatic flux from a star) and $C_2$  (arbitrary constant of integration),

\begin{equation}
    BC_V=2.5log\frac{f_V}{f_{Bol}}+C_2=2.5log\frac{\int_{0}^{\infty} S_\lambda(V)f_\lambda d\lambda}{\int_{0}^{\infty} f_\lambda d\lambda}+C_2,
	\label{eq:bc}
\end{equation}
here, we introduce a new method operating indirectly. This new method deserves to be called indirect because the $V$ filtered radiation 

\begin{equation}
    f_V = \int_{0}^{\infty} S_\lambda(V) f_\lambda d\lambda,
	\label{eq:Vflux}
\end{equation}
is calculated using the SED model with simplifications only for predicting relative light contributions (to be explained later) of component stars in binaries, while $f_{Bol}$ stands for the bolometric flux of a component reaching to telescope if there is no extinction. Only after, apparent magnitudes of the components calculated from apparent magnitude of the system using the light ratio of components, and absolute magnitudes are calculated from the apparent magnitudes as being corrected for interstellar extinction, $BC$ of each component is calculated as

\begin{equation}
    BC_V = M_{Bol} - M_V 
\end{equation}

The first simplification is that the  spectrum of a component ($f_\lambda$) is approximated by a Planck Function, rather than a model atmosphere characterized by a $T_{eff}$, log $g$,  micro-turbulent velocity ($\zeta$) and metal abundance $[Fe/H]$. Being independent of model atmosphere parameters, the new method is easier to use and more suitable for obtaining empirical $BC$s; e.g. for main-sequence stars, rather than series of $BC$ tables specified with a log $g$, $\zeta$ or $[Fe/H]$ for various passbands of different photometric systems. 

Assuming no interstellar extinction in the first approximation, a spherical star with a radius $R$ produces a flux continuum (SED) at a distance $d$ from its centre. The monochromatic flux could be expressed as: 

\begin{equation}
    f_\lambda = \frac{R^2}{d^2} \pi B_\lambda(Teff).
	\label{eq:flambda}
\end{equation}

Then, the second simplification becomes clear; the spectral lines and prominent spectral features are ignored since the Plank function represents the star's continuum. The equation implies that limb darkening is also ignored within the solid angle $\pi{R^2}/{d^2}$  where the isotropic intensity is  $B_\lambda(T_{eff})$. Eq.(4) would be adapted to a detached binary with spherical components as:

\begin{equation}
    f_\lambda (system)= \frac{\pi}{d^2} \left[R_1^2 B_\lambda(T_1)+R_2^2 B_\lambda(T_2)\right],
	\label{eq:systemflux}
\end{equation}
where $R_1$ \& $R_2$  and $T_1$ \& $T_2$  are radii and effective temperatures of the primary and secondary, respectively. If $d$ is also the distance from the Earth, then $f_\lambda$ in both equations would represent an unreddened SED in units of W $m^{-2}A^{-1}$. The unreddened SED of V618~Per, which is one of the systems in Table~1, is shown by a dashed continuous curve starting from the upper left and ending at the lower right in Fig.~2. The observed spectrophotometric flux data of V618~Per are taken from the SIMBAD database (Wenger et al. 2000). The observed flux data does not appear to fit the unreddened SED, especially toward shorter wavelengths. The deviation from the unreddened SED is expected because of interstellar extinction. 

\begin{figure*}
	\includegraphics[width=14cm]{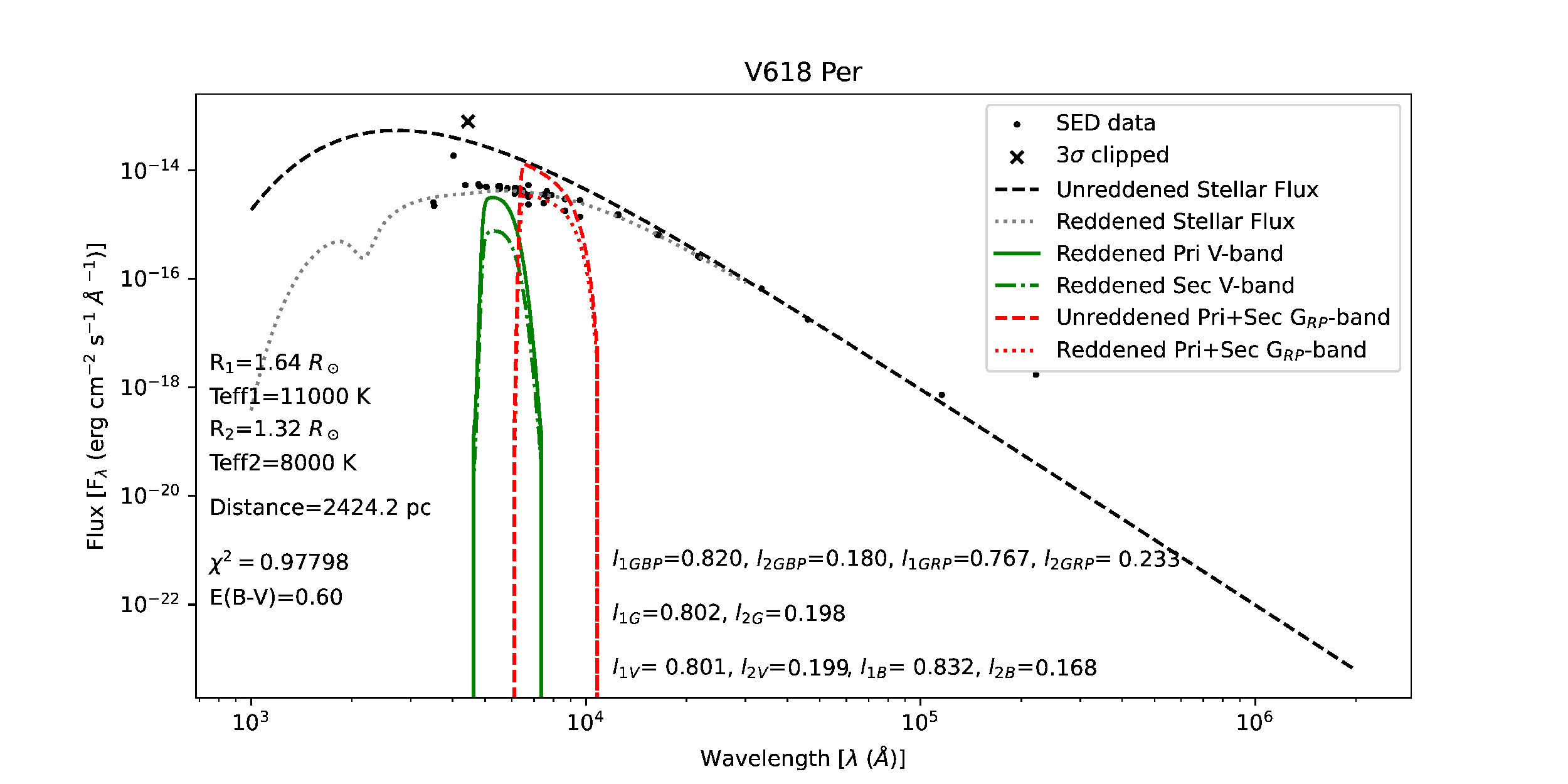}
    \caption{Unreddened (dashed) and reddened (dotted)  SED models for V618~Per. Flux data determining the reddened SED are shown by filled circles. Crosses are data deviated more than 3 sigmas. For simplicity, only the relative $V$ band contributions of primary (continuous) and secondary (dot-dashed) are shown by vertical profiles on the left. The $G$ band dimming ($A_{G_{RP}}$) is computed from the vertical profiles on the right which are the total (primary + secondary) light of the system from the reddened (dotted) and unreddened (dashed) SED after convolution by the $G$ filter.} 
    \label{fig:v618per}
\end{figure*}

For modelling the observed SED data, our unreddened SED model is reddened by adjusting $E(B-V)$ of the system until a best fitting reddened SED is obtained using the reddening model of Fitzpatrick (1999).  The parameter $R(V)$ is adopted as $R(V)= 3.1$. Since we have calculated the parameter $R(\lambda) = A_\lambda/E(B-V)$ for each filter individually, the initially selected value of $R(V)$ did not affect our analysis. We will discuss this issue later.  The $\chi^2$ fitting of the reddened SED is displayed by a dotted continuous curve just below the unreddened SED in Fig.~2.

Assuming no interstellar extinction, that is the previously computed unreddened SED (Eq.(5)), one may compute unreddened visual flux $f_V$ of a component, if a reliable trigonometric parallax (or distance) is available, using $f_\lambda$ from Eq.(4) to indicate the flux contribution of the component in the $V$ filtered radiation (SED) reaching above the Earth's atmosphere. However, by replacing the unreddened  $f_\lambda$  with the reddened $f_\lambda$,  which is obtained by $\chi^2$ fitting, one may compute the $V$-band flux contribution of the very same component in the reddened SED.

We have systematically calculated both reddened and unreddened component contributions at $B$, $V$, $G$, $G_{BP}$ and $G_{RP}$ passbands  using the filter profiles from Bessel (1990) and Evans et al. (2018) for Johnson $B$, $V$, and Gaia passbands, respectively. For the sake of clarity, only $V$-band contributions of the primary and the secondary of V618~Per corresponding to the reddened SED are shown in Fig.~2 as vertical profiles on the left where the solid and dotted-dash lines are for primary's and secondary's contribution, respectively. Unreddened contributions of primary and secondary are not shown for clarity. Notice that having a bigger radius and a hotter effective temperature; the primary's contribution is larger than the secondary's contribution. It is not the absolute contributions, but relative contributions of components are given in Table\,2 (columns 6-12). The relative contribution of a primary is computed as:

\begin{equation}
    Primary's\,cont. = \frac{f_\xi(pri)}{f_\xi(pri)+f_\xi(sec)},
	\label{eq:pri_contrib}
\end{equation}
where $\xi$ represents one of the passbands $B$, $V$, $G$, $G_{BP}$ and $G_{RP}$.  Then, the secondary's relative contribution is just one minus the primary's contribution.

\subsection{Multiband standard $BC_\xi$ by the new method}

The new method provides not only component contributions but also provides the amount of dimming due to interstellar extinction in magnitude scale ($A_\xi$) which is needed when computing absolute magnitude ($M_\xi$) of a component from its apparent magnitude $\xi$,

\begin{equation}
    M_\xi = \xi + 5log\varpi +5 - A_\xi,
	\label{eq:abs_mag}
\end{equation}
where $\varpi$ and $A_\xi$ are trigonometric parallax in arcseconds and interstellar extinction in magnitudes for the passband $\xi$, respectively. It would be clear according to Fig.~2 that

\begin{equation}
    A_\xi = 2.5log\frac{\int_{0}^{\infty} S_\lambda(\xi)f_\lambda^0(system) d\lambda}{\int_{0}^{\infty} S_\lambda(\xi)f_\lambda(system) d\lambda},
	\label{eq:A_dimming}
\end{equation}
where $f_\lambda^0$(system) is the unreddened SED of the system. The profile after convolution of $f_\lambda^0$(system) by the $G$ filter is shown as the dashed vertical profile on the right in Fig.~2. $f_\lambda$ (system) is the reddened SED of the system. The profile after convolution of $f_\lambda$(system) by the $G$ filter is shown as the dotted vertical profile in Fig.~2. 

The dimming ($A_\xi$) in each of the photometric bands $B$, $V$, $G$, $G_{BP}$, and $G_{RP}$ are calculated according to Eq.(8) and they are listed in Table\,3 together with the other parameters needed for computing absolute magnitudes of the components. The first three columns are the same as in Table\,2 (order, name, and p or s). Parallax and relative error of parallax are in the 4$^{th}$ and 5$^{th}$ columns. $L$ of the components according to Stefan-Boltzmann law in Solar and SI units and its associated relative error propagated from the uncertainties of radius and $T_{eff}$ are in the 6$^{th}$, 7$^{th}$, and 8$^{th}$ columns. After bolometric absolute magnitudes (column 9), which are computed directly from $L$, using the relation suggested by IAU General Assembly Resolution B2, hereafter (IAU 2015 GAR B2) 

\begin{equation}
    M_{bol} = -2.5logL+C_{bol},
	\label{eq:M_bol}
\end{equation}
where $C_{bol}=71.197 425 ...$  if $L$ uses SI units, $C_{bol}= 88.697 425 ...$, and if $L$ uses cgs units (IAU 2015 GAR B2, Eker et al. (2021a)), the interstellar extinctions (dimming) in $V$, $B$, $G$, $G_{BP}$, and $G_{RP}$ passbands are given in the 11th, 12th, 13th, 14th, and 15th columns. The rest of the columns of Table\,3 is reserved for the absolute magnitudes of components in $V$, $B$, $G$, $G_{BP}$, and $G_{RP}$ bands and their associated errors. Eventually, the multiband standard $BC$ values of the component stars according to the basic definition $BC_{\xi} = M_{bol} - M_\xi$  (multiband form of Eq.(3)) are listed in Table\,4, where the columns are self-explanatory; order, name, p or s, $BC_B$ and its error, $BC_V$ and its error, $BC_G$ and its error, $BC_{G_{BP}}$ and its error, $BC_{G_{RP}}$ and its error.


\begin{figure*}
	\includegraphics[width=12cm]{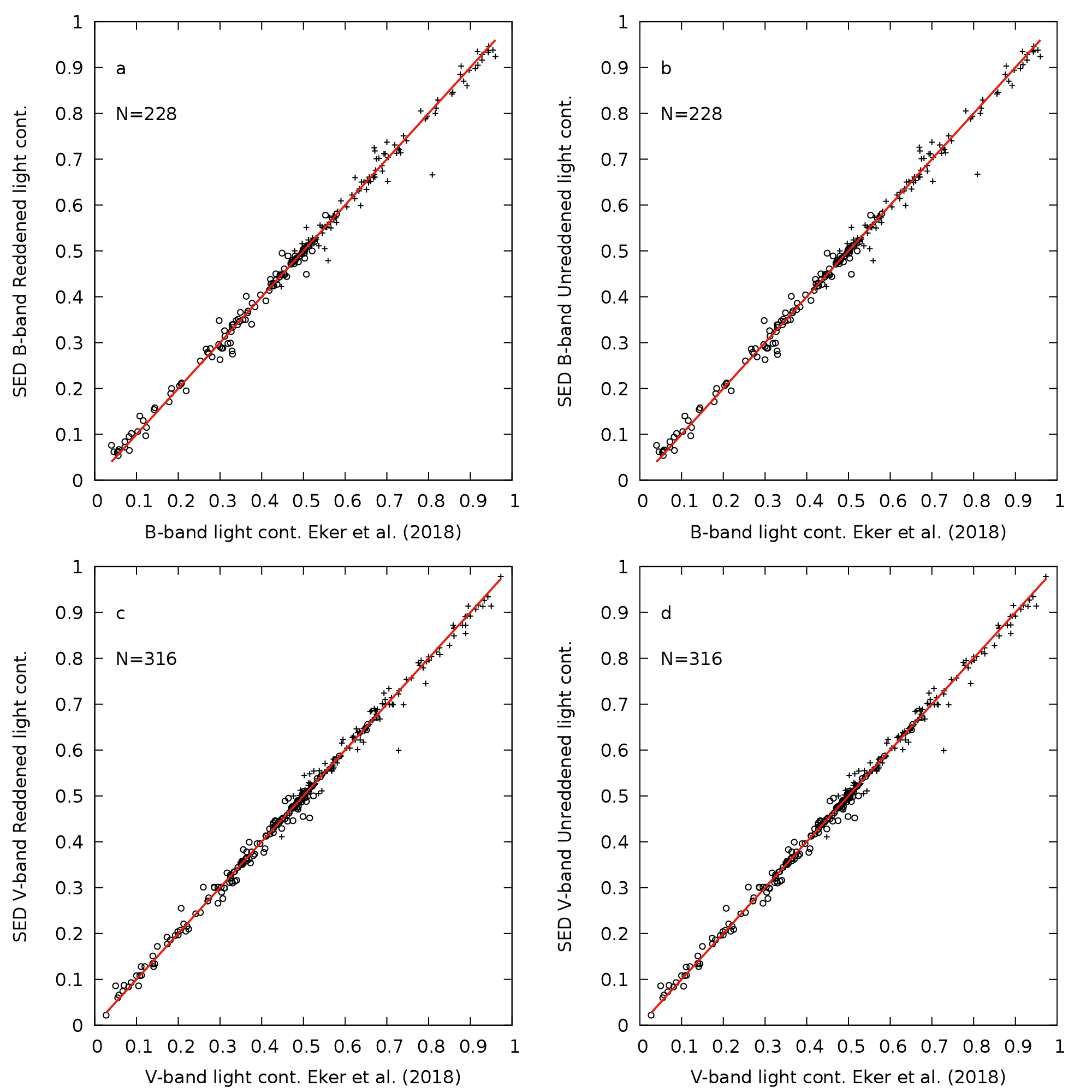}
    \caption{Comparison of $B$ and $V$ passband calculated primary (+) and secondary (o) component light contributions with those listed in Eker et al. (2020). a and c are from reddened SED, b and d are from unreddened SED. Solid line refers to $y=x$. Note that the number of data points ($N$) are not the same, this is because some systems have only a $V$-band light ratio predicted from a $V$-band light curve while the rest have both $B$ and $V$-band light ratios from $B$ and $V$-band light curves}.
    \label{fig:light_factors}
\end{figure*}

\subsection{Testing the new method}

Even if a zero-point error is absent besides the propagated errors originating from the random observational uncertainties, consequence errors would also be appended to a computed $BC$ if it is calculated directly from Eq.(1) with a simplified SED. The consequence errors are defined here to indicate errors in a computed $BC$  if it is predicted according to Eq.(1) where the SED of the component is not its observed spectrum with a sufficient resolution but a spectrum represented by a Planck function with $T_{eff}$ of the component.   The consequence errors are expected because simplifications introduced by Planck functions make some prominent spectral features lost, thus the computed $BC$ would be affected. Existence of a consequence or a zero-point error is sufficient to make a calculated $BC$ non-standard.

Zero-point errors are avoided if one uses Eqs. (3) and (9), as it was claimed by IAU 2015 GAR B2 and Eker et al. (2021a), when computing $BC$ of a component direcly from $BC_V = M_{Bol}$ - $M_V$. For the consequence errors, however, we claim: Unlike Eq.(1) with a simplified SED excluding certain prominent spectral features together with spectral lines, the method in this study, which uses Eq.(3) rather than Eq.(1), does not exclude any of the spectral features and lines despite using a simplified SED. This is because the total effect of all prominent features and lines on a spectrum is automatically included in through $M_V$ in Eq.(3) where a simplified SED is needed indirectly only for estimating relative light contributions of binary components from which $M_{Bol}$ and $M_V$ of the components are predicted.

Nevertheless, a test is necessary to make sure if the simplified SED provides reliable light contributions of the components. Fig.~3 compares fractional component contributions predicted in this study to a limited number of fractional light contributions  in $B$ and $V$ passbands from the eclipsing light curves collected by Eker et al. (2020) for producing $BC_V$ and $BC_V$ - $T_{eff}$ relation obtained from DDEB. The one-to-one correlation of almost all data is very clear. Not only component contributions from the reddened SED but also the unreddened SED of this study almost perfectly confirm the component contributions (columns 6-12, Table~2) obtained from the eclipsing binary light curves.

Fig.~3 confirms that even if light curve of an eclipsing binary system is highly reddened because of interstellar extinction, the light ratio of components or component contributions predicted from light curve solutions are the same as the light contributions predicted by the new method using the reddened and unreddened SED of the system.

\section{Calibrations of multiband $BC$ - Temperature Relations}

Once $BC$s according to Eq.(3) are available (Table\,4), then it is straightforward to calibrate $BC$-$T_{eff}$ relations using $T_{eff}$  of the component stars. The least-squares method is used to obtain the best-fitting curve of a calibrated $BC$-$T_{eff}$. Fig.~4 shows the empirical standard $BC$ values computed in this study and the best-fitting fourth-degree polynomials together with $1 \sigma$ deviations below each panel for $G$, $G_{BP}$ and $G_{RP}$. Empirical standard $BC$-$T_{eff}$ relations for $V$ and $B$ passbands of Johnson photometry are also produced for comparing $BC$-$T_{eff}$ of the $V$ band by Eker et al. (2020).  Fig.~5 shows the $BC$-$T_{eff}$ curves of $B$ and $V$ bands.

\begin{figure}
\begin{subfigure}{\columnwidth}
  \centering
  \includegraphics[width=0.6\columnwidth]{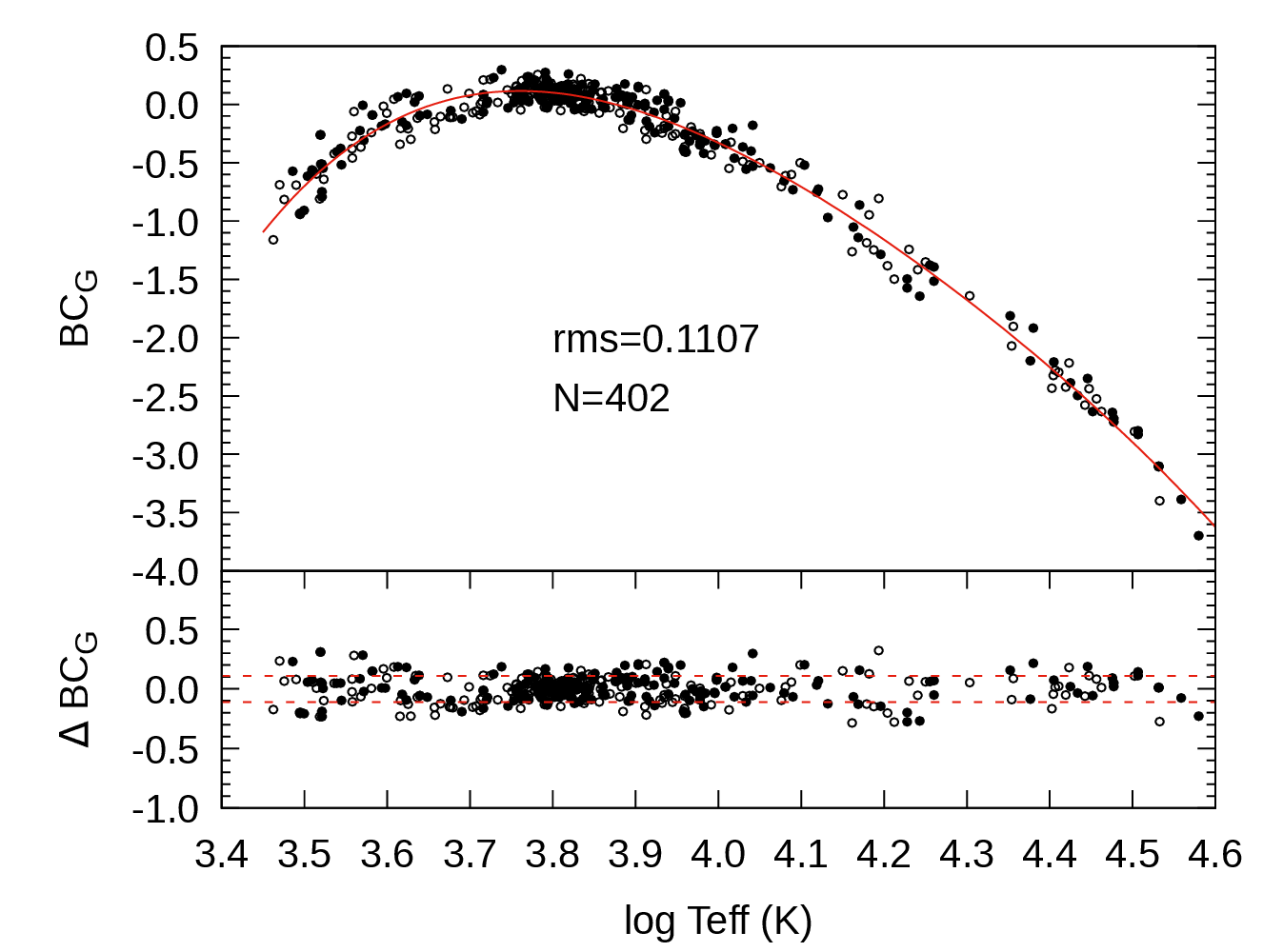}  
\end{subfigure}
\newline
\begin{subfigure}{\columnwidth}
  \centering
  \includegraphics[width=0.6\columnwidth]{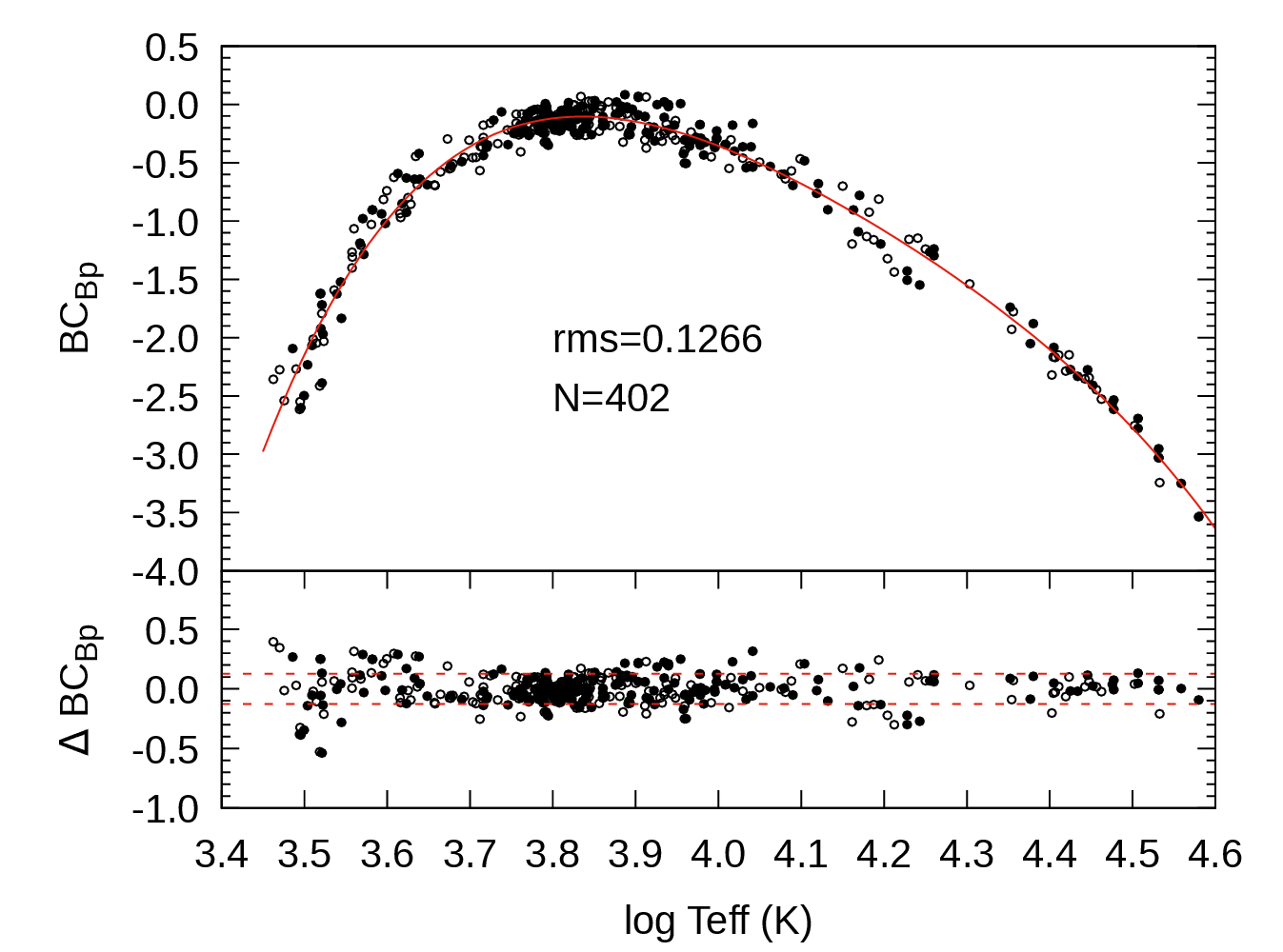}
\end{subfigure}
\newline
\begin{subfigure}{\columnwidth}
  \centering
  \includegraphics[width=0.6\columnwidth]{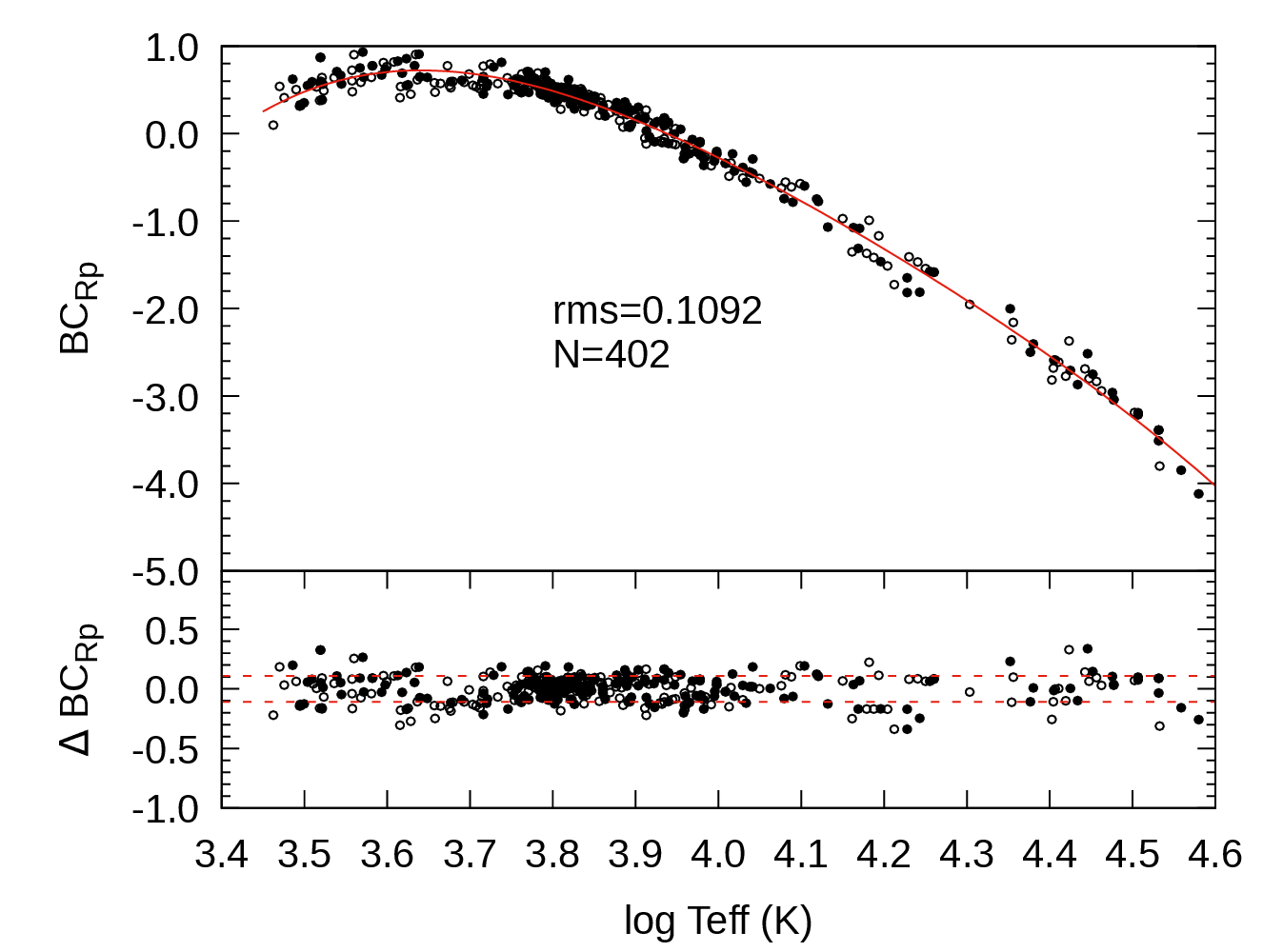}
\end{subfigure}
\caption{Empirical Standard $BC-T_{eff}$  relation for $G$ (\textit{top}), $G_{BP}$ (\textit{middle}) and ${G_{RP}}$ passbands (\textit{bottom}). The best-fitting curves (solid lines), the standard (RMS) and  $1\sigma$ deviations (dashed lines) are indicated. N is the total number of standard $BC$ used in the fit. Filled and empty circles refers to primary and secondary components, respectively.}
\label{fig:BC_G_Bp_Rp}
\end{figure}

\begin{figure}
\begin{subfigure}{\columnwidth}
  \centering
  \includegraphics[width=0.6\columnwidth]{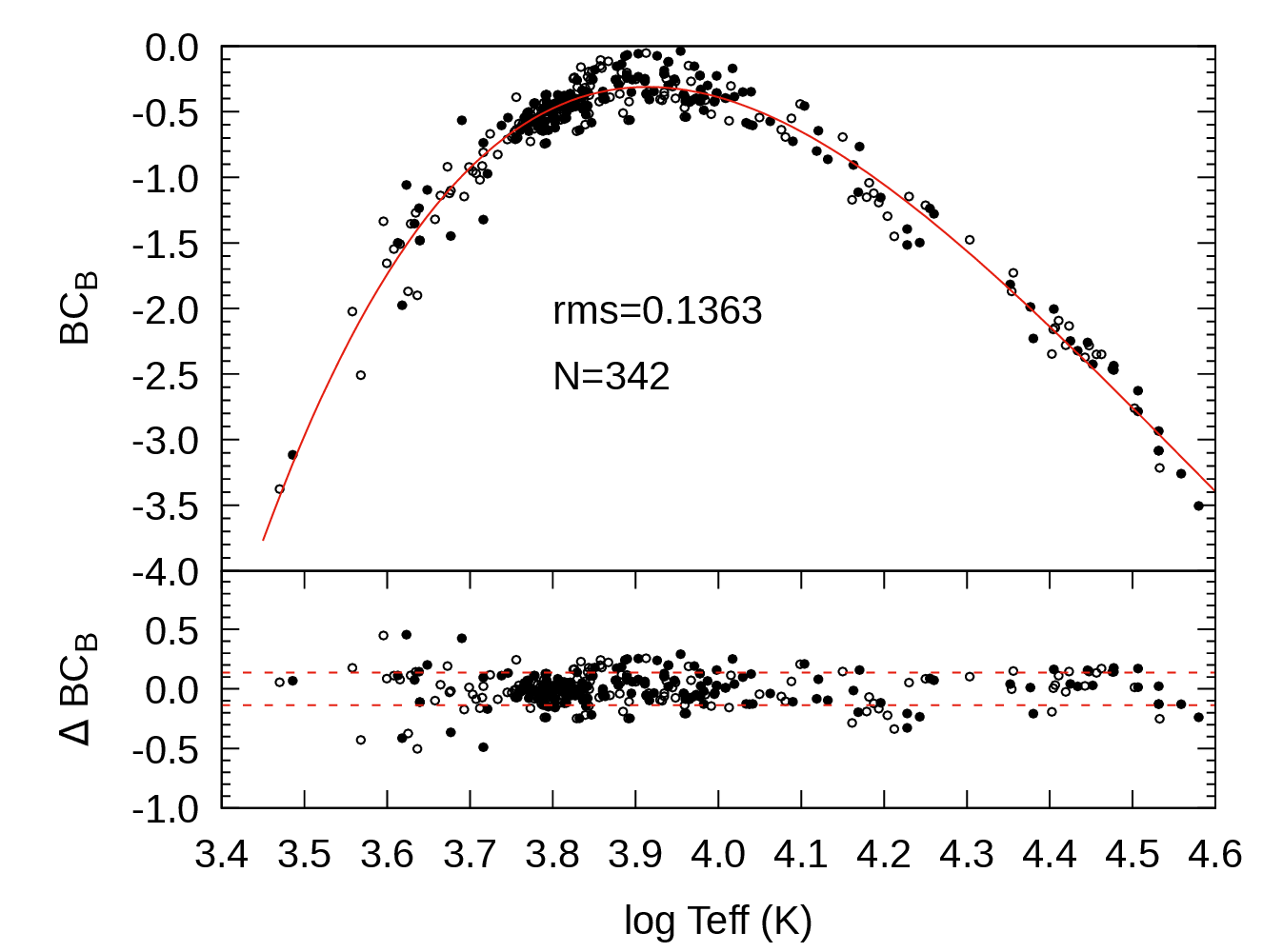}  
\end{subfigure}
\newline
\begin{subfigure}{\columnwidth}
  \centering
  \includegraphics[width=0.6\columnwidth]{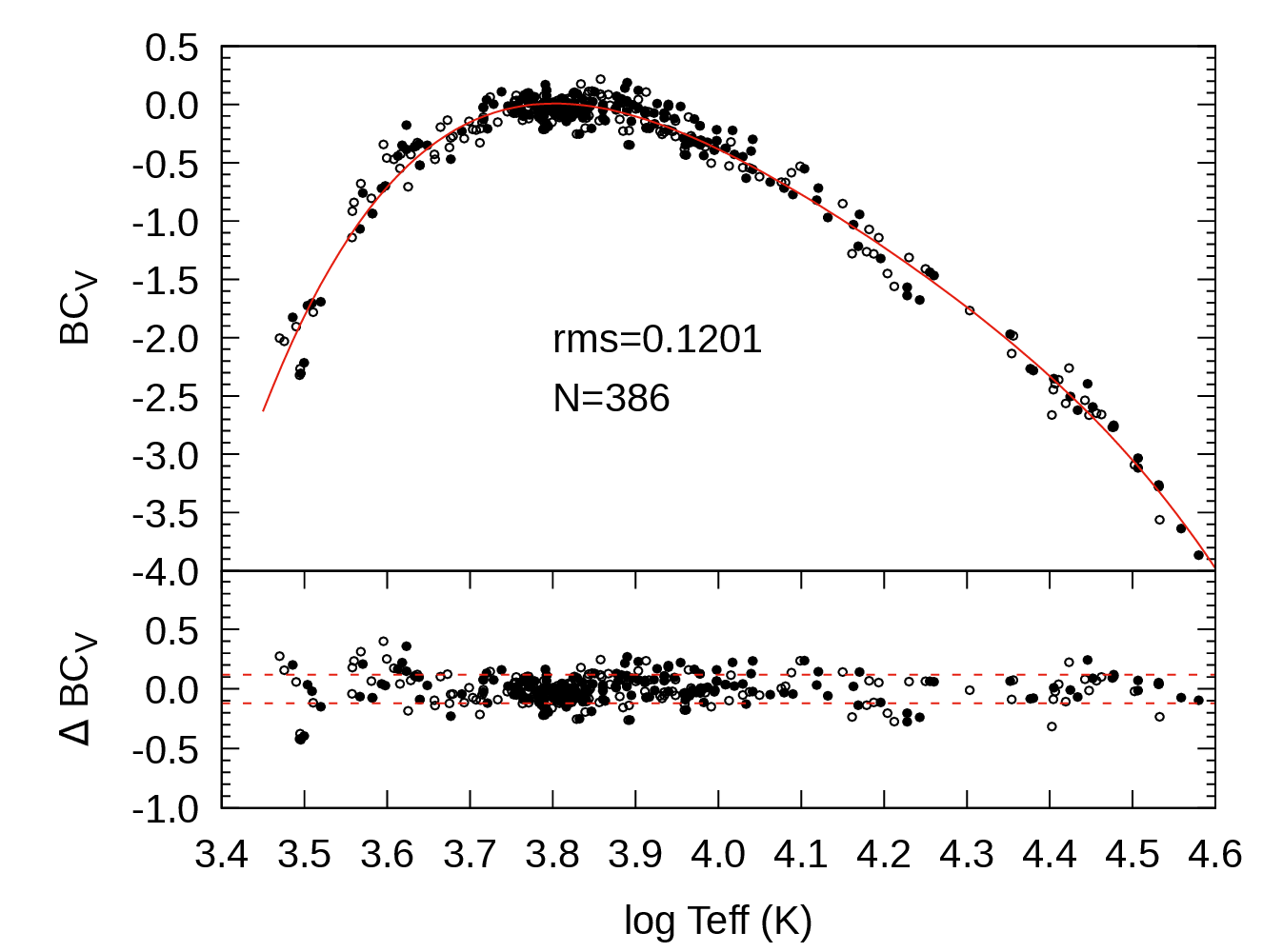}
\end{subfigure}
\caption{Empirical Standard $BC-T_{eff}$  relation for $B$ (\textit{top}) and $V$ passbands (\textit{bottom}). The best-fitting curves (solid lines), the standard (RMS) and  $1\sigma$ deviations (dashed lines) are indicated. N is the total number of standard $BC$. Filled and empty circles refer to primary and secondary components, respectively.}
\label{fig:BC_V_B}
\end{figure}

Comparing component light contributions produced in this study to the ones from light curve solutions as in Figure 3 is a preliminary test of the new method if reliable component contributions are produced or not. Having consistent $BC$-$T_{eff}$ relations in all the passbands $B$, $V$, $G$, $G_{BP}$ and $G_{RP}$ is the second test of the new method, which is successful again. Successfully produced $BC$-$T_{eff}$ relations confirm not only the validity of component light contributions but also confirm the reliability of dimming ($A_\xi$) provided using the new method. 

Coefficients and uncertainties for the $BC$-$T_{eff}$ functions from the least-square method are listed in Table\,5 where the columns are for the photometric bands used in this study including a comparison column (column 4) of Eker et al. (2020). The rows are for the coefficients of the fitting polynomial, where associated errors are indicated by $\pm$ just below the value of a coefficient. The lower part of the table compares standard deviation (RMS), correlation ($R^2$), and the standard $BC$ of a main-sequence star with $T_{eff}$ = 5772 K (a solar twin). Maximum $BC$ and corresponding effective temperatures are given below the absolute and apparent magnitudes. The lowest part of the table is for indicating the range of the positive $BC$ values if exist. $T_1$ and $T_2$ are the two temperatures, which make $BC = 0$. The $BC$ values between $T_1$ and $T_2$ are all positive else negative. If $T_1$ and $T_2$ are not given, then all $BC$ values are all negative or all positive. If only $T_2$ is given, then $BC$ is positive for $T_{eff} < T_2$, $BC = 0$ if $T_{eff} = T_2$ else $BC$ is negative.

Table\,5 indicates $BC$-$T_{eff}$ curve of this study has a smaller $RMS$ and higher correlation ($R^2$ value) compared to the $RMS$ and correlation coefficient obtained in Eker et al. (2020). Fig.\,6a  compares the $BC_V$-$T_{eff}$ curve of this study to the $BC_V$-$T_{eff}$ curves by Flower (1996), Eker et al. (2020), Mamajec (2021) and Cox (2000).  $BC$s of Cox (2000) are all overestimated (more negative) compared to other $BC$ displayed in the figure except Flower's (1996) towards the coolest part of the temperature scale. The $BC$ values of Mamajec (2021) appear overestimated (more negative) compared to the $BC$ values of Eker et al. (2020) for the stars hotter than 10000 K, also for a very small range of coolest stars but the rest appears as the same. The $BC$ values of Flower (1996) deviate from the curve of Eker et al. (2020) and Mamajec (2021) as being bigger absolute value towards the coolest temperatures. Nevertheless, except for a limited temperature range near 10000 K, all other $BC$ values appear overestimated when compared to the standard $BC$s of this study.  

Several tables exist which provide $BC$ in different photometric passbands including Gaia photometry (Martins and Plez 2006; Jordi et al. 2010 and Pedersen et al. 2020) as a function of atmospheric parameters $T_{eff}$, log $g$, $\zeta$ or $[Fe/H]$ and are commonly used to derive isochrones in different colours; Girardi et al. (2002) is one example. However, only Andrae et al. (2018) combined $BC_G$ of various atmospheric parameters (log $g$, $\zeta$, metallicity) and produced a single $BC_G$ - $T_{eff}$ relation for the main-sequence stars. Fig.\,6b compares empirical $BC$-$T_{eff}$ curve in the $G$ passband of this study to the $G$ band $BC$-$T_{eff}$  curve of Andrae et al. (2018), which appears overestimating $G$ band $BC$ for the stars cooler than 6500 K. The $BC$s of Andrae et al. (2018) cover a temperature range $3300 - 8000 K$. Other $BC$-$T_{eff}$ relations representing a specific $\zeta$ or $[Fe/H]$ predicted from model atmospheres are not suitable for comparison to the empirical relations of this study.

\begin{figure}
\centering
\includegraphics[width=\columnwidth]{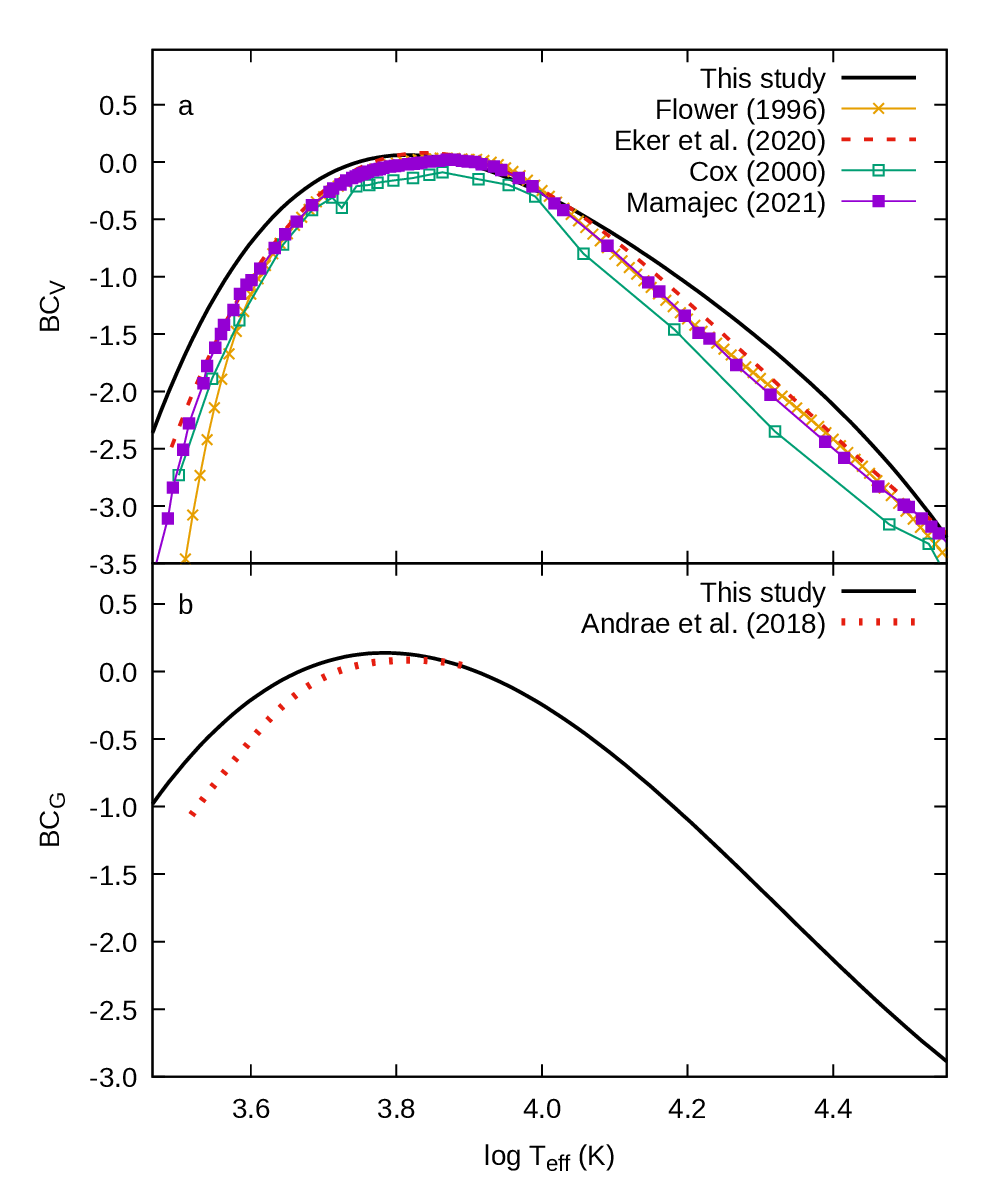}  
\caption{Comparing $BC$-$T_{eff}$ relations of this study to others for $V$ (a) and $G$-bands (b).}
\label{fig:BC_comparison}
\end{figure}

\begin{figure}
	\includegraphics[width=\columnwidth]{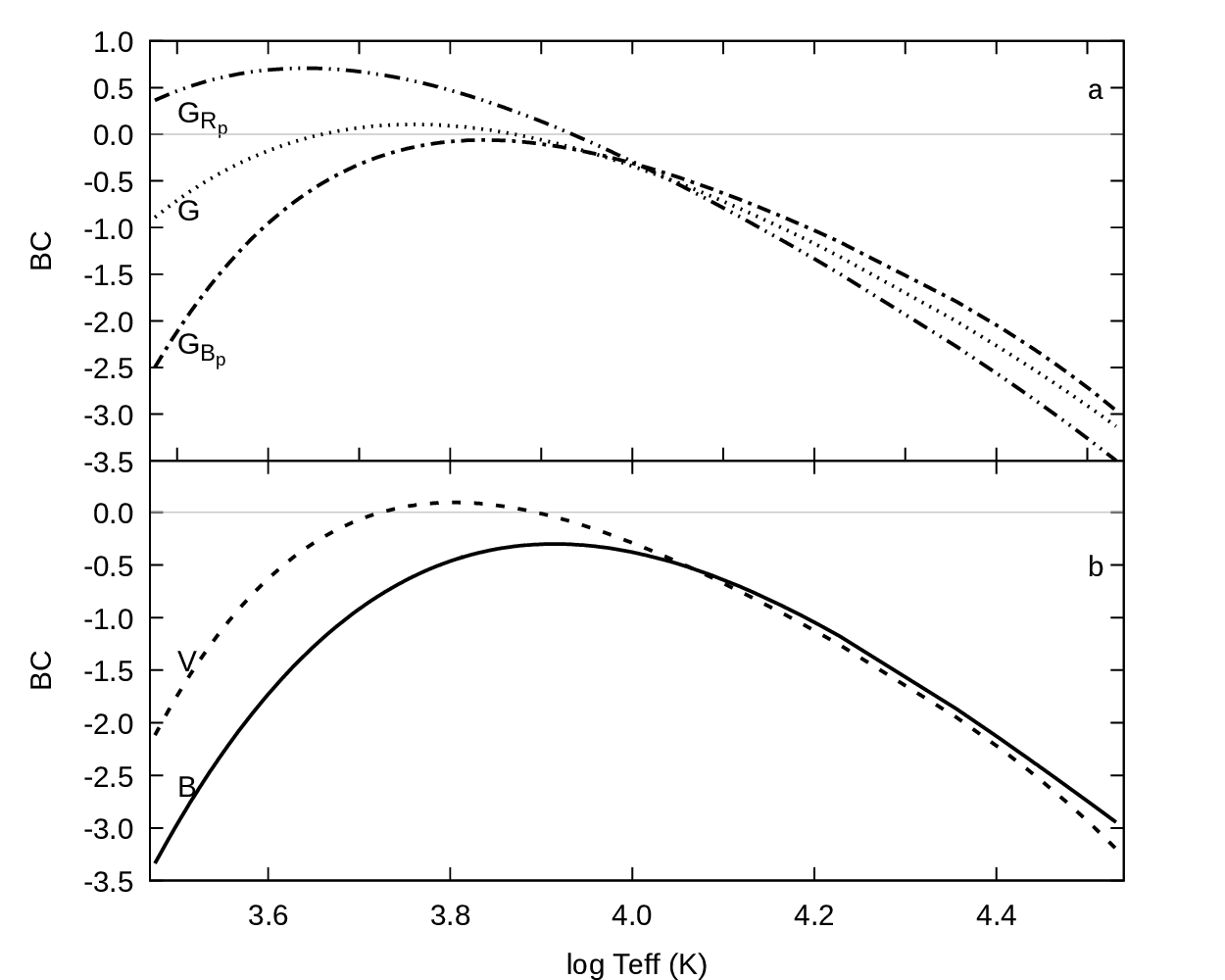}
    \caption{Standard $BC-T_{eff}$ relations in all passbands indicated on the left of each line.}
    \label{fig:all_BC_func} 
\end{figure}

Empirical $BC$-$T_{eff}$ relations are not like fundamental relations; e.g. Stefan-Boltzmann law. They are rather statistical relations like classical $MLR$ (Eker et al. 2018, 2021b). They could be used only under correct conditions set statistically. Because of stellar evolution (Clayton 1968), there could be many stars with the same $M$ but various $L$ due to different ages, different chemical compositions and internal mixing. Therefore, in reality, there is no unique luminosity ($L$) for a typical main-sequence star of a given mass ($M$). However, with a large uncertainty covering $L$ values of all main-sequence stars, the classical $MLR$'s may provide only a mean $L$ for a typical main-sequence star of a given $M$. The same is true that the $BC$-$T_{eff}$ relations may provide only a mean or a typical $BC$ for a typical main-sequence star of a given typical $T_{eff}$. 

Therefore, it is an astrophysical interest to have a table indicating typical $T_{eff}$  and typical $BC$ of main-sequence stars.  Table\,6 is the extension of an original table given by Eker et al. (2018) and Eker et al. (2020), where typical fundamental astrophysical parameters of main-sequence stars are presented with $BC$, $(B-V)_0$ and $M_V$  as a function of typical effective temperatures associated with the spectral types. Table\,6, here, is kept short to contain only spectral types, typical $T_{eff}$ and mean $BC$s  and intrinsic colors of nearby Galactic main-sequence stars with 0.008 $<$ Z $<$  0.040  for the bands $B$, $V$, $G$, $G_{BP}$ and $G_{RP}$. 

An interesting feature of multiband $BC$-$T_{eff}$ relations would be revealed if the $BC$ values on Table\,6 are plotted on a single frame (Fig.\,7) as a function of effective temperature. It is not a surprise to see $BC$-$T_{eff}$ relations of Gaia pass-bands (Fig.\,7a)  cut each other at a common point near $\sim$10000 K (log $T_{eff}=$ 4). This must be due to using the Vega system of magnitudes, which assumes zero intrinsic colours for a hypothetical star of $A0V$ spectral type with an effective temperature 10000 K. The Vega system of magnitudes uses Vega as the standard star and assumes that all intrinsic colours of Vega are equal to 0. While Vega is classified as an A0V star, its effective temperature is ~9600K while its apparent brightness is V = 0.03 mag. The crossing point of the $BC$-$T_{eff}$ relations of Johnson $B$, $V$  occurring erroneously at a higher temperature  (Fig.\,7b) indicates lower  precision and accuracy of Johnson magnitudes with respect to Gaia magnitudes used in this study. This is not a surprise because the total $B$ and $V$ brightness of DDEB systems are collected from various sources (see Table 1) unlike Gaia magnitudes, which are taken from a single source, EDR3. The maximum $BC$ and/or ranges of positive $BC$ values given in Table\,5 appear to be visualised in Fig.\,7.


\section{How to improve standard luminosities}

Improving the accuracy of standard $L$ could be achieved in two ways. One way is to improve the accuracy of existing standard $BC - T_{eff}$ relations and the second way is to increase the diversity of the standard $BC - T_{eff}$ relations. The former way is already achieved by calibrating multiband standard $BC - T_{eff}$ relations using the most accurate stellar astrophysical data available. The improved relations and their related statistics are given in Table\,5. A user would produce a standard $L$ of a star almost twice more accurate now if he/she uses $BC_V - T_{eff}$ relation of this study rather than $BC_V - T_{eff}$ relation of Eker et al. (2020) if propagated uncertainty of $M_V$ of the star is dominated by relative parallax error $\left(\frac{\sigma_\varpi}{\varpi}\right)$ and if it is $\frac{\sigma_\varpi}{\varpi}$ $\ll$ 5.5 per cent. This is because the standard deviation of the $BC_V - T_{eff}$ curve of this study is reduced to $SD=$ 0.12 mag implying 11.05 per cent for $\Delta L/L$ if $M_V$ is errorless, while it was $SD=$ 0.215 mag (see Table\,5) implying 19.8 per cent for $\Delta L/L$ correspondingly. Otherwise (if $\frac{\sigma_\varpi}{\varpi} \gg$ 5.5 per cent), the uncertainty of computed $L$ would naturally be dominated by the parallax error. Then, the propagated uncertainty of the standard $L$ would have been much bigger  ($\Delta L/L$ $\gg$ 11.05 per cent).

\begin{figure}
\begin{subfigure}[b]{.48\textwidth}
  \centering
  \includegraphics[width=\textwidth]{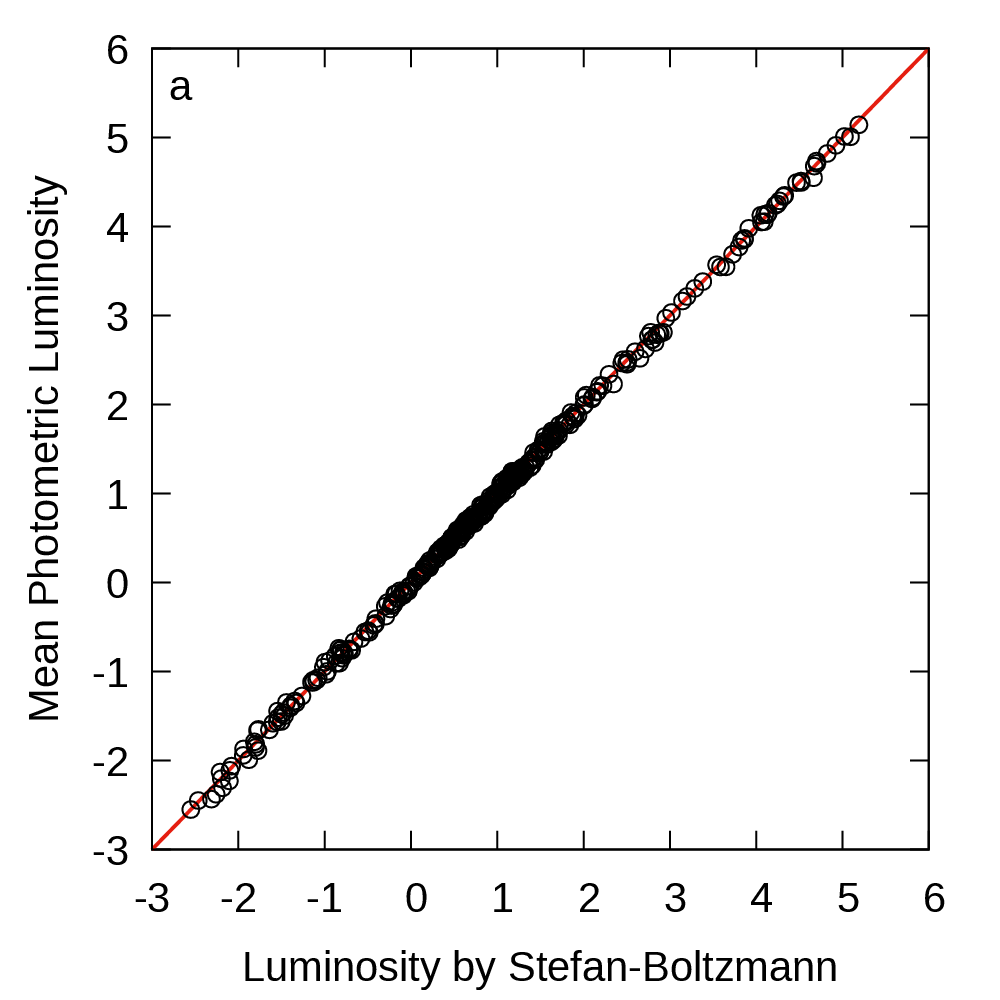}
\end{subfigure}
\hfill
\begin{subfigure}[b]{.48\textwidth}
  \centering
  \includegraphics[width=\textwidth]{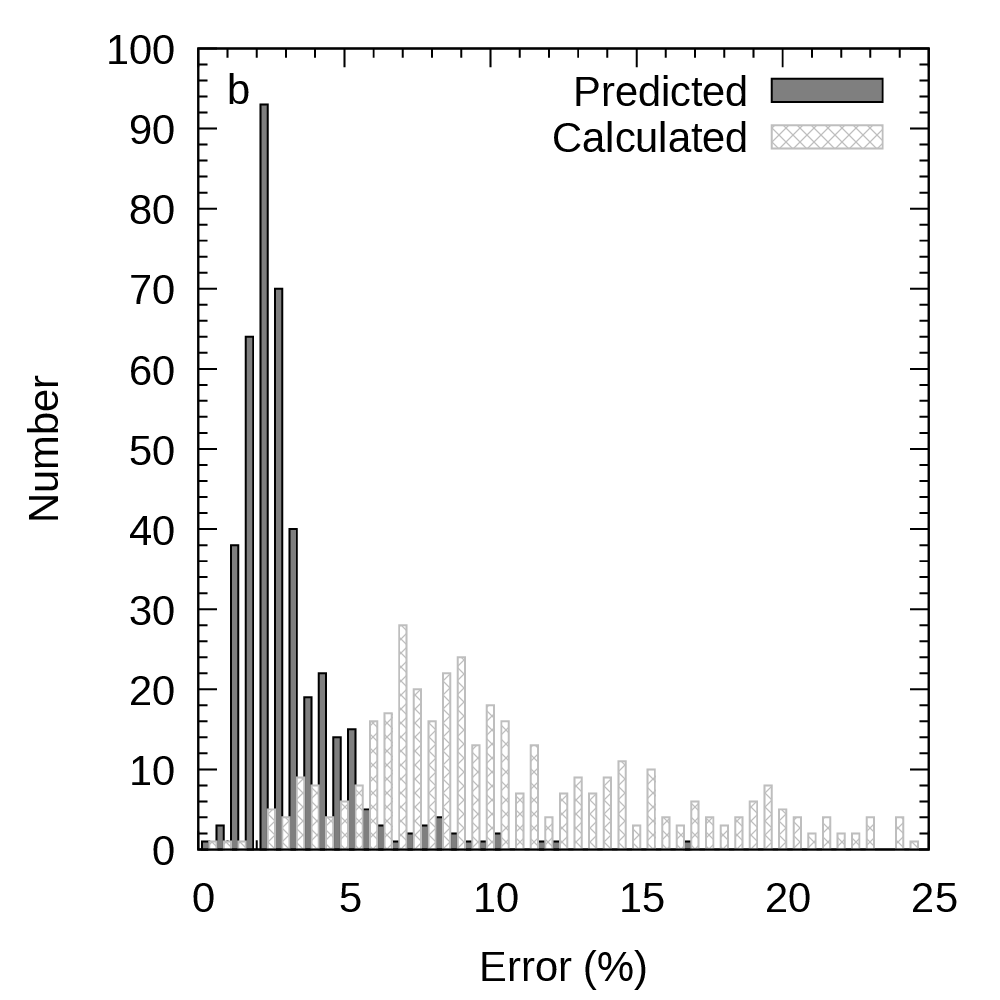}
\end{subfigure}
\caption{a) Comparing predicted (from photometry) and calculated (from $R$ and $T_{eff}$) $L$ of the sample stars. b) Histogram distribution of the uncertainties associated with predicted (dark) $L$ is compared to the histogram distribution of the uncertainties associated with calculated (grey) $L$.}
\label{fig:luminosity_errors}
\end{figure}

 The standard deviation of a $BC_V - T_{eff}$ curve determines the limiting accuracy of the standard $L$ (Eker et al. 2021b). Therefore, Table\,5 implies that a user can get a standard $L$ with an error as small as 12.5 per cent, 11.05 per cent, 10.2 per cent, 11.7 per cent, or 10.05 per cent correspondingly if he/she uses one of the $BC - T_{eff}$ relations at $B$, $V$, $G$, $G_{BP}$, or $G_{RP}$ photometric bands, respectively. However, if $\frac{\sigma_\varpi}{\varpi}$ $\gg$ 6.3 per cent, 5.5 per cent, 5.1 per cent, 5.8 per cent or 5.02 per cent, in accord with the photometric bands, the standard error of the L is bigger. At the limit, when the uncertainty of $M_V$ dominates over the uncertainty of $BC$ and the distance errors dominates over brightness and extinction errors, it becomes twice the relative error of the parallax according to the formulation of Eker et al. (2021b).

Providing independent $BC - T_{eff}$ relations at $B$, $V$, $G$, $G_{BP}$ or $G_{RP}$ passbands determined independently by the least-squares method from the independent observational photometric and astrometric data of DDEB, which are known to provide the most accurate stellar astrophysical parameters, Table\,5 allowed us to investigate the second way of improving the accuracy of a standard $L$ as well. One does not need to calculate the actual value of the standard $L$ for speculating about its relative uncertainty ($\Delta L/L$) if it comes from a single $BC - T_{eff}$ relation. Similar speculations, however, are not possible in the second way of improving. Calculation of actual $L$ from each of the $BC - T_{eff}$ relations is needed.

Then, there are many standard $L$ values for a star representing each of the photometric bands, like many independent measurements of a quantity. However, we prefer not to calculate many different standard $L$ for a star and then take an average. Instead, we prefer to predict five different $M_{Bol}$ together with their associated uncertainty propagated from the uncertainty of $M_\xi$, and the uncertainty of $BC_\xi$ (Table\,5) first. Then combine them according to 
\begin{equation}
M_{bol} = \frac{1}{N}\sum_i^N M_{bol,i}
\label{eq:Mbol}
\end{equation}
to get a single $M_{bol}$ for a star, where $M_{bol,i} = M_i + BC_i$, provided with $i=$ $B$, $V$, $G$, $G_{BP}$, and $G_{RP}$ passbands. N is a number between 2 (if $M_{Bol}$ is predicted from $B$, $V$ only) and 5 (if $M_{Bol}$ is predicted from all the passbands) because some systems do not have total apparent brightness measured at certain photometric bands. At last, the most improved standard $L$ of a star is predicted directly from its mean $M_{bol}$ value according to Eq.(9). To estimate its relative uncertainty ($\Delta L/L$), we have preferred to calculate a standard error for the $M_{Bol}$ first and then propagate it to the standard $L$. A similar approach of using an average bolometric magnitude calculated based on several different photometric passbands to derive $L$ was used by Pedersen et al. (2020) to derive $L$ of $B$ dwarfs, but for apparent instead of absolute bolometric magnitudes.     

Predicted (from photometry) and calculated (from $R$ and $T_{eff}$) $L$ of the sample stars in this study are compared in Fig.~8a. A very high correlation ($R^2>0.999$) between the predicted and calculated luminosities is seen clearly. Fig.~8b compares histogram distributions of their uncertainties. Uncertainties of the predicted $L$ have a sharp well-defined peak at 2 per cent with a smaller dispersion, while the uncertainties of the calculated $L$ have a fussy peak at 8 per cent with a much wider dispersion. Fig.~8 shows that a  prominent improvement in predicting a standard L of a star occurs if all the existing independent $BC - T_{eff}$ relations are used according to the method introduced in this study. The improvement is remarkable and real (not speculative) that there is a method, now, which could provide a standard luminosity of a star more accurate than the classical method using observed radii and effective temperatures according to the Stefan-Boltzmann law.

We summarise the data produced by the method used in this study in Table\,7. The columns are self-explanatory to show order, system name, the component (primary or secondary) in the first three columns. Then next ten columns are reserved for the predicted $M_{bol}$ values from its definition ($M_{bol} = M_\xi + BC_\xi$) and associated propagated errors at $B$, $V$, $G$, $G_{BP}$, and $G_{RP}$ passbands. Then, the combined (mean) $M_{bol}$ and its standard error are illustrated in columns 14 and 15. The logarithm of the predicted $L$ in solar units and its relative uncertainty are listed in columns 16 and 17. The last two columns are for the calculated $L$ in solar units and their relative uncertainty. Fig.~8 is produced from the last four columns of Table\,7. Therefore, the last four columns are ideal for a reader who is interested in comparing actual numerical values of the predicted and computed $L$ and to see how small the relative errors of the predicted $L$ are compared to the errors of the computed $L$.                 

\section{Discussions}

\subsection{The Sun and a solar twin for testing}

In a first thought, one may think the Sun is not a good candidate for testing how good its luminosity would be predicted according to the method described in this study because it is the reference star that IAU 2015 GAR B2 used to determine the zero point constant of the bolometric magnitude scale $C_{Bol}=$ 71.197 425 ... mag from $L=$ 3.8275($\pm$0.0014)  $x10^{26}$ W and $M_{Bol,\odot}\cong$ 4.739 996...mag.   The $BC$ values (--0.600, 0.069, 0.106, --0.134, 0.567 respectively at $B$, $V$, $G$, $G_{BP}$ and $G_{RP}$) given in Table\,5, which are marked with $\odot$ symbol, should not be understood as the $BC$ of the Sun. They are the predicted $BC$ values for a typical main-sequence star having $T_{eff}=$ 5772 K. Consequently, absolute and apparent magnitudes given in Table\,5 just below those $BC$ values are typical absolute and apparent magnitudes if this star is replaced as the Sun. Now the question: is it possible to estimate the luminosity of the Sun just from its effective temperature ($T_{eff,\odot}=$ 5772 K) using the $BC$ given in Table\,5? To proceed towards this aim, one needs apparent magnitudes and distance as in the other stars used in this study.

It is for sure, one cannot find measured Gaia apparent magnitudes for the Sun despite it is possible one could calculate what it would be using a measured/calculated solar composite spectrum (Willmer, 2018) while it is possible for one to run into a value $V_\odot=$ --26.76 $\pm$ 0.03 mag (Torres, 2010).  Same and slightly different (in the second digit after the decimal) values seem to be preferred by various authors (see the references in Torres (2010) and Eker et al. (2021b)). Astronomers handbook "Allen's Astrophysical Quantities" (Cox, 2000) gives it 0.01 mag dimmer, where apparent and absolute magnitudes of the Sun in $U$, $B$, $V$, $R$, $I$ and $K$ bands could be found. Taking $(B-V)_\odot=$ 0.65 mag, one can start from $V_\odot=$ --26.76 mag and $B_\odot=$ --26.11 mag and continue predicting the solar luminosity together with its uncertainty by applying the method described in this study. Here, we assume $B$ and $V$ apparent magnitudes of the Sun have the same uncertainty ($\pm0.03$ mag).

No interstellar extinction for the Sun and because its distance is also known with a great precision compared to other stars; only the observational uncertainty $\Delta V \approx \Delta B \approx 0.03$ propagates to the solar absolute magnitudes: $M_{V,\odot}=$ 4.812 $\pm$ 0.03 mag, and $M_{B,\odot}=$ 5.462 $\pm$ 0.03 mag. Using the $BC$ and RMS values (Table\,5) for a typical main-sequence star with $T_{eff}=$ 5772 K, the predicted bolometric magnitudes of the Sun would be  $M_{Bol,\odot}$ (1)= 4.881 $\pm$ 0.136 mag and $M_{Bol,\odot}$ (2)= 4.862 $\pm$ 0.12 mag, respectively from its $V$ and $B$ magnitudes. After combining them by taking a simple average $M_{Bol,\odot}$ = 4.872 mag. The differences from the mean indicate an $\pm$ 0.095 mag uncertainty. At last, using Eq.(9), the predicted solar luminosity is  $L_P (\odot)= 3.39\times10^{26}$ W, and the relative uncertainty is $\Delta L/L\approx$ 8.7 per cent.

Comparing this value to the nominal Solar luminosity $L= 3.838\times10^{26}$ W, one can see how successful the method is.  A luminosity, which is about 11.7 per cent smaller than actual $L_\odot$, is predicted.  A single channel prediction from $M_{Bol,\odot}$(1) or $M_{Bol,\odot}$(2), would have given us a prediction with a relative error of 12.5 per cent or 11.05 per cent respectively.  All predicted $L$ values agree to the read $L$ within the error limits estimated.

\subsection{A solar Twin for a test}

The primary of HP Aur system with a mass $M=$ 0.9543 $\pm$ 0.0041 $M_\odot$, a radius $R=$ 1.0278 $\pm$ 0.0042 $R_\odot$, and an effective temperature $T_{eff}=$ 5810 $\pm$ 120 K (Lacy et al. 2014) could be considered as a solar twin. According to the Stefan-Boltzmann law, its luminosity is $L=$ 1.084 $L_\odot= 4.162\times10^{26}$ W. Propagation of the observational uncertainties shows its relative error ($\Delta L/L$) would be about 8.302 per cent. Now the question is: Using the method of this study, how accurately would its luminosity be predicted?

Its apparent brightness' are $V=$ 11.489$\pm$0.07, $G=$ 11.283$\pm$0.003, $G_{BP}=$ 11.628$\pm$0.004, $G_{RP}=$  10.761$\pm$0.004 mag.  According to simplified SED, the primary contributes 75.7 per cent of the total light radiated by the system in $V$, 75 per cent in $G$, 76.5 per cent in $G_{BP}$ and 72.9 percent in $G_{RP}$. The simplified SED also implies $A_V=$ 0.335 mag, $A_G=$ 0.298 mag, $A_{G_{BP}}=$ 0.366 mag, and $A_{G_{RP}}=$ 0.207 mag as the interstellar dimming.   Parallax of the system is 5.2432$\pm$0.0306 mas. Consequently, its absolute magnitudes are $M_V=$ 4.753$\pm$ 0.033, $M_G=$ 4.583$\pm$0.033, $M_{G_{BP}}=$  4.860$\pm$0.042, and $M_{G_{RP}}=$ 4.152$\pm$0.024 mag. Table\,5 gives $BC_V=$ 0.072$\pm$0.12, $BC_G=$  0.106$\pm$ 0.11, $BC_{G_{BP}}=$ --0.129$\pm$0.13, and $BC_{G_{RP}}=$ 0.562$\pm$ 0.11 mag.  Here, we notice that the errors in $BC$ values are bigger than errors in fiter based absolute magnitudes. Thus, they colud be ignored. That is, $BC$ errors would be the dominant factor when calculating the uncertainty of its bolometric absolute magnitudes. This means if single pass-band used in predicting its $L$, the relative error of its $L$ ($\Delta L/L$) will not be smaller than 10 per cent.

According to Eq.(3), the $M_{Bol}$  are calculated as  4.824, 4.688, 4.73, and 4.713 mag, respectively at four photometric bands. At last, one obtains a mean $M_{Bol}=$ 4.739$\pm$0.03 mag for the solar twin, where its uncertainty is assumed to be the standard error by definition. After computing its $L$ using Eq.(9), $L= 3.831\times10^{26}$ W is found. At last its uncertainty $\pm$0.03 translates to $\Delta L/L\sim$2.5 per cent.

Predicted and calculated $L$ for the primary of HP Aur agree to each other within the error limits. The calculated luminosity from $R$ and $T_{eff}$ using the Stefan-Boltzmann law seems to be overestimated. The predicted standard $L= 3.831\times10^{26}$ W appears more reliable because of its predicted uncertainty. At last, it can be concluded that the method of computing standard L of stars using multiband photometry and $BC-T_{eff}$ relations involving SED is very successful in predicting luminosities much more accurately than the direct method using observed $R$ and $T_{eff}$. Using four $BC-T_{eff}$ curves ended up predicting a standard $L$ at least four times more accurately than in the case if one uses only one of the $BC-T_{eff}$ relations existing. 

\subsection{Standard or non-standard}

What makes $BC$s (and $BC_V$ - $T_{eff}$ relation) of this study standard? What makes BCs of Cassagrande \& VandenBerg (2018), Cox (2020), Andrae et al. (2018), and Mamajek's personal online $BC$ Table accessible on the internet\footnote{www.pas.rochester.edu/$\sim$emamajek/EEM$\_$dwarf$\_$UBVIJHK$\_$colors$\_$Teff.txt} non-standard?

In the first look, IAU 2015 GAR B2 was issued only for solving the long-lasting problem of arbitrariness attributed to the zero point of bolometric magnitudes. However, the arbitrariness of the bolometric magnitude scale is not independent of the arbitrariness of the $BC$ scale according to Eq.(3). Articles such as Cassagrande \& VandenBerg (2018), Andrae et al. (2018), which are still defending the arbitrariness of the $BC$ scale, cause confusion.

Eker et al. (2021a) have shown that fixing the zero point of the bolometric magnitude scale also fixes the zero point of the bolometric correction scale. To avoid $BC$ determinations with different zero points, Eker et al. (2021a) have defined the concept of standard $BC$. The standard $BC$ is not only for the $V$ band; the definition covers all bands of all photometric systems. Eker et al. (2021b) explained how to recognise non-standard $BC$ values.

Briefly, using Eq.(9) with a definite $C_{Bol}$ makes the computed $M_{Bol}$ unique. Since stellar absolute magnitudes at well-defined passbands of various photometric systems are also unique (absolute magnitude of a star cannot have two or more values for a specified band), the product of Eq.(3) was defined as the standard $BC$ because subtracting a unique number from another unique number is also unique. 

The nominal value of $C_{Bol} =$ 71.197 425 ... corresponds to the nominal values of $M_{Bol,\odot} =$ 4.74 mag (a rounded value. The true value is 4.739 996 ...) and $L_{\odot} = 3.828 x10^{26}$ W, thus $C_{Bol} = M_{Bol,\odot} +2.5 log L_{\odot}$ (see IAU 2015 GAR B2, Eker et al. (2021a,b)).  Consequently, using a different $C_{Bol}$ than the nominal $C_{Bol}$ in Eq.(9) or using a non-nominal value of $M_{Bol,\odot}$ or $L_{\odot}$ in the following equation:
\begin{equation}
M_{bol} = M_{bol,\odot} - 2.5log\frac{L}{L_\odot}
\label{eq:Mbol_sun}
\end{equation}
is sufficient to make a computed $BC$ non-standard. Moreover, a $BC$ is also not standard if it is computed through Eq.(1) with an arbitrary $C_2$,

Cassagrande \& VandenBerg (2018) used $M_{Bol,\odot} =$ 4.75 mag for the absolute bolometric magnitude for the Sun rather than the nominal value $M_{Bol,\odot} =$ 4.74 mag suggested by (IAU 2015 GAR B2). On the other hand, the $BC$s of Cox (2000) are also not standard because the nominal $M_{Bol,\odot} =$ 4.74 mag was used together with a non-nominal $L_{\odot} =$ $3.845 \times10^{26}$ W corresponding to a non-nominal $C_{Bol}$ (see Eker et al. 2021b). Despite using the nominal values $M_{Bol,\odot}$ or $L_{Bol,\odot}$, the $BC$ values of Andrae et al. (2018) are also not standard. This is because Andrae et al. (2018) preferred to use Eq.(1) with an assumed arbitrary $C_2$ when computing $BC$ values for the Gaia photometric passbands. A slightly different case seems to have occurred on the $BC$ tables given by Cox (2000), who took the arbitrariness of $C_2$ granted. Consequently, to be consistent with the paradigm, "bolometric corrections must be negative" (see page 381 of Cox (2000)), the zero point of the $BC$ scale was set to make all $BC$ negative. This is the second reason why $BC$ values of Cox (2000) are not standard.

Eker et al. (2021a) have shown that the zero point constant in Eq.(1) has different values at different filters such: $C_2 = C_{Bol} - C_\xi$, where $C_\xi$ is the zero point for a passband, thus $C_2$ is also not arbitrary but has a definite value. Although $C_2$ appears like an integration constant in Eq.(1), actually it is not an integration constant or algebraic sum of integration constants required by integrals appearing in Eq.(1). It is well known that definite integrals do not take constants. Therefore, $C_2$ must be a constant imposed by the absolute photometry, such as
\begin{equation}
M_{\xi} = -2.5log L_\xi+C_\xi.
\label{eq:Mksi}
\end{equation}
Subtracting this from Eq.(9), which is imposed  by Eq.(3), gives Eq.(1), where the definite integrals are for producing the surface fluxes of the star for the bolometric: $f_{Bol}=\int^{\infty}_{0}f_\lambda d\lambda$ and for a photometric band $f_{\xi}=\int^{\infty}_{0}S_\lambda (\xi) f_\lambda d\lambda$. The definite integrals do not take constants, but the absolute photometry (Eqs 9 and 12) requires $C_2 = C_{Bol} - C_\xi$.

Since the value of $C_{Bol}$ was unknown before IAU 2015 GAR B2, and no telescope or a detector exist to observe $M_{Bol}$, it was natural to assume both $C_{Bol}$ and $C_{2}$ arbitrary. Therefore, authors such as Cox (2000), and Pecaut \& Mamajec (2013) have excuse to assume the $BC$ scale is arbitrary and then impose a personal condition to set up a private absolute scale. Cox (2000) took $BC_V = 0$ for F2 supergiants and Mamajec (2021) uses $BC_V = -0.085$ for a G2 main-sequence star in his online $BC$ table to set up a private zero point for the $BC$ scale.

Similarly, Andrae et al. (2018) also set his absolute scale by taking $BC_{V,\odot}= -0.07$ and stating "bolometric correction needs a reference point to set the absolute scale". Setting up a private absolute scale for $BC$ as done by Mamajec (2021) and Andrae et al. (2018) is not acceptable anymore since 2015. Despite IAU 2015 GAR B2, such private absolute scales do not mean recognizing the absolute $BC$ scale set by IAU 2015 GAR B2, or IAU 2015 GAR B2 is not understood properly.
Therefore, any $BC$ value which is according to a private $BC$ scale as implied by Cassagrande \& VandenBerg (2018), Cox (2000), Andrae et al. (2018), Mamajec (2021) is not standard.


\subsection{Colour-Temperature and Temperature-Colour Relations}

It is a great advantage to have already calibrated $BC$--$T_{eff}$ relations at various bands of a photometric system. This way, intrinsic colour-temperature relations would automatically be set. Flower (1996) and Eker et al. (2020) had to compute observed ($B-V$) colours of the components first from the light ratio ($l_2/l_1$) of components if they were available from light curve solutions in both $B$ and $V$ bands. Then, intrinsic $(B-V)_0$ colours are obtained using the reddening law  $A_V/E(B-V) = R_V$ and definition of $E(B-V)= (B-V) - (B-V)_0$.  Only after obtaining $(B-V)_0$ of components, then $(B-V) - T_{eff}$  relation is calibrated using published component effective temperatures. 

Here, in this study, first the intrinsic colours of the component stars (data) are computed directly as the difference between the absolute magnitudes in Table\,3. The computed intrinsic colours are then plotted in Fig.\,9 where solid lines represent  colour -- temperature relations as $(G_{BP}-G_{RP})_0 - T_{eff}$, $(G-G_{RP})_0 - T_{eff}$, $(G-G_{BP})_0 - T_{eff}$, $(V-G)_0 - T_{eff}$, and $(B-V)_0 - T_{eff}$, respectively from top to bottom. At last, the five colour -- effective temperature relations are directly computable as the difference of $BC - T_{eff}$ relations. For example: $(B-V)_0 - T_{eff}$ relation is obtained as $BC_V (T_{eff}) - BC_B (T_{eff})$ from the functions presented in Table\,5. Similarly, for the other colours: $BC_G (T_{eff}) - BC_V (T_{eff})$ gives $(V-G)_0 - T_{eff}$ relation, $BC_{G_{BP}} (T_{eff}) - BC_G (T_{eff})$ gives $(G-G_{BP})_0 - T_{eff}$  relation,  $BC_{G_{RP}} (T_{eff}) - BC_G (T_{eff})$ gives $(G-G_{RP})_0 - T_{eff}$  relation, and finally $BC_{G_{RP}} (T_{eff}) - BC_{G_{BP}} (T_{eff})$ gives $(G_{BP}-G_{RP})_0 - T_{eff}$ relations, respectively. 

It is for sure that the solid lines (colour -- temperature relations) follow the trend of the data quite nicely. Especially, the upper two panels in Fig.~9, $(G_{BP}-G_{RP})_0$ and $(G-G_{RP})_0$ are represented very nicely by the solid lines while the middle panel [$(G-G_{BP})_0$] could only be considered successful for the medium hot and cooler stars (log $T_{eff}$ < 4.2 ). Nevertheless, a  small but a clear offset between the solid lines and data is obvious in the lowest two panels [$(V-G)_0$ and $(B-V)_0$]  in Fig.~9. It has been already discussed at the end of section 4 that $B$ and $V$ data is less reliable compared the Gaia data. Moreover, if the number of data towards the cooler and hotter ends in Figs.~4 and 5 are compared, the Gaia bands appear relatively more crowded. Being less reliable and having less number of data compared to Gaia bands towards both ends of the temperature scale, the $BC - T_{eff}$ relations of the $B$ and $V$ bands appear to be the most probable cause of the offset seen in the lowest two panels of Fig.~9. This is because the solid lines are just the differences of the $BC$ -- $T_{eff}$ relations and the bias caused by the less number of data appears not only effecting both ends but also changing the mean value of the $BC$ values thus the solid lines appear to be noticeably shifted causing the offset seen especially for $B$ and $V$ bands. 

Therefore, only the solid lines in the upper two panels [$(G_{BP}-G_{RP})_0$ and $(G-G_{RP})_0$]  are found suitable to represent colour -- temperature relations which are to be included in Table 6, where they are presented together with the $BC$ values produced from the polynomials in Table 5 as a function of spectral types and typical effective temperatures for main-sequence stars having metallicity 0.008 $<$ Z $<$ 0.040 are presented. On the other hand, it is more practical for a user to have an effective temperature - colour relation in order to estimate the effective temperature of a main-sequence star from an intrinsic colour. For this, we have calibrated inverse relations only for $B-V$ for Johnson photometry and $G_{BP}$ -- $G_{RP}$ for Gaia photometry.

\begin{figure}
  \centering
  \includegraphics[width=0.6\columnwidth]{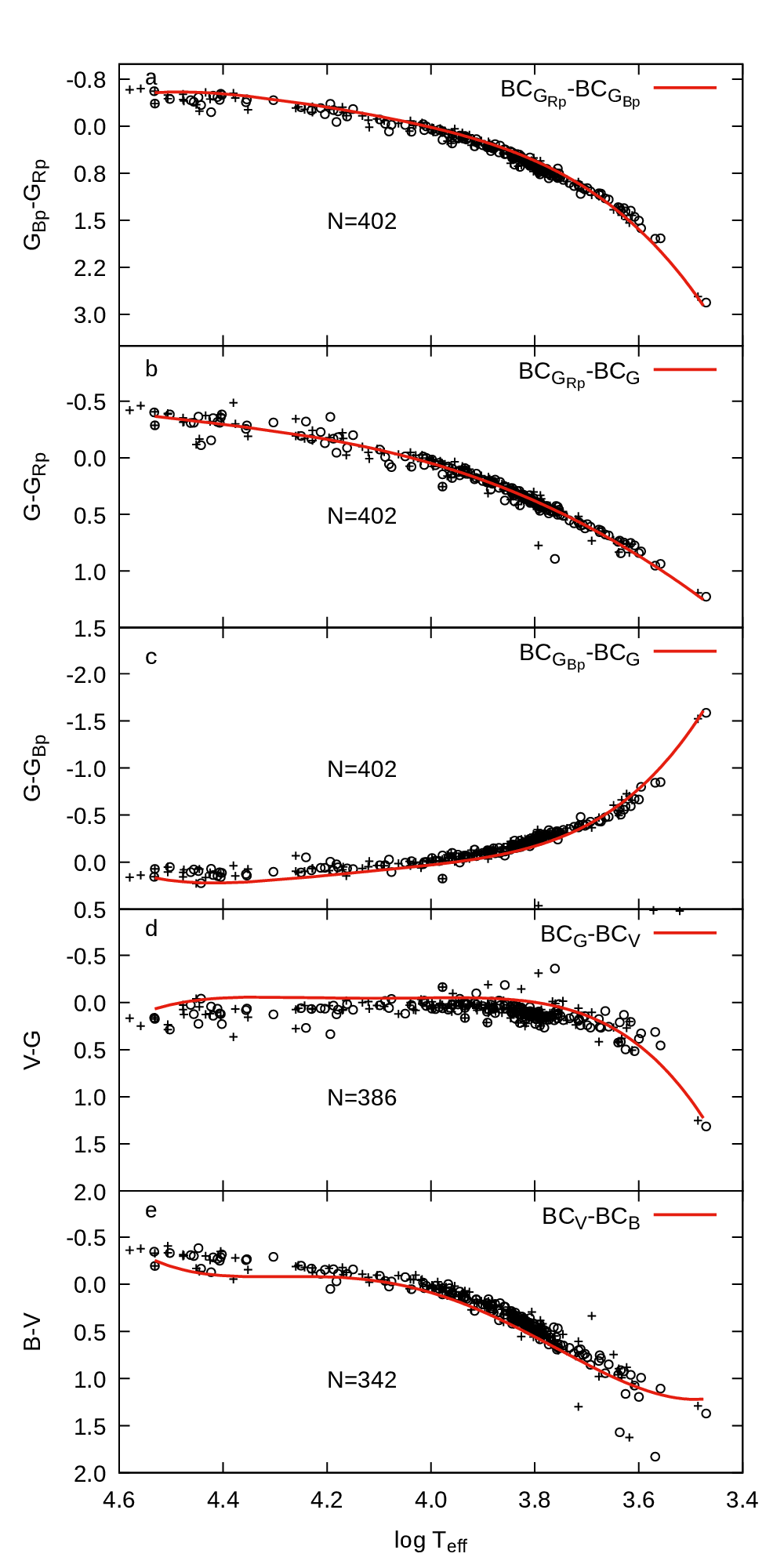}
\caption{Intrinsic colours of DDEB components as a function of log $T_{eff}$, (+) for primary and (o) for secondary. Solid lines are main-sequence Colour $- T_{eff}$ relations imposed by $BC - T_{eff}$ relations. From (a) to (e) $(G_{BP}-G_{RP})_0$, $(G-G_{RP})_0$, $(G-G_{BP})_0$, $(V-G)_0$, and $(B-V)_0$.}
\label{fig:logTeff_color}
\end{figure}

Effective temperatures of the DDEB sample of this study are plotted as a function of $(B-V)_0$ and $(G_{BP}-G_{RP})_0$ in Fig.\,10. Data points are the same as the lowest and uppermost panels in Fig.\,9, but the vertical and horizontal axis are interchanged and re-organised. The solid lines in Fig.\,10 are the temperature - colour relations which are re-predicted from intrinsic colours of DDEB marked on Fig.\,10 unlike the colour - temperature relations shown in Fig.\,9 which are obtained from the differences of $BC$-$T_{eff}$ relations. The temperature -- colour relations as polynomials are given in Table\,8. Fourth-degree polynomials are found best to explain  $(B-V)$  and $(G_{BP} -G_{RP})$  intrinsic colours of the main-sequence stars chosen from components of the DDEB sample of this study. Coefficients and errors associated are determined by the least-squares method and listed in Table\,8 together with the ranges of their validity expressed in intrinsic colours as $-0.5 \leq (B-V)_0 \leq 1.5$ and $-0.6 \leq (G_{BP} -G_{RP})_0 \leq 1.7$.
 
Except for the four stars (components of V881 Per and 2MASS J19071662+ 463932), which are marked with their order number in Fig.\,10a, the intrinsic $(B-V)$ colours are well represented by the predicted $T_{eff}$ - colour relation (solid line). The $T_{eff}$ - colour relations by Eker et al. (2020), marked as  dotted curves, and Mamajec (2021), marked as a  dashed curve, are also plotted on the same figure just for comparison. It is clear in Fig.\,10a that the solid line (this study) is more successful in representing  data than the dotted and the dashed curves. Although both the dotted and the dashed curves are drawn to represent intrinsic $(B-V)$ colours up to 2.00, the reddest stars ($(B-V)_0$ > 0.80) are also not well represented by the dotted and dashed curves.

Unfortunately, there are no other full-range intrinsic $(G_{BP}-G_{RP})$ colours published for comparing  temperature - colour relation predicted in this study. The Main-sequence $(G_{BP}-G_{RP})$  intrinsic colours of Mamajec (2021) cover a range of temperatures  $2350 \leq T_{eff} \leq 10700 $ K,  spectral types B9.5V  to M9.5V and $(G_{BP}-G_{RP})$ from --1.2 to 4.86. The full range of $(G_{BP}-G_{RP})$ data is represented better by the $T_{eff}$ - colour relation of this study. The dashed curve (Mamajec 2021) does not reach the hottest stars. Agreement between solid and dashed curves the middle temperatures are clear. The coolest stars are again better represented by $T_{eff}$ - colour relation of this study than the dashed curve of (Mamajec 2021).

\begin{figure}
\begin{subfigure}{\columnwidth}
  \centering
  \includegraphics[width=0.5\columnwidth]{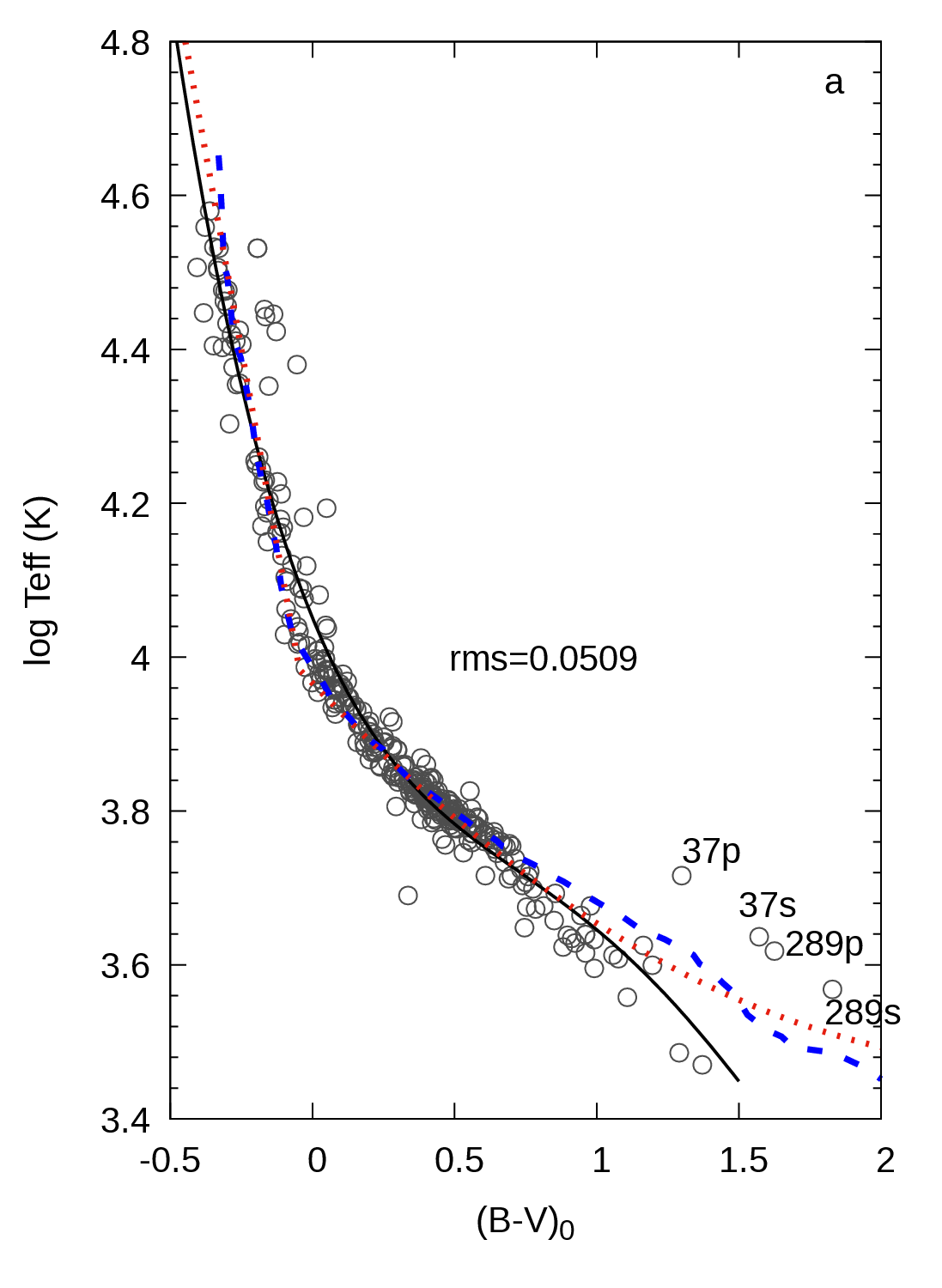}  
\end{subfigure}
\newline
\begin{subfigure}{\columnwidth}
  \centering
  \includegraphics[width=0.5\columnwidth]{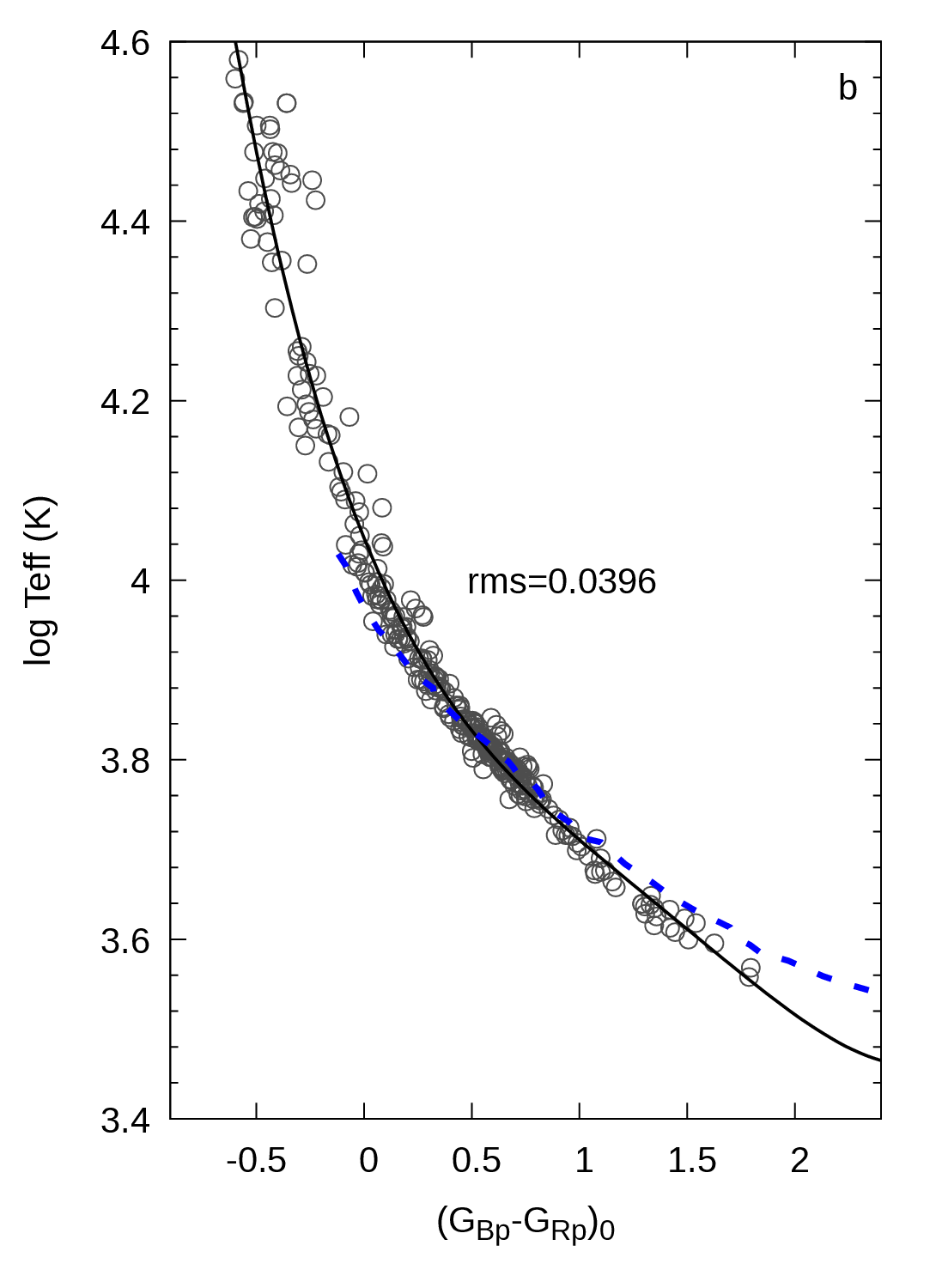}
\end{subfigure}
\caption{$T_{eff}$ -- $(B-V)_0$ ($\textit{a}$) and $T_{eff}$ -- $(G_{BP}-G_{RP})_0$  ($\textit{b}$) relations. The best-fitting models (Table\,8) are shown by solid lines where the dotted curve ($\textit{only top panel}$) is from Eker et al. (2020) and dashed curves in both upper and lower panels are from Mamajek (2021).}
\label{fig:color_logTeff}
\end{figure}


Intrinsic colours as a function of spectral types and effective temperatures computed according to the two $T_{eff}$ - color relations shown in Fig. 10 and listed in Table 8 are also included in Table 6 . Now, it is important to notice that Table 6 has four columns with three intrinsic colours; first two [$(G-G_{RP})_0$ and $(G_{BP}-G_{RP})_0$] of them (columns 8 and 9) are produced by subtracting the proper $BC$ columns in the same table as described before, that is directly from the $BC$ - $T_{eff}$ relations listed in Table 5 and the last two [$(G_{BP}-G_{RP})_0$ and $(B - V)_0$] of them (columns 10 and 11) are produced from the $T_{eff}$ - Colour relations listed in Table 8, which are produced directly from the intrinsic colours of DDEB stars. 

Having the same intrinsic colour [$(G_{BP}-G_{RP})_0$] produced by the two different methods described above is good for fine testing of the new method on producing intrinsic colors since in the first approximation the three intrinsic colours [$(B-V)_0$, $(V-G)_0$, and $(G-G_{BP})_0$] produced by the new method were already eliminated by eye inspections in Fig.~9. Eliminations of these intrinsic colors indicate that $BC - T_{eff}$ relation in Table 5 are not sufficiently accurate enough to produce intrinsic colors while they are shown reliable for estimating (if one of them is used) a standard $L$ and improving its accuracy (if multiples of them are used) as demonstrated in Fig.~ 8.

The mean difference between the two columns in Table 6 giving the same intrinsic color [$(G_{BP}-G_{RP})_0$] produced by the different methods (columns 9 and 10) could be used as a parameter to indicate reliability of the new method with respect to the classical method. The existing numbers in Table 6 indicate a 0.06 mag difference for this study. Minimizing this value in a future study would definitely indicate a noticeable improvement of the new method for producing reliable intrinsic colors from $BC - T_{eff}$ relations. Not only the  intrinsic colours, but also the predicted standard $L$ vould be improved because $BC - T_{eff}$ relations themselves would also be improved automatically. An ideal case is that both methods are producing the same numbers, that is the mean difference between the two columns producing the same colour should be zero or negligible. For that we encourage future researchers not only to increase the number of filters, photometric systems and the number of DDEB stars to be used (especially towards both ends of the temperature scale) in their study but also to find a method for homogenizing the systemic brightness' or set up an observing program to obtain consistent total brightness of systems for inconsistent bands to improve their consistency like Gaia bands.

\subsection{Reddening Law and Colour-Colour Relations}

It is possible to check the best value of the passband-based parameter $R(\xi)$ by modelling the $A(\xi)$-$E(B-V)$ relation, which is given as: 

\begin{equation}
R_{\xi}=\frac{A(\xi)}{E(B-V)}
\label{eq:Rfilter}
\end{equation}


\begin{equation}
E(B-V)=A(B)-A(V)
\label{eq:EBV}
\end{equation}

Using passband based $A(\xi)$ values from Table\,3 and using Eq.(14), colour excess - interstellar dimming relations for Johnson and GAIA passbands have been constructed and shown in Figs.\,11 and 12 together with standard deviation (RMS) from the correlation equations on each plot. The correlation is actually in the form $f(x)=a + bx$, where $a$ is the constant term and $b$ is $R(\xi)$. In all correlations, the constant term is zero under Eq.(13). The most commonly used parameter, $R(V)$ is found to be $3.012\pm0.002$, which is slightly smaller than the common average value for the solar neighbourhood in the  Milky Way ($R(V)=$3.1). For the Johnson-$B$ filter, this relation is predicted as $R(B)=4.012\pm0.002$ and for the GAIA passbands, they are found as $R(G)=2.872\pm0.013$,  $R(G_{BP})=3.494\pm0.009$ and $R(G_{RP})=1.885\pm0.001$. It is noteworthy that the errors of the parameters ($R(\xi)$) are found to be relatively small ($<0.1$ per cent for $A(V)$ and $A(B)$-$E(B-V)$ relations, $<0.7$ per cent for $A(G)$, $A(G_{BP})$ and $A(G_{RP})$-$E(B-V)$ relations). The accuracy of the correlation parameter is relatively better for the interstellar dimming in GAIA passbands versus GAIA colour excess except for $A(G_{RP})$-$E(G_{BP}-G_{RP})$ which is $\sim0.8$ per cent.

\begin{figure}
  \centering
  \includegraphics[width=0.75\columnwidth]{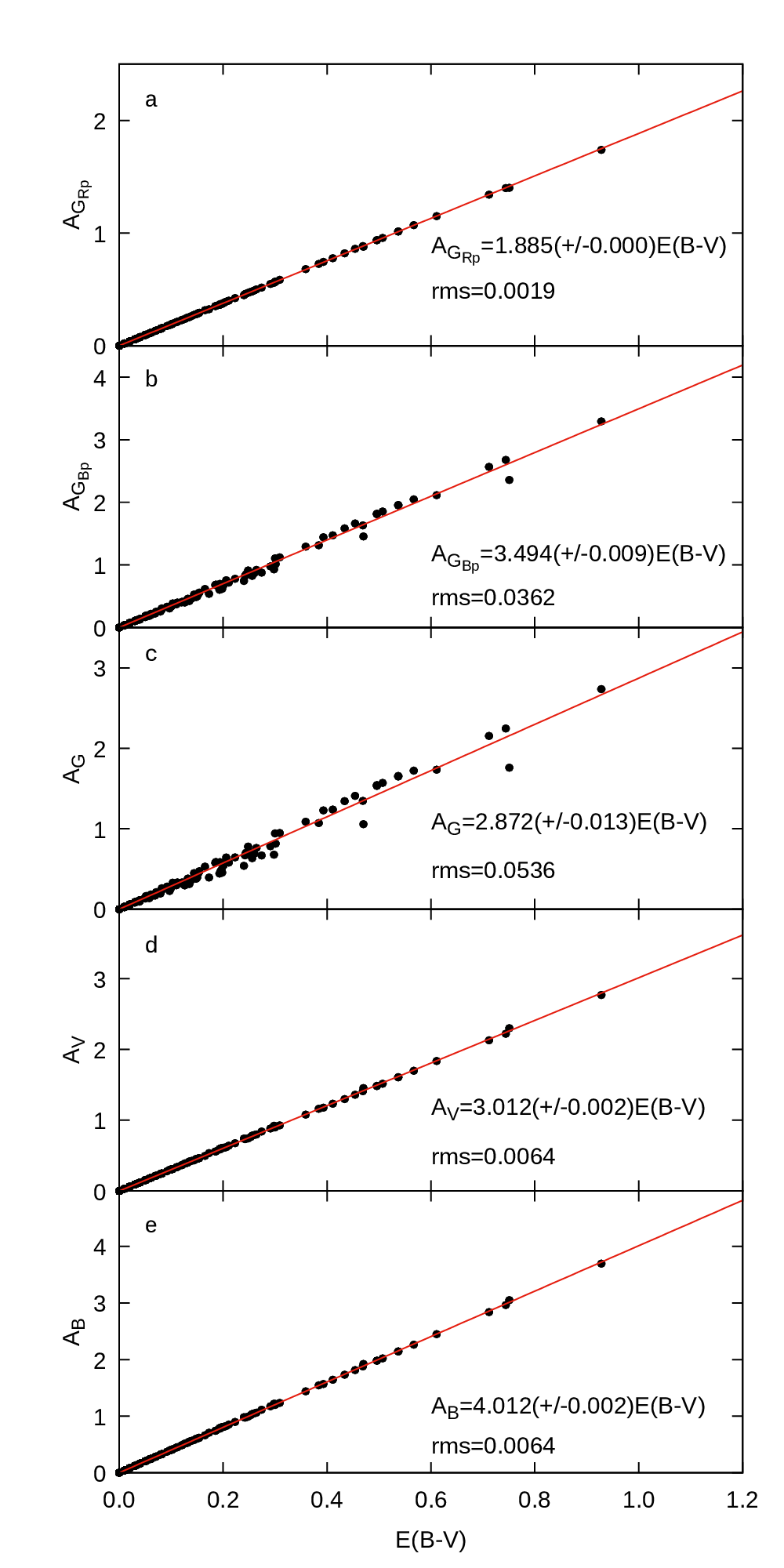}
\caption{Correlation of interstellar dimming $(A_\xi)$ with $E(B-V)$ colour excess in Johnson and GAIA passbands.}
\label{fig:EBVextinctions}
\end{figure}

\begin{figure}
  \centering
  \includegraphics[width=0.75\columnwidth]{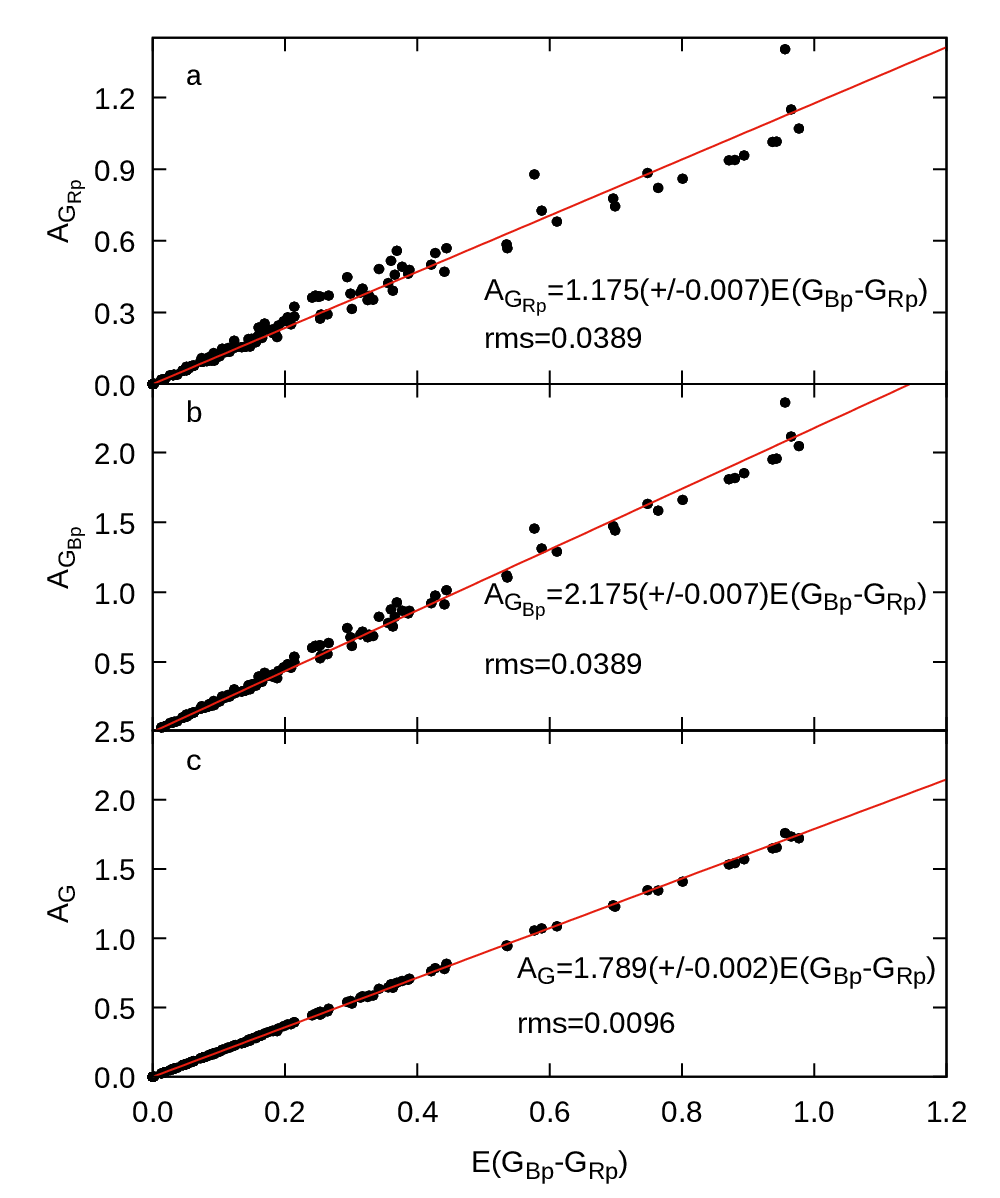}
\caption{Correlation of interstellar dimming $(A_\xi)$ with $E(G_{BP}-G_{RP})$ colour excess in GAIA passbands.}
\label{fig:EBpRpextinctions}
\end{figure}

Other useful relations used in photometry are the colour-colour diagrams and colour excess relations between colours based on certain photometric systems. Colour excess relations between different colours may show the direction of interstellar extinction on the diagram. Having this information on the colour-colour diagrams permits users to define unreddened colours of stars by tracing back the extinction direction up to the intersecting point on the line of unreddened main-sequence stars.

Fig.\,13 shows colour-colour and colour excess - colour excess diagrams for the nearby main-sequence stars as predicted from the DDEB sample of this study for various GAIA passbands. Fig.\,13a compares  $(G-G_{BP})$-$(G_{BP}-G_{RP})$ relation of this study to the one given by Arenou et al. (2018). A very good agreement between them is very clear. Nevertheless,  the colour-colour curves for GAIA passbands of main-sequence stars are almost parallel to the direction of interstellar reddening which creates difficulties in the determination of unreddened colours by going back along the reddening direction. Among the colour-colour relations in Fig.\,13, $(G-G_{RP})$-$(G_{BP}-G)$ (panel c) seem to be more suitable for searching intrinsic colours towards cooler stars since the reddest part is un-parallel to the reddening direction. 

The ratio of color excess of  $E(G-G_{BP})$/$E(G_{BP}-G_{RP})$, $E(G-G_{RP})$/$E(G_{BP}-G_{RP})$ and $E(G-G_{BP})$/$E(G_{BP}-G_{RP})$ in panels d, e and f of the figure gives the direction of extinction in the color-color diagrams shown in panels a, b and c, respectively. The solid lines shown with red color are the best fits to all data while the dashed lines represent the borders of the data. These borders refer to the limits. Since a unique reddening direction in a color-color diagram for all stars in our sample is not expected, it is normal to see a certain interval of the reddening direction values in different Gaia EDR3 passbands. The slope of the solid lines refers to the average value for the relevant color excess ratio defining the direction of reddening in color-color diagrams. The slope of the dashed lines changes between --0.31 and --0.68, 0.44 and 2.2, and 0.31 and 0.69 for $E(G-G_{BP})$/$E(G_{BP}-G_{RP})$ (panel d), $E(G-G_{RP})$/$E(G_{BP}-G_{RP})$ (panel e) and $E(G-G_{BP})$/$E(G_{BP}-G_{RP})$ (panel f) ratios, respectively. Therefore, the reddening direction shown by an arrow in the left panels is not unique. Direction of the arrow may change for different galactic directions.

\begin{figure*}
\begin{subfigure}[b]{0.5\textwidth}
  \centering
  \includegraphics[width=\textwidth]{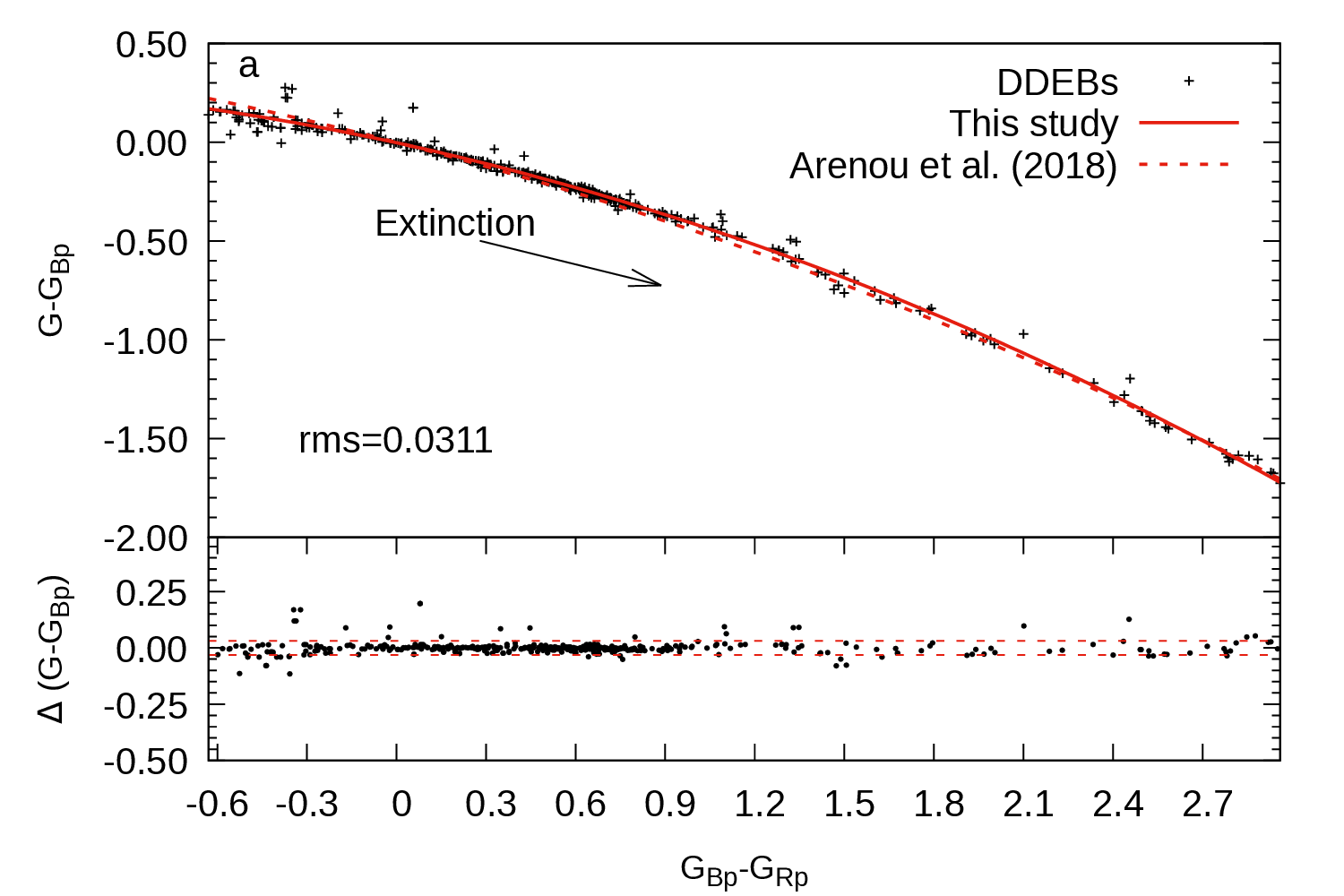}  
\end{subfigure}
\hfill
\begin{subfigure}[b]{0.5\textwidth}
  \centering
  \includegraphics[width=\textwidth]{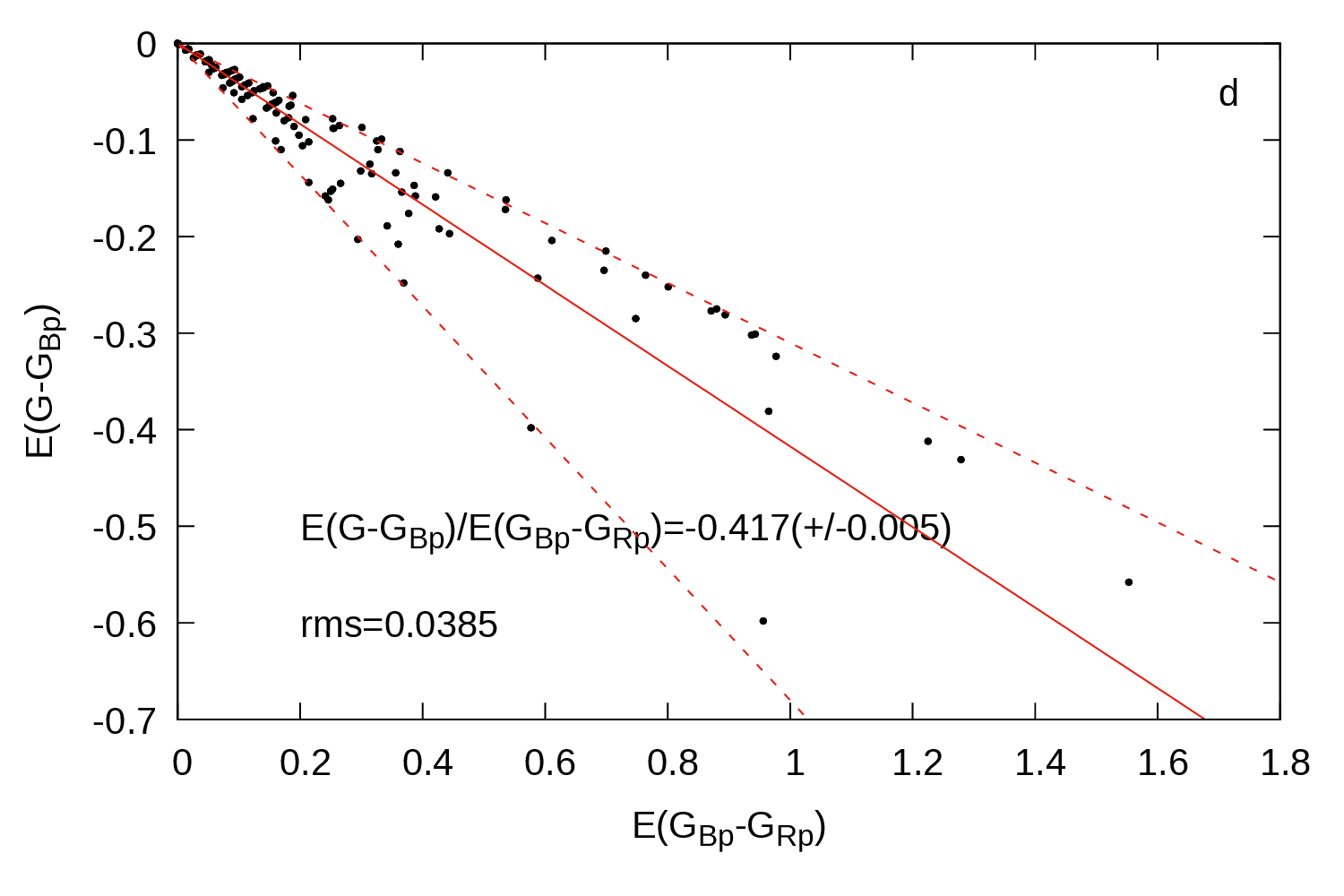}  
\end{subfigure}
\newline
\begin{subfigure}[b]{0.5\textwidth}
  \centering
  \includegraphics[width=\textwidth]{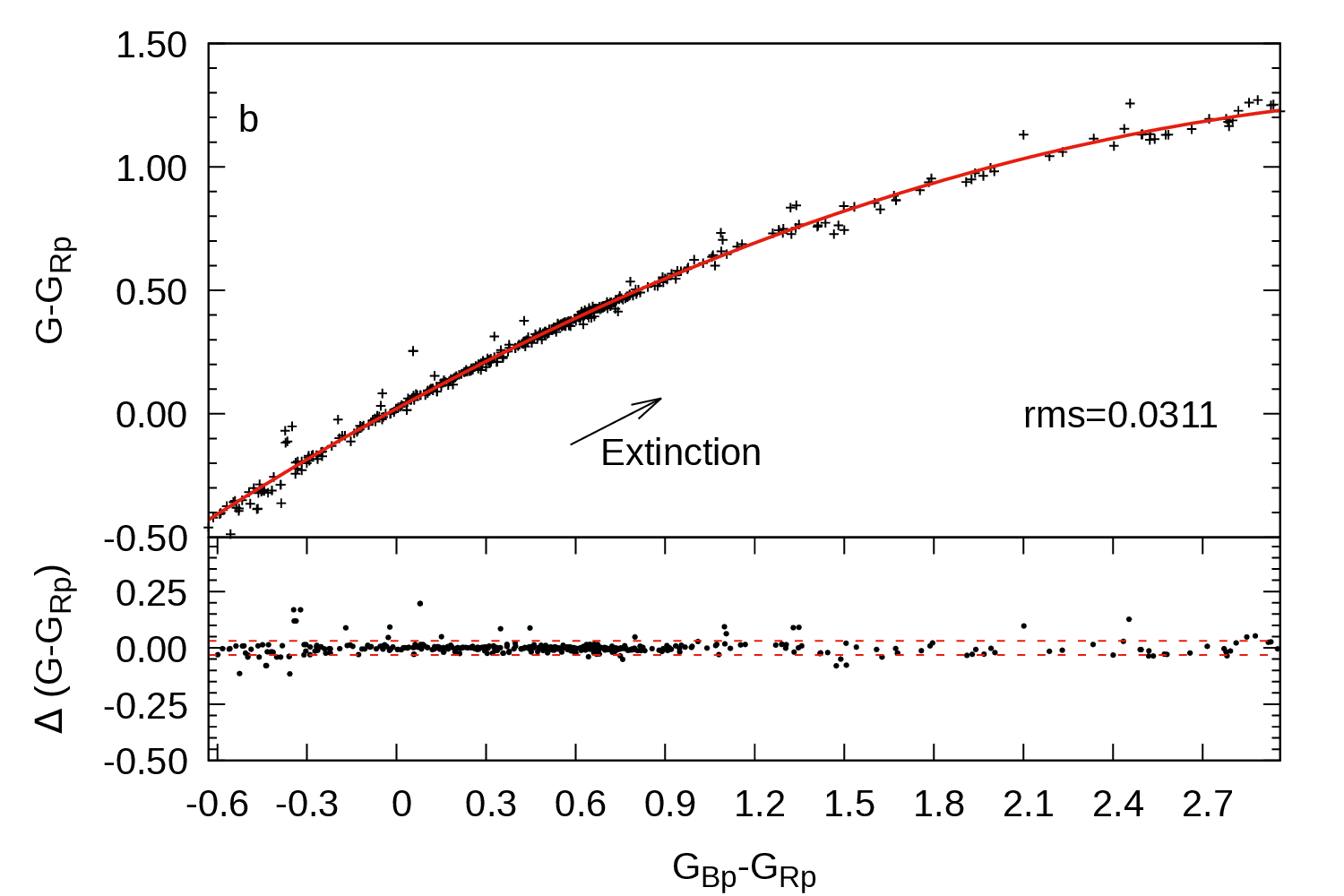}  
\end{subfigure}
\hfill
\begin{subfigure}[b]{0.5\textwidth}
  \centering
  \includegraphics[width=\textwidth]{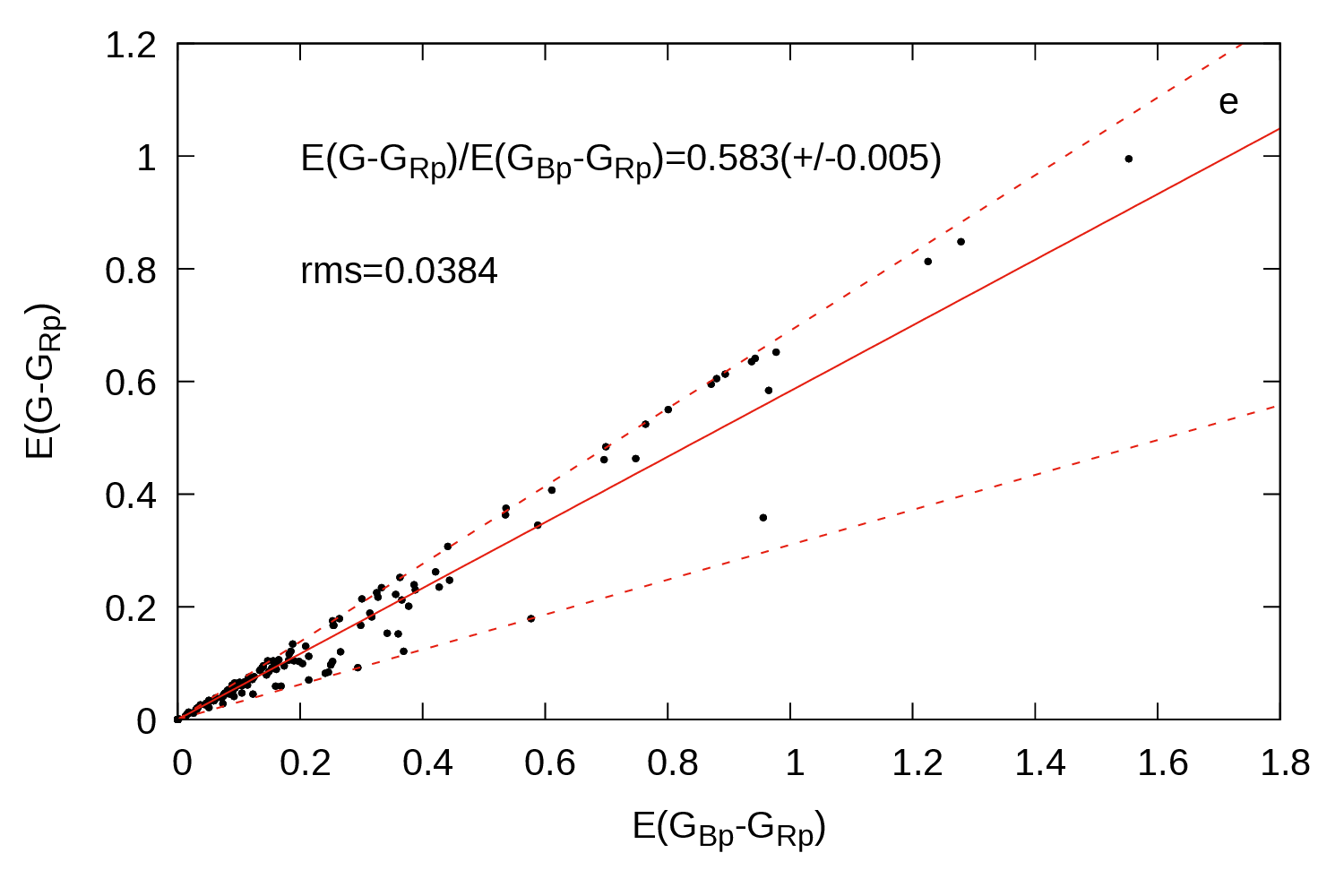}  
\end{subfigure}
\newline
\begin{subfigure}[b]{0.5\textwidth}
  \centering
  \includegraphics[width=\textwidth]{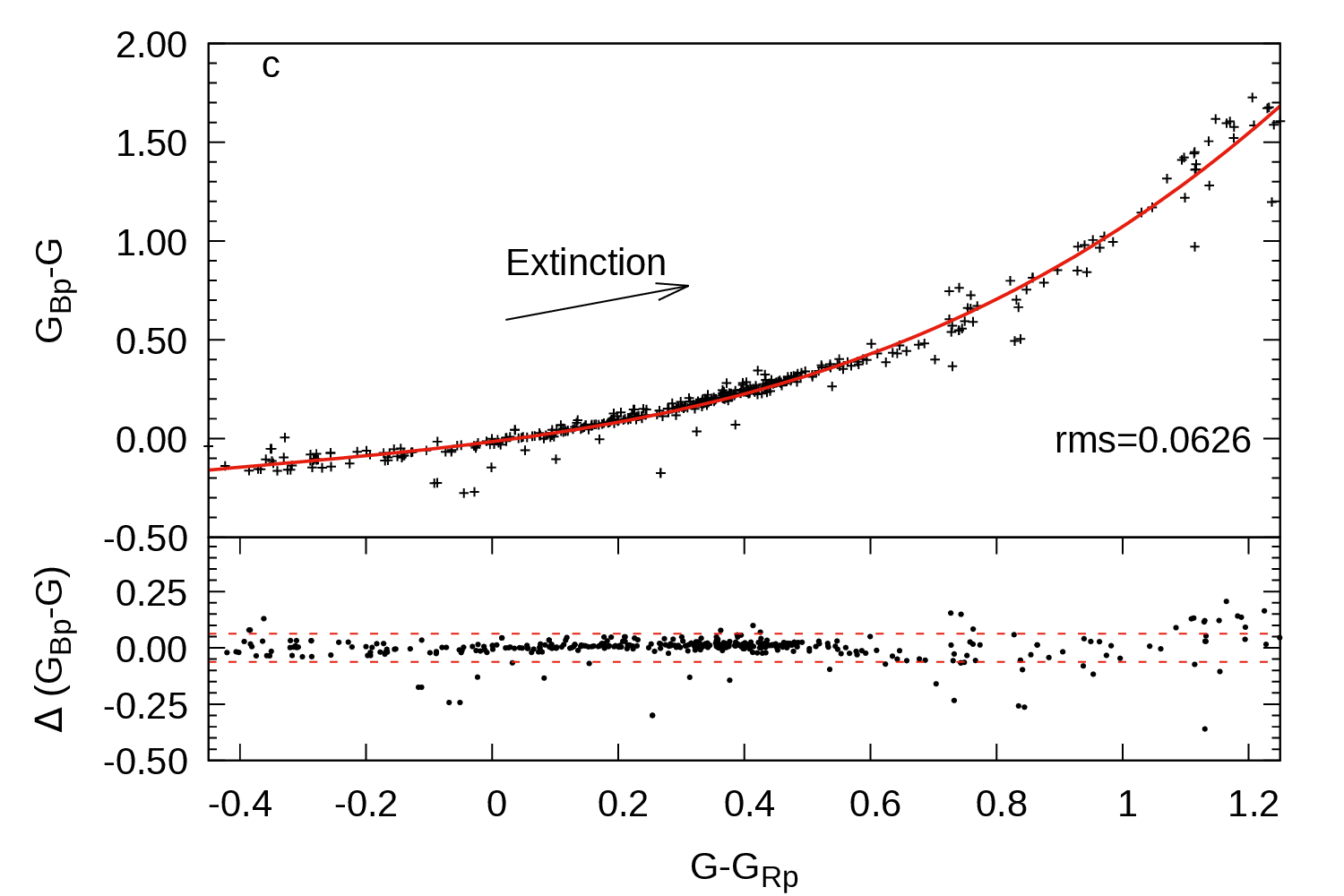}  
\end{subfigure}
\hfill
\begin{subfigure}[b]{0.5\textwidth}
  \centering
  \includegraphics[width=\textwidth]{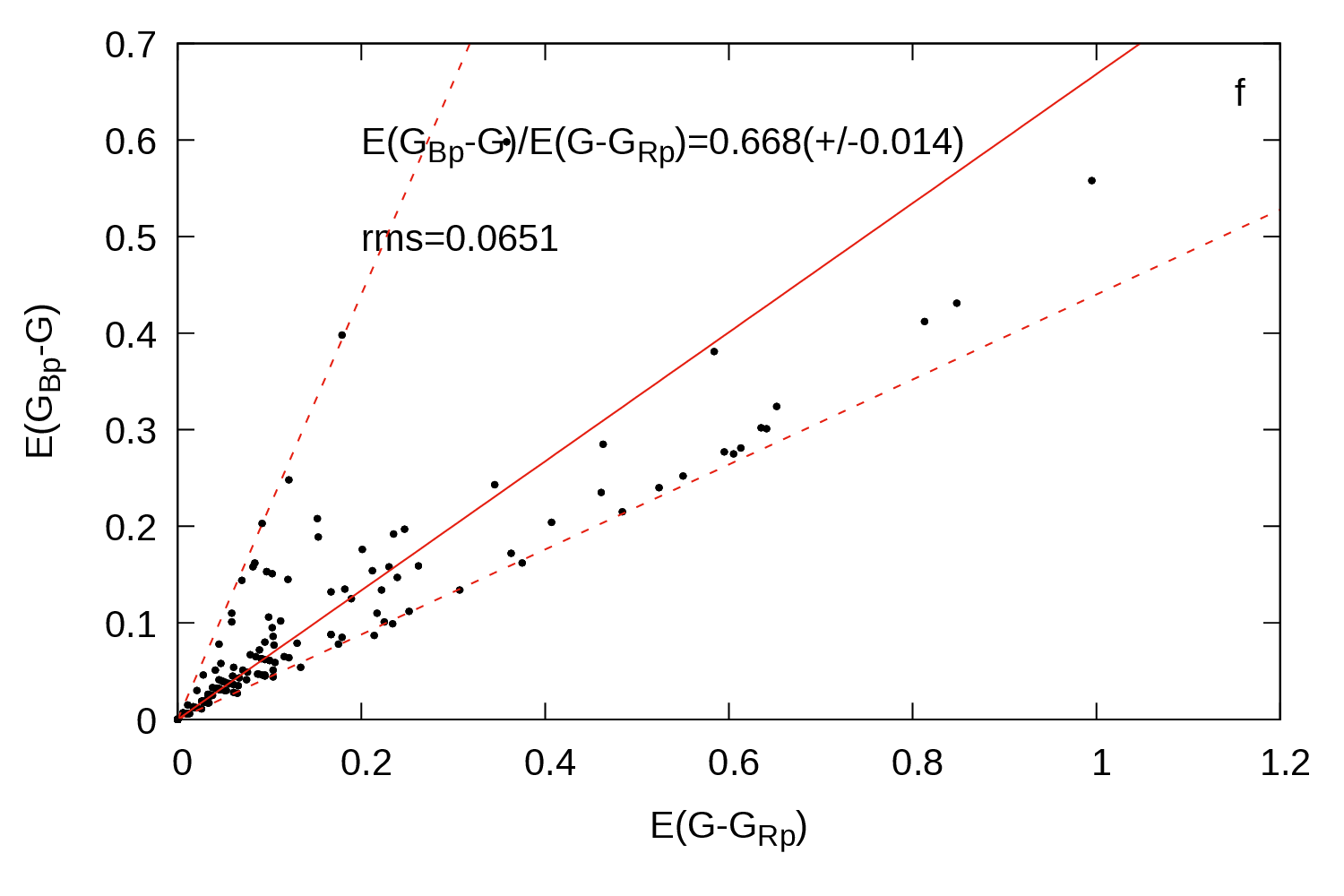}
\end{subfigure}
\caption{Colour-colour (\textit{panels on the left}) and colour excess-colour excess relations (\textit{panels on the right}) in Gaia-bands for nearby main-sequence stars. Equations and RMS values in the right panel refer to the solid lines. Dashed lines below each panel in the left are $1\sigma$ deviations while the dashed lines on the right indicate the borders of the most extreme values.}
\label{fig:color_excess}
\end{figure*}

\section{Conclusions}

\begin{itemize}
    \item A simplified SED model is established for predicting component light contributions of binaries and their interstellar extinctions.
    
    \item The component light contributions predicted by the simplified SED model in $B$ and $V$ bands of Johnson photometry are tested by comparing to the $B$ and $V$ band light contributions predicted from the light curve solutions of DDEB. The simplified SED model is found very successful and reliable in predicting component light contributions according to test in this study.
    
    \item 209 DDEB are found eligible to provide a binary SED model without complexities (third light or any excess flux) which may spoil the SED of the binary. Then, using component contributions, which are produced from the simplified SED model, empirical standard $BC$s are produced by a method described by Eker et al. (2020).
    
    \item The empirical standard $BC$ values are used in calibrating empirical standard $BC$-$T_{eff}$ relations in $B$, $V$, $G$, $G_{BP}$ and $G_{RP}$ bands. The most accurate $BC$-$T_{eff}$ relation ever discussed is produced and presented for interested readers.
    
    \item Empirical standard $BC$-$T_{eff}$ relations of five passbands are used for predicting standard stellar $L$. They  are compared to the $L$ calculated from the observed $R$ and $T_{eff}$. If a standard $L$ is predicted from a single $BC$-$T_{eff}$ relation of a given band, propagated errors indicate that it cannot be more accurate than about 10 cent. Accuracy of the predicted $L$ increases by increasing the number of $BC$-$T_{eff}$ relations at various passbands. A standard $L$ with an uncertainty as high as one per cent (peak at $\sim$ 2.5 per cent), is possible.
    
    \item Multi-band $BC$-$T_{eff}$ relations are shown to be practical to obtain intrinsic color-temperature relations. Intrinsic colour - temperature relations could be produced directly from differences of $BC$-$T_{eff}$ relations.
    
    \item Inverse color-temperature relations involving  $(B-V)_0$ and $(G_{BP}-G_{RP})_0$ are produced for interested readers who wants to calculate effective temperature of a main-sequence star from its $(B-V)_0$ and $(G_{BP}-G_{RP})_0$.
    
    \item Reddening laws, colour-colour and colour excess - colour excess relations involving Johnson $B$, $V$ and Gaia passbands covering all spectral classes of the main-sequence from the DDEB sample of this study are demonstrated.
    
\end{itemize}

\clearpage

\newpage

\Acknow{This work uses the VizieR catalogue access tool, CDS, Strasbourg, France; the SIMBAD database, operated at CDS, Strasbourg, France. This work presents results from the European Space Agency (ESA) space mission, Gaia. Gaia data are being processed by the Gaia Data Processing and Analysis Consortium (DPAC). Funding for the DPAC is provided by national institutions, in particular the institutions participating in the Gaia MultiLateral Agreement (MLA). The Gaia mission website is https://www.cosmos.esa.int/gaia. The Gaia archive website is https://archives.esac.esa.int/gaia. We thank to Assoc.Prof. Mustafa Caner for proof reading and correcting English grammar and linguistics. We are also grateful to the anonymous referee who carefully read our paper and made very meaningful comments that made us improve its clearance.}

\newpage

\begin{table*}
\label{tab:physicalpars}
\centering
\caption{Physical parameters and total magnitudes of selected systems in $B$, $V$, $G$, $G_{BP}$ and $G_{RP}$. The full table is available online.}
\resizebox{\textwidth}{!}{\begin{tabular}{cccccccccccccccccccccc}
\hline
Order	&	Name	&	pri/sec	&	$M$	& err &	$R$ 	& err &	Reference	&	$T$	& err &	Reference	&	$B$ & err & $V$ & err & Reference & $G$ & err & $G_{BP}$ & err & $G_{RP}$ & err \\
& & & ($M_\odot$) & ($M_\odot$) & ($R_\odot$) & ($R_\odot$) &  & ($K$) & ($K$) & & (mag) & (mag) & (mag) & (mag) & & (mag) & (mag) & (mag) & (mag) & (mag) & (mag) \\
\hline
1	&	V421 Peg	&	p	&	1.594	&	0.029	&	1.584	&	0.028	&	 2016NewA...46...47O	&	7250	&	80	&	2016NewA...46...47O	&	8.650	&	0.010	&	8.280	&	0.010	&	2016NewA...46...47O	&	8.208	&	0.003	&	8.384	&	0.003	&	7.890	&	0.004	\\
2	&	V421 Peg	&	s	&	1.356	&	0.029	&	1.328	&	0.029	&	 2016NewA...46...47O	&	6980	&	120	&	2016NewA...46...47O	&	8.650	&	0.010	&	8.280	&	0.010	&		&	8.208	&	0.003	&	8.384	&	0.003	&	7.890	&	0.004	\\
3	&	DV Psc	&	p	&	0.677	&	0.019	&	0.685	&	0.030	&	 2014PASA...31...24E	&	4450	&	8	&	 2007MNRAS.382.1133Z	&	11.604	&	0.010	&	10.621	&	0.010	&	2000A\&A...355L..27H	&	10.219	&	0.005	&	10.997	&	0.015	&	9.334	&	0.013	\\
4	&	DV Psc	&	s	&	0.475	&	0.010	&	0.514	&	0.020	&	 2014PASA...31...24E	&	3614	&	8	&	 2007MNRAS.382.1133Z	&	11.604	&	0.010	&	10.621	&	0.010	&		&	10.219	&	0.005	&	10.997	&	0.015	&	9.334	&	0.013	\\
5	&	MU Cas	&	p	&	4.657	&	0.100	&	4.192	&	0.050	&	2014PASA...31...24E	&	14750	&	500	&	2004AJ....128.1840L	&	11.112	&	0.009	&	10.808	&	0.007	&	2019ApJ...872...85G	&	10.742	&	0.003	&	10.894	&	0.003	&	10.452	&	0.004	\\
6	&	MU Cas	&	s	&	4.575	&	0.090	&	3.671	&	0.040	&	2014PASA...31...24E	&	15100	&	500	&	2004AJ....128.1840L	&	11.112	&	0.009	&	10.808	&	0.007	&		&	10.742	&	0.003	&	10.894	&	0.003	&	10.452	&	0.004	\\
7	&	TYC 4019-3345-1	&	p	&	1.920	&	0.010	&	1.760	&	0.050	&	2013PASA...30...26B	&	8600	&	310	&	2013PASA...30...26B	&	12.550	&	0.009	&	12.150	&	0.008	&	 2013PASA...30...26B	&	11.952	&	0.003	&	12.164	&	0.003	&	11.597	&	0.004	\\
8	&	TYC 4019-3345-1	&	s	&	1.920	&	0.010	&	1.760	&	0.050	&	2013PASA...30...26B	&	8600	&	570	&	2013PASA...30...26B	&	12.550	&	0.009	&	12.150	&	0.008	&		&	11.952	&	0.003	&	12.164	&	0.003	&	11.597	&	0.004	\\
9	&	YZ Cas	&	p	&	2.263	&	0.012	&	2.525	&	0.011	&	2014MNRAS.438..590P	&	9520	&	120	&	2014MNRAS.438..590P	&	5.715	&	0.026	&	5.660	&	0.015	&	 2019ApJ...872...85G	&	5.630	&	0.003	&	5.659	&	0.003	&	5.546	&	0.005	\\
10	&	YZ Cas	&	s	&	1.325	&	0.007	&	1.331	&	0.006	&	2014MNRAS.438..590P	&	6880	&	240	&	2014MNRAS.438..590P	&	5.715	&	0.026	&	5.660	&	0.015	&		&	5.630	&	0.003	&	5.659	&	0.003	&	5.546	&	0.005	\\
... & ... & ... & ... & ... & ... & ... & ... & ... & ... & ... & ... & ... & ... & ... & ... & ... & ... & ... & ... & ... & ... \\
409	&	IT Cas	&	p	&	1.330	&	0.009	&	1.603	&	0.015	&	2014PASA...31...24E	&	6470	&	110	&	 1997AJ....114.1206L	&	11.640	&	0.019	&	11.150	&	0.039	&	1997AJ....114.1206L	&	11.042	&	0.003	&	11.284	&	0.003	&	10.628	&	0.004	\\
410	&	IT Cas	&	s	&	1.328	&	0.008	&	1.569	&	0.040	&	2014PASA...31...24E	&	6470	&	110	&	 1997AJ....114.1206L	&	11.640	&	0.019	&	11.150	&	0.039	&		&	11.042	&	0.003	&	11.284	&	0.003	&	10.628	&	0.004	\\
411	&	BK Peg	&	p	&	1.414	&	0.007	&	1.985	&	0.008	&	2014PASA...31...24E	&	6265	&	85	&	 2010A\&A...516A..42C	&	10.540	&	0.041	&	9.982	&	0.010	&	2019ApJ...872...85G	&	9.835	&	0.003	&	10.116	&	0.003	&	9.386	&	0.004	\\
412	&	BK Peg	&	s	&	1.257	&	0.005	&	1.472	&	0.017	&	2014PASA...31...24E	&	6320	&	30	&	 2010A\&A...516A..42C	&	10.540	&	0.041	&	9.982	&	0.010	&		&	9.835	&	0.003	&	10.116	&	0.003	&	9.386	&	0.004	\\
413	&	AP And	&	p	&	1.277	&	0.004	&	1.234	&	0.006	&	2014AJ....147..148L	&	6565	&	150	&	2014AJ....147..148L	&	11.606	&	0.057	&	11.074	&	0.085	&	2015AAS...22533616H	&	10.910	&	0.003	&	11.195	&	0.003	&	10.457	&	0.004	\\
414	&	AP And	&	s	&	1.251	&	0.004	&	1.195	&	0.005	&	2014AJ....147..148L	&	6495	&	150	&	2014AJ....147..148L	&	11.606	&	0.057	&	11.074	&	0.085	&		&	10.910	&	0.003	&	11.195	&	0.003	&	10.457	&	0.004	\\
415	&	AL Scl	&	p	&	3.617	&	0.110	&	3.241	&	0.050	&	2014PASA...31...24E	&	13550	&	350	&	1987A\&A...179..141H	&	5.985	&	0.014	&	6.070	&	0.009	&	 2000A\&A...355L..27H	&	6.073	&	0.003	&	6.017	&	0.004	&	6.145	&	0.004	\\
416	&	AL Scl	&	s	&	1.703	&	0.040	&	1.401	&	0.020	&	2014PASA...31...24E	&	10300	&	360	&	1987A\&A...179..141H	&	5.985	&	0.014	&	6.070	&	0.009	&		&	6.073	&	0.003	&	6.017	&	0.004	&	6.145	&	0.004	\\
417	&	V821 Cas	&	p	&	2.025	&	0.066	&	2.308	&	0.028	&	 2014PASA...31...24E	&	9400	&	400	&	 2009MNRAS.395.1649C	&	8.402	&	0.029	&	8.286	&	0.017	&	2019ApJ...872...85G	&	8.227	&	0.003	&	8.265	&	0.003	&	8.121	&	0.004	\\
418	&	V821 Cas	&	s	&	1.620	&	0.058	&	1.390	&	0.022	&	 2014PASA...31...24E	&	8600	&	400	&	 2009MNRAS.395.1649C	&	8.402	&	0.029	&	8.286	&	0.017	&		&	8.227	&	0.003	&	8.265	&	0.003	&	8.121	&	0.004	\\
\hline
\end{tabular}}
\end{table*}

\newpage

\begin{table*}
\label{tab:lightcontrib}
\centering
\caption{Component apparent magnitudes of DDEB in $B$, $V$, $G$, $G_{BP}$ and $G_{RP}$ according to component contributions estimated by the new method using simplified SED. The full table is available online.}
\resizebox{\textwidth}{!}{\begin{tabular}{cccccccccccccccccccccc}
\hline
Order & Name & pri/sec & \multicolumn{2}{c}{Cross Reference} & \multicolumn{2}{c}{Cont. (Eker et al. 2020)} & \multicolumn{5}{c}{Unreddened light contribution (this study)} &  \multicolumn{10}{c}{Component apparent brightness (mag)} \\
&	&		&	Xref 1	&	Xref 2 	&	$l_{B}$	&	$l_{V}$	&	$l_{B}$	&	$l_{V}$	&	$l_{G}$	&	$l_{G_{BP}}$	&	$l_{G_{RP}}$	&	$B$	& err & $V$ & err & $G$ & err & $G_{BP}$ & err & $G_{RP}$ & err \\
\hline
1	&	V421 Peg	&	p	&	1	&	1	&	---	&	0.624	&	0.630	&	0.622	&	0.621	&	0.625	&	0.614	&	9.152	&	0.010	&	8.796	&	0.010	&	8.725	&	0.003	&	8.894	&	0.003	&	8.421	&	0.004	\\
2	&	V421 Peg	&	s	&	2	&	1	&	---	&	0.376	&	0.370	&	0.378	&	0.379	&	0.375	&	0.386	&	9.729	&	0.010	&	9.335	&	0.010	&	9.261	&	0.003	&	9.449	&	0.003	&	8.923	&	0.004	\\
3	&	DV Psc	&	p	&	5	&	2	&	0.918	&	0.889	&	0.906	&	0.873	&	0.852	&	0.878	&	0.826	&	11.711	&	0.010	&	10.768	&	0.010	&	10.393	&	0.005	&	11.138	&	0.015	&	9.541	&	0.013	\\
4	&	DV Psc	&	s	&	6	&	2	&	0.082	&	0.111	&	0.094	&	0.127	&	0.148	&	0.122	&	0.174	&	14.166	&	0.010	&	12.861	&	0.010	&	12.294	&	0.005	&	13.284	&	0.015	&	11.233	&	0.013	\\
5	&	MU Cas	&	p	&	7	&	3	&	0.556	&	0.557	&	0.551	&	0.554	&	0.553	&	0.552	&	0.556	&	11.758	&	0.009	&	11.450	&	0.007	&	11.385	&	0.003	&	11.539	&	0.003	&	11.090	&	0.004	\\
6	&	MU Cas	&	s	&	8	&	3	&	0.444	&	0.443	&	0.449	&	0.446	&	0.447	&	0.448	&	0.444	&	11.982	&	0.009	&	11.684	&	0.007	&	11.617	&	0.003	&	11.766	&	0.003	&	11.333	&	0.004	\\
7	&	TYC 4019-3345-1	&	p	&	9	&	4	&	0.497	&	0.496	&	0.500	&	0.500	&	0.500	&	0.500	&	0.500	&	13.303	&	0.009	&	12.903	&	0.008	&	12.705	&	0.003	&	12.917	&	0.003	&	12.350	&	0.004	\\
8	&	TYC 4019-3345-1	&	s	&	10	&	4	&	0.503	&	0.504	&	0.500	&	0.500	&	0.500	&	0.500	&	0.500	&	13.303	&	0.009	&	12.903	&	0.008	&	12.705	&	0.003	&	12.917	&	0.003	&	12.350	&	0.004	\\
9	&	YZ Cas	&	p	&	13	&	5	&	0.943	&	0.919	&	0.934	&	0.916	&	0.915	&	0.925	&	0.895	&	5.790	&	0.026	&	5.756	&	0.015	&	5.726	&	0.003	&	5.744	&	0.003	&	5.667	&	0.005	\\
10	&	YZ Cas	&	s	&	14	&	5	&	0.058	&	0.081	&	0.066	&	0.084	&	0.085	&	0.075	&	0.105	&	8.658	&	0.026	&	8.347	&	0.015	&	8.310	&	0.003	&	8.465	&	0.003	&	7.988	&	0.005	\\
 ... & ... & ... & ... & ... & ... & ... & ... & ... & ... & ... & ... & ... & ... & ... & ... & ... & ... & ... & ... & ... & ... \\
409	&	IT Cas	&	p	&	577	&	202	&	0.508	&	0.509	&	0.511	&	0.511	&	0.511	&	0.511	&	0.511	&	12.370	&	0.019	&	11.880	&	0.039	&	11.771	&	0.003	&	12.014	&	0.003	&	11.357	&	0.004	\\
410	&	IT Cas	&	s	&	578	&	202	&	0.492	&	0.491	&	0.489	&	0.489	&	0.489	&	0.489	&	0.489	&	12.416	&	0.019	&	11.926	&	0.039	&	11.818	&	0.003	&	12.060	&	0.003	&	11.404	&	0.004	\\
411	&	BK Peg	&	p	&	579	&	203	&	0.634	&	0.636	&	0.635	&	0.637	&	0.637	&	0.636	&	0.639	&	11.034	&	0.041	&	10.472	&	0.010	&	10.324	&	0.003	&	10.607	&	0.003	&	9.873	&	0.004	\\
412	&	BK Peg	&	s	&	580	&	203	&	0.366	&	0.364	&	0.365	&	0.363	&	0.363	&	0.364	&	0.361	&	11.633	&	0.041	&	11.081	&	0.010	&	10.935	&	0.003	&	11.213	&	0.003	&	10.492	&	0.004	\\
413	&	AP And	&	p	&	581	&	204	&	0.530	&	0.529	&	0.529	&	0.527	&	0.526	&	0.528	&	0.524	&	12.297	&	0.057	&	11.770	&	0.085	&	11.607	&	0.003	&	11.889	&	0.003	&	11.158	&	0.004	\\
414	&	AP And	&	s	&	582	&	204	&	0.470	&	0.471	&	0.471	&	0.473	&	0.474	&	0.472	&	0.476	&	12.424	&	0.057	&	11.886	&	0.085	&	11.721	&	0.003	&	12.010	&	0.003	&	11.263	&	0.004	\\
415	&	AL Scl	&	p	&	583	&	205	&	0.960	&	0.950	&	0.924	&	0.914	&	0.916	&	0.920	&	0.904	&	6.070	&	0.014	&	6.167	&	0.009	&	6.169	&	0.003	&	6.107	&	0.004	&	6.255	&	0.004	\\
416	&	AL Scl	&	s	&	584	&	205	&	0.040	&	0.050	&	0.076	&	0.086	&	0.084	&	0.080	&	0.096	&	8.790	&	0.014	&	8.738	&	0.009	&	8.760	&	0.003	&	8.764	&	0.004	&	8.685	&	0.004	\\
417	&	V821 Cas	&	p	&	585	&	206	&	0.797	&	0.779	&	0.794	&	0.784	&	0.785	&	0.789	&	0.774	&	8.652	&	0.029	&	8.550	&	0.017	&	8.490	&	0.003	&	8.522	&	0.003	&	8.400	&	0.004	\\
418	&	V821 Cas	&	s	&	586	&	206	&	0.203	&	0.221	&	0.206	&	0.216	&	0.215	&	0.211	&	0.226	&	10.120	&	0.029	&	9.951	&	0.017	&	9.893	&	0.003	&	9.957	&	0.003	&	9.736	&	0.004	\\
\hline
\end{tabular}}
\end{table*}

\begin{table*}
\centering
\caption{Component absolute magnitudes of DDEB in $B$, $V$, $G$, $G_{BP}$ and $G_{RP}$ and propagated uncertainties.  Eleven DR2 and one Hipparcos parallaxes are shown in square brackets and parenthesis, respectively. The full table is available online.}
\label{tab:absolute_mag}
\resizebox{\textwidth}{!}{\begin{tabular}{ccccccccccccccccccccccccc}
\hline
Order	&	Name	&	pri/sec	&	Parallax	& $\frac{\sigma_\varpi}{\varpi}$ &	$L/L_\odot$	&	L(SI)	&	$\frac{\Delta L}{L}$ &	$M_{Bol}$	&	err	&	$A_B$	&	$A_V$	&	$A_G$	&	$A_{G_{BP}}$	&	$A_{G_{RP}}$	&	$M_B$ & err & $M_V$ & err & $M_{G}$ & err & $M_{G_{BP}}$ & err & $M_{G_{RP}}$ & err \\
 & & & (mas) & (\%)  & &$\times10^{27}$ & (\%) & (mag) & (mag) &	(mag) &	(mag) &	(mag) &	(mag) &	(mag) &	(mag) &	(mag) & (mag) & (mag) & (mag) & (mag) & (mag) & (mag)	& (mag) & (mag) \\
\hline
1	&	V421 Peg	&	p	&	6.5051	&	0.4	&	6.245	&	2.390	&	6	&	2.751	&	0.061	&	0.122	&	0.092	&	0.087	&	0.104	&	0.057	&	3.096	&	0.042	&	2.770	&	0.042	&	2.704	&	0.041	&	2.857	&	0.041	&	2.430	&	0.041	\\
2	&	V421 Peg	&	s	&	6.5051	&	0.4	&	3.771	&	1.444	&	8	&	3.299	&	0.088	&	0.122	&	0.092	&	0.087	&	0.104	&	0.057	&	3.673	&	0.042	&	3.310	&	0.042	&	3.240	&	0.041	&	3.412	&	0.041	&	2.932	&	0.041	\\
3	&	DV Psc	&	p	&	23.7216	&	0.1	&	0.166	&	0.063	&	9	&	6.691	&	0.095	&	0.800	&	0.603	&	0.492	&	0.633	&	0.368	&	7.787	&	0.041	&	7.041	&	0.041	&	6.777	&	0.040	&	7.381	&	0.043	&	6.049	&	0.042	\\
4	&	DV Psc	&	s	&	23.7216	&	0.1	&	0.041	&	0.016	&	8	&	8.219	&	0.085	&	0.800	&	0.603	&	0.492	&	0.633	&	0.368	&	10.242	&	0.041	&	9.134	&	0.041	&	8.678	&	0.040	&	9.528	&	0.043	&	7.741	&	0.042	\\
5	&	MU Cas	&	p	&	0.5133	&	3.7	&	749.386	&	286.827	&	14	&	-2.447	&	0.149	&	1.644	&	1.233	&	1.243	&	1.446	&	0.776	&	-1.333	&	0.091	&	-1.231	&	0.090	&	-1.305	&	0.090	&	-1.356	&	0.090	&	-1.133	&	0.090	\\
6	&	MU Cas	&	s	&	0.5133	&	3.7	&	631.206	&	241.594	&	13	&	-2.260	&	0.146	&	1.644	&	1.233	&	1.243	&	1.446	&	0.776	&	-1.109	&	0.091	&	-0.997	&	0.090	&	-1.074	&	0.090	&	-1.128	&	0.090	&	-0.891	&	0.090	\\
7	&	TYC 4019-3345-1	&	p	&	0.8918	&	1.4	&	15.266	&	5.843	&	15	&	1.781	&	0.168	&	1.062	&	0.797	&	0.763	&	0.908	&	0.499	&	1.992	&	0.050	&	1.856	&	0.050	&	1.693	&	0.050	&	1.760	&	0.050	&	1.602	&	0.050	\\
8	&	TYC 4019-3345-1	&	s	&	0.8918	&	1.4	&	15.266	&	5.843	&	27	&	1.781	&	0.294	&	1.062	&	0.797	&	0.763	&	0.908	&	0.499	&	1.992	&	0.050	&	1.856	&	0.050	&	1.693	&	0.050	&	1.760	&	0.050	&	1.602	&	0.050	\\
9	&	YZ Cas	&	p	&	10.6528	&	0.5	&	47.181	&	18.058	&	5	&	0.556	&	0.056	&	0.041	&	0.031	&	0.031	&	0.036	&	0.019	&	0.886	&	0.049	&	0.862	&	0.044	&	0.833	&	0.041	&	0.845	&	0.041	&	0.785	&	0.042	\\
10	&	YZ Cas	&	s	&	10.6528	&	0.5	&	3.576	&	1.369	&	14	&	3.357	&	0.152	&	0.041	&	0.031	&	0.031	&	0.036	&	0.019	&	3.754	&	0.049	&	3.453	&	0.044	&	3.417	&	0.041	&	3.567	&	0.041	&	3.106	&	0.042	\\
...	&	...	& ...	&	...	&	...	&	...	&	...	&	...	&	...	&	...	&	...	&	...	&	...	&	...	&	...	&	...	&	...	&	...	&	...	&	...	&	...	&	...	&	...	&	...	&	...	\\
409	&	IT Cas	&	p	&	1.9419	&	0.8	&	4.057	&	1.553	&	7	&	3.220	&	0.077	&	0.203	&	0.153	&	0.142	&	0.171	&	0.095	&	3.607	&	0.048	&	3.168	&	0.059	&	3.071	&	0.044	&	3.284	&	0.044	&	2.703	&	0.044	\\
410	&	IT Cas	&	s	&	1.9419	&	0.8	&	3.886	&	1.488	&	8	&	3.266	&	0.092	&	0.203	&	0.153	&	0.142	&	0.171	&	0.095	&	3.654	&	0.048	&	3.214	&	0.059	&	3.117	&	0.044	&	3.331	&	0.044	&	2.750	&	0.044	\\
411	&	BK Peg	&	p	&	3.2643	&	0.5	&	5.469	&	2.093	&	5	&	2.895	&	0.060	&	0.203	&	0.153	&	0.141	&	0.170	&	0.095	&	3.400	&	0.058	&	2.888	&	0.043	&	2.753	&	0.042	&	3.006	&	0.042	&	2.347	&	0.042	\\
412	&	BK Peg	&	s	&	3.2643	&	0.5	&	3.114	&	1.192	&	3	&	3.507	&	0.032	&	0.203	&	0.153	&	0.141	&	0.170	&	0.095	&	3.999	&	0.058	&	3.497	&	0.043	&	3.364	&	0.042	&	3.612	&	0.042	&	2.966	&	0.042	\\
413	&	AP And	&	p	&	2.9143	&	0.7	&	2.546	&	0.975	&	9	&	3.725	&	0.100	&	0.406	&	0.306	&	0.282	&	0.341	&	0.190	&	4.213	&	0.071	&	3.787	&	0.095	&	3.648	&	0.043	&	3.871	&	0.043	&	3.291	&	0.043	\\
414	&	AP And	&	s	&	2.9143	&	0.7	&	2.291	&	0.877	&	9	&	3.840	&	0.101	&	0.406	&	0.306	&	0.282	&	0.341	&	0.190	&	4.340	&	0.071	&	3.903	&	0.095	&	3.762	&	0.043	&	3.991	&	0.043	&	3.395	&	0.043	\\
415	&	AL Scl	&	p	&	4.6006	&	3.6	&	319.014	&	122.103	&	11	&	-1.519	&	0.117	&	0.041	&	0.031	&	0.032	&	0.037	&	0.020	&	-0.657	&	0.090	&	-0.550	&	0.089	&	-0.549	&	0.088	&	-0.616	&	0.089	&	-0.451	&	0.089	\\
416	&	AL Scl	&	s	&	4.6006	&	3.6	&	19.903	&	7.618	&	14	&	1.493	&	0.155	&	0.041	&	0.031	&	0.032	&	0.037	&	0.020	&	2.063	&	0.090	&	2.021	&	0.089	&	2.042	&	0.088	&	2.042	&	0.089	&	1.979	&	0.089	\\
417	&	V821 Cas	&	p	&	3.4262	&	0.6	&	37.469	&	14.341	&	17	&	0.806	&	0.187	&	0.123	&	0.092	&	0.092	&	0.107	&	0.058	&	1.203	&	0.051	&	1.132	&	0.045	&	1.072	&	0.042	&	1.089	&	0.042	&	1.016	&	0.042	\\
418	&	V821 Cas	&	s	&	3.4262	&	0.6	&	9.522	&	3.644	&	19	&	2.293	&	0.205	&	0.123	&	0.092	&	0.092	&	0.107	&	0.058	&	2.671	&	0.051	&	2.532	&	0.045	&	2.475	&	0.042	&	2.524	&	0.042	&	2.352	&	0.042	\\
\hline
\end{tabular}}
\end{table*}

\begin{table*}
\centering
\caption{Empirical standard component $BC$s of DDEB in $B$, $V$, $G$, $G_{BP}$ and $G_{RP}$ and propagated uncertainties. The full table is available online.}
\label{tab:BC_table}
\resizebox{\textwidth}{!}{\begin{tabular}{cccccccccccccccccc}
\hline
Order	&	Name	&	pri/sec	& $BC_B$ & err & $BC_V$ & err & $BC_G$ & err & $BC_{G_{BP}}$ & err & $BC_{G_{RP}}$ & err & $(B-V)_0$	&	$(V-G)_0$	&	$(G-G_{BP})_0$	&	$(G-G_{RP})_0$	&	$(G_{BP}-G_{RP})_0$	\\
& & & (mag)& (mag)  & (mag)  & (mag)  & (mag)  & (mag)  & (mag)  & (mag)  & (mag)  & (mag) & (mag)  & (mag)  & (mag)  & (mag)  & (mag) \\  
\hline
1	&	V421 Peg	&	p	&	-0.345	&	0.074	&	-0.019	&	0.070	&	0.047	&	0.069	&	-0.105	&	0.074	&	0.322	&	0.065	&	0.326	&	0.066	&	-0.152	&	0.275	&	0.427	\\
2	&	V421 Peg	&	s	&	-0.374	&	0.098	&	-0.011	&	0.094	&	0.059	&	0.094	&	-0.113	&	0.097	&	0.367	&	0.091	&	0.363	&	0.070	&	-0.172	&	0.308	&	0.480	\\
3	&	DV Psc	&	p	&	-1.096	&	0.104	&	-0.350	&	0.101	&	-0.085	&	0.100	&	-0.689	&	0.105	&	0.642	&	0.098	&	0.746	&	0.265	&	-0.604	&	0.728	&	1.332	\\
4	&	DV Psc	&	s	&	-2.023	&	0.095	&	-0.916	&	0.091	&	-0.460	&	0.090	&	-1.309	&	0.095	&	0.478	&	0.088	&	1.108	&	0.456	&	-0.849	&	0.938	&	1.787	\\
5	&	MU Cas	&	p	&	-1.113	&	0.175	&	-1.216	&	0.173	&	-1.141	&	0.173	&	-1.091	&	0.175	&	-1.313	&	0.171	&	-0.102	&	0.075	&	0.050	&	-0.172	&	-0.222	\\
6	&	MU Cas	&	s	&	-1.151	&	0.172	&	-1.263	&	0.169	&	-1.187	&	0.169	&	-1.132	&	0.171	&	-1.369	&	0.168	&	-0.112	&	0.077	&	0.054	&	-0.183	&	-0.237	\\
7	&	TYC 4019-3345-1	&	p	&	-0.212	&	0.176	&	-0.076	&	0.174	&	0.088	&	0.173	&	0.021	&	0.175	&	0.179	&	0.172	&	0.136	&	0.164	&	-0.068	&	0.090	&	0.158	\\
8	&	TYC 4019-3345-1	&	s	&	-0.212	&	0.299	&	-0.076	&	0.297	&	0.088	&	0.297	&	0.021	&	0.299	&	0.179	&	0.297	&	0.136	&	0.164	&	-0.068	&	0.090	&	0.158	\\
9	&	YZ Cas	&	p	&	-0.330	&	0.074	&	-0.306	&	0.066	&	-0.277	&	0.064	&	-0.290	&	0.069	&	-0.230	&	0.060	&	0.024	&	0.029	&	-0.012	&	0.048	&	0.060	\\
10	&	YZ Cas	&	s	&	-0.398	&	0.159	&	-0.097	&	0.156	&	-0.060	&	0.155	&	-0.210	&	0.157	&	0.250	&	0.154	&	0.301	&	0.037	&	-0.150	&	0.310	&	0.460	\\
...	&	...	&	...	&	...	&	...	&	...	&	...	&	...	&	...	&	...	&	...	&	...	&	...	&	...	&	...	&	...	&	...	&	...	\\
409	&	IT Cas	&	p	&	-0.388	&	0.090	&	0.052	&	0.093	&	0.149	&	0.084	&	-0.064	&	0.088	&	0.517	&	0.081	&	0.440	&	0.097	&	-0.213	&	0.368	&	0.581	\\
410	&	IT Cas	&	s	&	-0.388	&	0.104	&	0.052	&	0.106	&	0.149	&	0.099	&	-0.064	&	0.102	&	0.517	&	0.096	&	0.440	&	0.097	&	-0.213	&	0.368	&	0.581	\\
411	&	BK Peg	&	p	&	-0.504	&	0.083	&	0.007	&	0.068	&	0.143	&	0.068	&	-0.111	&	0.073	&	0.549	&	0.064	&	0.511	&	0.136	&	-0.253	&	0.406	&	0.660	\\
412	&	BK Peg	&	s	&	-0.492	&	0.067	&	0.009	&	0.047	&	0.143	&	0.046	&	-0.105	&	0.053	&	0.541	&	0.040	&	0.501	&	0.134	&	-0.248	&	0.398	&	0.646	\\
413	&	AP And	&	p	&	-0.488	&	0.123	&	-0.062	&	0.135	&	0.078	&	0.105	&	-0.146	&	0.109	&	0.435	&	0.103	&	0.426	&	0.139	&	-0.223	&	0.357	&	0.580	\\
414	&	AP And	&	s	&	-0.500	&	0.123	&	-0.063	&	0.136	&	0.078	&	0.106	&	-0.151	&	0.110	&	0.445	&	0.104	&	0.437	&	0.141	&	-0.229	&	0.367	&	0.597	\\
415	&	AL Scl	&	p	&	-0.863	&	0.147	&	-0.970	&	0.145	&	-0.970	&	0.144	&	-0.903	&	0.147	&	-1.069	&	0.143	&	-0.107	&	0.000	&	0.067	&	-0.099	&	-0.165	\\
416	&	AL Scl	&	s	&	-0.570	&	0.179	&	-0.528	&	0.177	&	-0.549	&	0.176	&	-0.549	&	0.178	&	-0.486	&	0.175	&	0.042	&	-0.021	&	0.000	&	0.063	&	0.063	\\
417	&	V821 Cas	&	p	&	-0.397	&	0.194	&	-0.326	&	0.190	&	-0.267	&	0.190	&	-0.283	&	0.191	&	-0.210	&	0.188	&	0.071	&	0.059	&	-0.016	&	0.057	&	0.073	\\
418	&	V821 Cas	&	s	&	-0.377	&	0.211	&	-0.239	&	0.208	&	-0.182	&	0.208	&	-0.230	&	0.209	&	-0.059	&	0.206	&	0.138	&	0.057	&	-0.048	&	0.123	&	0.172	\\
\hline
\end{tabular}}
\end{table*}

\begin{table}
\centering
\caption{Parameters for passband based $BC$ - temperature relations shown in Figs.~4 and 5 in the form $BC_{\xi}$ = a + b $\times$ (log $T_{eff}$) + c $\times$ (log $T_{eff}$)$^2$ + d $\times$ (log $T_{eff}$)$^3$ + e $\times$ (log $T_{eff}$)$^4$. For the calculation of passband based solar absolute magnitudes, solar absolute bolometric magnitude is adopted to be $M_{bol,\odot}=4.74$ and the $BC_{\odot}$ refers to $T_{eff,\odot}=5772$ K. The relations are valid in the temperature range of 2900-38000 K.}
\label{tab:BCpar_table}
\resizebox{\columnwidth}{!}{\begin{tabular}{cccccccc}
\hline
Coefficient & $BC_{B}$ & $BC_{V}$ & $BC_{V}*$ & $BC_{G}$ & $BC_{G_{BP}}$ & $BC_{G_{RP}}$ \\
\hline
a & --1272.43 & --3767.98 & --2360.69565 & --1407.14 & --3421.55 & --1415.67 \\
  & $\pm$394.2 & $\pm$288.8 & $\pm$519.80058 & $\pm$256.7 & $\pm$293.6 & $\pm$253.3 \\
b & 1075.85   & 3595.86  & 2109.00655  & 1305.08  & 3248.19 & 1342.38 \\
  & $\pm$394.4 & $\pm$290.9 & $\pm$519.47090  & $\pm$258.9 & $\pm$296.1 & $\pm$255.4 \\
c & --337.831 & --1286.59 & --701.96628  & --453.605 & --1156.82 & --475.827 \\
  & $\pm$147.7 & $\pm$109.6 & $\pm$194.29038 & $\pm$97.67 & $\pm$111.7 & $\pm$96.34 \\
d & 46.8074 & 204.764  & 103.30304 & 70.2338 & 183.372 & 74.9702 \\
  & $\pm$24.53 & $\pm$18.32 & $\pm$32.23187 & $\pm$16.34 & $\pm$18.68  & $\pm$16.12  \\
e & --2.42862 & --12.2469 & --11.5957 & --4.1047 & --10.9305 & --4.44923\\
  & $\pm$1.552 & $\pm$1.146  & $\pm$1.441  & $\pm$1.023 & $\pm$1.169   & $\pm$1.009 \\
\hline
rms & 0.136257 & 0.120071 & 0.215 & 0.11068 & 0.126577 & 0.109179 \\
$R^2$ & 0.9616 & 0.9789 & 0.941 & 0.9793 & 0.9738 & 0.9884 \\
$BC_\odot$ (mag) & --0.600 & 0.069 & --0.016 & 0.106 & --0.134 & 0.567\\
$M_\odot$ (mag) & 5.340 & 4.671 & 4.756 & 4.634 & 4.874 & 4.173\\
$m_\odot$ (mag) & --26.232 & --26.901 & --26.816 & --26.938 & --26.698 & -27.399 \\
\hline
$BC_{max}$ (mag) & --0.301 & 0.094 & 0.095 & 0.106 & --0.062 & 0.709\\
$T_{max}$ (K) & 8222 & 6397 & 6897 & 5715 & 6879 & 4345\\
\hline
$T_{1}$ (K) &-& 5300 & 5859 & 4565 &-&-\\
$T_{2}$ (K) &-& 7830 & 8226 & 7420 &-&8590\\
\hline
*: Eker et al. (2020).
\end{tabular}}
\end{table}

\begin{table*}
\centering
\caption{Mean bolometric corrections and intrinsic colours of nearby main-sequence stars as a function of typical effective temperature and spectral types having metallicities 0.008 $<$ $Z$ $<$ 0.040 for the passbands $B$, $V$, $G$, $G_{BP}$ and $G_{RP}$.}
\resizebox{\textwidth}{!}{\begin{tabular}{lcccccccccc}
\toprule
  &   & \multicolumn{7}{|c}{From Table 5} & \multicolumn{2}{|c}{From Table 8} \\
\midrule
SpT	&	Teff	&	$BC_{B}$	&	$BC_{V}$	&	$BC_{G}$	&	$BC_{G_{BP}}$	&	$BC_{G_{RP}}$ & $(G-G_{RP})_0$ & $(G_{BP}-G_{RP})_0$  & $(G_{BP}-G_{RP})_0$ & $(B-V)_0$ \\
& (K) & (mag) & (mag) & (mag) & (mag) & (mag) & (mag) & (mag) & (mag) & (mag) \\
\midrule
O7	&	35810	&	-3.088	&	-3.399	&	-3.291	&	-3.150	&	-3.671	&	-0.380	&	-0.521	&	-0.562	&	-0.364	\\
O8	&	33963	&	-2.941	&	-3.192	&	-3.126	&	-2.957	&	-3.492	&	-0.367	&	-0.536	&	-0.543	&	-0.352	\\
O9	&	32211	&	-2.795	&	-2.997	&	-2.965	&	-2.775	&	-3.319	&	-0.354	&	-0.544	&	-0.525	&	-0.340	\\
B0	&	29512	&	-2.556	&	-2.700	&	-2.709	&	-2.497	&	-3.042	&	-0.334	&	-0.545	&	-0.493	&	-0.320	\\
B1	&	25119	&	-2.127	&	-2.218	&	-2.266	&	-2.045	&	-2.561	&	-0.296	&	-0.516	&	-0.432	&	-0.279	\\
B2	&	21135	&	-1.690	&	-1.769	&	-1.827	&	-1.626	&	-2.079	&	-0.252	&	-0.453	&	-0.359	&	-0.231	\\
B3	&	18408	&	-1.364	&	-1.446	&	-1.501	&	-1.327	&	-1.714	&	-0.212	&	-0.387	&	-0.295	&	-0.190	\\
B5	&	15136	&	-0.955	&	-1.029	&	-1.076	&	-0.946	&	-1.222	&	-0.146	&	-0.276	&	-0.194	&	-0.123	\\
B6	&	13964	&	-0.808	&	-0.869	&	-0.914	&	-0.803	&	-1.029	&	-0.115	&	-0.226	&	-0.148	&	-0.093	\\
B7	&	13032	&	-0.694	&	-0.737	&	-0.782	&	-0.687	&	-0.868	&	-0.086	&	-0.181	&	-0.106	&	-0.065	\\
B8	&	12023	&	-0.576	&	-0.591	&	-0.636	&	-0.560	&	-0.685	&	-0.049	&	-0.125	&	-0.053	&	-0.031	\\
B9	&	10666	&	-0.436	&	-0.389	&	-0.437	&	-0.390	&	-0.425	&	0.012	&	-0.035	&	0.032	&	0.026	\\
A0	&	9886	&	-0.371	&	-0.274	&	-0.323	&	-0.297	&	-0.269	&	0.055	&	0.028	&	0.093	&	0.067	\\
A1	&	9419	&	-0.340	&	-0.206	&	-0.257	&	-0.243	&	-0.173	&	0.084	&	0.070	&	0.135	&	0.095	\\
A2	&	9078	&	-0.322	&	-0.158	&	-0.209	&	-0.206	&	-0.102	&	0.107	&	0.104	&	0.168	&	0.118	\\
A3	&	8750	&	-0.309	&	-0.113	&	-0.164	&	-0.173	&	-0.033	&	0.130	&	0.140	&	0.202	&	0.142	\\
A5	&	8222	&	-0.301	&	-0.045	&	-0.094	&	-0.125	&	0.078	&	0.172	&	0.204	&	0.266	&	0.185	\\
A6	&	7980	&	-0.303	&	-0.016	&	-0.064	&	-0.107	&	0.130	&	0.194	&	0.237	&	0.298	&	0.209	\\
A7	&	7745	&	-0.308	&	0.009	&	-0.036	&	-0.091	&	0.179	&	0.215	&	0.271	&	0.332	&	0.233	\\
A8	&	7534	&	-0.317	&	0.030	&	-0.012	&	-0.079	&	0.224	&	0.236	&	0.304	&	0.365	&	0.257	\\
F0	&	7161	&	-0.343	&	0.062	&	0.027	&	-0.066	&	0.302	&	0.275	&	0.367	&	0.427	&	0.303	\\
F1	&	6966	&	-0.363	&	0.075	&	0.045	&	-0.062	&	0.342	&	0.297	&	0.404	&	0.466	&	0.331	\\
F2	&	6792	&	-0.384	&	0.084	&	0.059	&	-0.062	&	0.377	&	0.318	&	0.440	&	0.501	&	0.357	\\
F3	&	6637	&	-0.406	&	0.089	&	0.071	&	-0.064	&	0.408	&	0.337	&	0.473	&	0.535	&	0.382	\\
F5	&	6397	&	-0.447	&	0.094	&	0.086	&	-0.073	&	0.455	&	0.369	&	0.529	&	0.591	&	0.429	\\
F6	&	6310	&	-0.464	&	0.093	&	0.091	&	-0.078	&	0.472	&	0.381	&	0.550	&	0.613	&	0.447	\\
F7	&	6223	&	-0.483	&	0.092	&	0.095	&	-0.084	&	0.488	&	0.393	&	0.573	&	0.636	&	0.465	\\
F8	&	6152	&	-0.499	&	0.091	&	0.098	&	-0.090	&	0.501	&	0.403	&	0.591	&	0.655	&	0.481	\\
G0	&	6026	&	-0.530	&	0.086	&	0.102	&	-0.102	&	0.524	&	0.422	&	0.626	&	0.690	&	0.510	\\
G1	&	5957	&	-0.548	&	0.082	&	0.104	&	-0.110	&	0.536	&	0.433	&	0.646	&	0.710	&	0.526	\\
G2	&	5888	&	-0.567	&	0.078	&	0.105	&	-0.118	&	0.548	&	0.443	&	0.667	&	0.730	&	0.543	\\
G3	&	5848	&	-0.579	&	0.075	&	0.105	&	-0.124	&	0.555	&	0.450	&	0.679	&	0.743	&	0.554	\\
G5	&	5741	&	-0.612	&	0.065	&	0.106	&	-0.140	&	0.573	&	0.467	&	0.713	&	0.776	&	0.582	\\
G6	&	5689	&	-0.629	&	0.060	&	0.106	&	-0.148	&	0.581	&	0.476	&	0.730	&	0.793	&	0.596	\\
G7	&	5649	&	-0.642	&	0.055	&	0.105	&	-0.155	&	0.588	&	0.483	&	0.743	&	0.806	&	0.607	\\
G8	&	5559	&	-0.674	&	0.044	&	0.104	&	-0.172	&	0.602	&	0.498	&	0.774	&	0.837	&	0.633	\\
K0	&	5248	&	-0.801	&	-0.010	&	0.090	&	-0.247	&	0.645	&	0.556	&	0.892	&	0.951	&	0.729	\\
K1	&	5070	&	-0.888	&	-0.054	&	0.075	&	-0.302	&	0.666	&	0.591	&	0.969	&	1.028	&	0.788	\\
K2	&	4898	&	-0.982	&	-0.106	&	0.056	&	-0.366	&	0.683	&	0.628	&	1.049	&	1.102	&	0.846	\\
K3	&	4732	&	-1.085	&	-0.167	&	0.031	&	-0.439	&	0.696	&	0.665	&	1.135	&	1.177	&	0.902	\\
K5	&	4345	&	-1.375	&	-0.362	&	-0.053	&	-0.660	&	0.709	&	0.762	&	1.368	&	1.365	&	1.025	\\
M0	&	3802	&	-1.939	&	-0.803	&	-0.258	&	-1.138	&	0.666	&	0.924	&	1.804	&	1.659	&	1.193	\\
M1	&	3648	&	-2.143	&	-0.977	&	-0.341	&	-1.323	&	0.636	&	0.978	&	1.959	&	1.750	&	1.240	\\
M2	&	3499	&	-2.363	&	-1.174	&	-0.435	&	-1.528	&	0.598	&	1.033	&	2.126	&	1.845	&	1.286	\\
M3	&	3350	&	-2.610	&	-1.402	&	-0.545	&	-1.765	&	0.547	&	1.092	&	2.312	&	1.947	&	1.331	\\
M4	&	3148	&	-2.991	&	-1.770	&	-0.722	&	-2.143	&	0.457	&	1.179	&	2.600	&	2.111	&	1.393	\\
M5	&	2999	&	-3.314	&	-2.094	&	-0.879	&	-2.473	&	0.370	&	1.249	&	2.843	&	2.264	&	1.439	\\
\bottomrule
\end{tabular}}
\label{tab:empiricalBC}
\end{table*}

\begin{table*}
\centering
\caption{Component bolometric magnitudes and luminosities of DDEB in $B$, $V$, $G$, $G_{BP}$ and $G_{RP}$ passbands. The full table is available online.}
\label{tab:bol_mag_DDEB}
\resizebox{\textwidth}{!}{\begin{tabular}{ccccccccccccccccccc}
\hline
Order	&	Name	&	pri/sec	&	$M_{bol}(B)$	& err & $M_{bol}(V)$ & err  & $M_{bol}(G)$ & err & $M_{bol}(G_{BP})$ & err & $M_{bol}(G_{RP})$ & err & $\left<M_{bol}\right>$ & Mean err & log $\left<L/L_\odot\right>$ & $\frac{\Delta L}{L}$ & log ($L/L_\odot$)(SB) & $\frac{\Delta L}{L}$ \\
& & & (mag) & (mag) & (mag) & (mag) & (mag) & (mag) & (mag) & (mag) & (mag) & (mag) & (mag) & (mag) & & (\%) & & (\%) \\
\hline
1	&	V421 Peg	&	p	&	2.760	&	0.143	&	2.825	&	0.124	&	2.722	&	0.115	&	2.789	&	0.133	&	2.713	&	0.111	&	2.762	&	0.021	&	0.791	&	1.924	&	0.796	&	5.655	\\
2	&	V421 Peg	&	s	&	3.312	&	0.143	&	3.383	&	0.124	&	3.284	&	0.115	&	3.349	&	0.133	&	3.271	&	0.111	&	3.320	&	0.021	&	0.568	&	1.919	&	0.577	&	8.146	\\
3	&	DV Psc	&	p	&	6.499	&	0.142	&	6.741	&	0.124	&	6.751	&	0.115	&	6.789	&	0.134	&	6.757	&	0.112	&	6.707	&	0.053	&	-0.787	&	4.857	&	-0.780	&	8.789	\\
4	&	DV Psc	&	s	&	8.051	&	0.142	&	8.115	&	0.124	&	8.317	&	0.115	&	8.160	&	0.134	&	8.369	&	0.112	&	8.202	&	0.061	&	-1.385	&	5.580	&	-1.391	&	7.832	\\
5	&	MU Cas	&	p	&	-2.240	&	0.164	&	-2.208	&	0.148	&	-2.329	&	0.140	&	-2.255	&	0.155	&	-2.293	&	0.137	&	-2.265	&	0.021	&	2.802	&	1.943	&	2.875	&	13.768	\\
6	&	MU Cas	&	s	&	-2.060	&	0.164	&	-2.021	&	0.148	&	-2.145	&	0.140	&	-2.070	&	0.155	&	-2.108	&	0.137	&	-2.081	&	0.021	&	2.728	&	1.950	&	2.800	&	13.423	\\
7	&	TYC 4019-3345-1	&	p	&	1.687	&	0.145	&	1.764	&	0.127	&	1.549	&	0.118	&	1.601	&	0.136	&	1.601	&	0.115	&	1.640	&	0.038	&	1.240	&	3.496	&	1.184	&	15.498	\\
8	&	TYC 4019-3345-1	&	s	&	1.687	&	0.145	&	1.764	&	0.127	&	1.549	&	0.118	&	1.601	&	0.136	&	1.601	&	0.115	&	1.640	&	0.038	&	1.240	&	3.496	&	1.184	&	27.114	\\
9	&	YZ Cas	&	p	&	0.540	&	0.145	&	0.642	&	0.125	&	0.562	&	0.115	&	0.591	&	0.133	&	0.591	&	0.112	&	0.585	&	0.017	&	1.662	&	1.569	&	1.674	&	5.117	\\
10	&	YZ Cas	&	s	&	3.382	&	0.145	&	3.533	&	0.125	&	3.469	&	0.115	&	3.505	&	0.133	&	3.466	&	0.112	&	3.471	&	0.025	&	0.508	&	2.340	&	0.553	&	13.983	\\
...	&	...	&	...	&	...	&	...	&	...	&	...	&	...	&	...	&	...	&	...	&	...	&	...	&	...	&	...	&	...	&	...	&	...	&	...	\\
409	&	IT Cas	&	p	&	3.174	&	0.145	&	3.261	&	0.131	&	3.153	&	0.116	&	3.214	&	0.134	&	3.144	&	0.113	&	3.189	&	0.022	&	0.620	&	1.988	&	0.608	&	7.053	\\
410	&	IT Cas	&	s	&	3.220	&	0.145	&	3.307	&	0.131	&	3.200	&	0.116	&	3.261	&	0.134	&	3.191	&	0.113	&	3.236	&	0.022	&	0.602	&	1.988	&	0.590	&	8.500	\\
411	&	BK Peg	&	p	&	2.926	&	0.148	&	2.981	&	0.125	&	2.846	&	0.115	&	2.925	&	0.133	&	2.827	&	0.112	&	2.901	&	0.028	&	0.736	&	2.617	&	0.738	&	5.487	\\
412	&	BK Peg	&	s	&	3.537	&	0.148	&	3.591	&	0.125	&	3.454	&	0.115	&	3.534	&	0.133	&	3.436	&	0.112	&	3.510	&	0.029	&	0.492	&	2.645	&	0.493	&	2.990	\\
413	&	AP And	&	p	&	3.796	&	0.154	&	3.879	&	0.151	&	3.724	&	0.116	&	3.805	&	0.134	&	3.713	&	0.112	&	3.783	&	0.030	&	0.383	&	2.772	&	0.406	&	9.194	\\
414	&	AP And	&	s	&	3.911	&	0.154	&	3.996	&	0.151	&	3.843	&	0.116	&	3.923	&	0.134	&	3.831	&	0.112	&	3.901	&	0.030	&	0.336	&	2.750	&	0.360	&	9.280	\\
415	&	AL Scl	&	p	&	-1.414	&	0.163	&	-1.361	&	0.147	&	-1.405	&	0.139	&	-1.368	&	0.154	&	-1.409	&	0.136	&	-1.391	&	0.011	&	2.453	&	1.032	&	2.504	&	10.783	\\
416	&	AL Scl	&	s	&	1.659	&	0.163	&	1.686	&	0.147	&	1.658	&	0.139	&	1.696	&	0.154	&	1.627	&	0.136	&	1.665	&	0.012	&	1.230	&	1.116	&	1.299	&	14.269	\\
417	&	V821 Cas	&	p	&	0.865	&	0.146	&	0.929	&	0.126	&	0.819	&	0.115	&	0.847	&	0.133	&	0.846	&	0.112	&	0.861	&	0.018	&	1.552	&	1.694	&	1.574	&	17.193	\\
418	&	V821 Cas	&	s	&	2.365	&	0.146	&	2.440	&	0.126	&	2.332	&	0.115	&	2.365	&	0.133	&	2.350	&	0.112	&	2.371	&	0.018	&	0.948	&	1.688	&	0.979	&	18.872	\\
\hline
\end{tabular}}
\end{table*}

\begin{table*}
	\centering
	\caption{Parameters of temperature - intrinsic colour relations shown in Fig.~10.}
	\label{tab:logTeff_par_table}
	\begin{tabular}{cccccc}
		\hline
		\hline
		\multicolumn{6}{c}{log $T_{eff}$ = a + b $\times$ $(B-V)_0$ + c $\times$ $(B-V)_0^2$ + d $\times$ $(B-V)_0^3$ + e $\times$ $(B-V)_0^4$}\\
		\hline
		& a & b & c & d & e \\
		\hline
		& 4.05136 & --0.902404 & 1.03912 & --0.686631 & 0.144272 \\
		& $\pm$0.005228 & $\pm$0.01865 & $\pm$0.06344 & $\pm$0.1399 &  $\pm$0.07865 \\
		\multicolumn{6}{c}{rms= 0.05091} \\
		\multicolumn{6}{c}{valid in the range --0.5 $ \leq (B-V)_0 \leq $ 1.5 mag.} \\
		\hline
		\hline
		
	    \multicolumn{6}{c}{log $T_{eff}$ = a + b $\times$ $(G_{BP}-G_{RP})_0$ + c $\times$ $(G_{BP}-G_{RP})_0^2$ + d $\times$  $(G_{BP}-G_{RP})_0^3$ + e $\times$ $(G_{BP}-G_{RP})_0^4$ } \\
		\hline
		& a & b & c & d & e \\
		\hline
		& 4.04695 & --0.595137 & 0.42341 & --0.199622 & 0.0351755 \\
		& $\pm$0.004102 & $\pm$0.007874 & $\pm$0.0211 & $\pm$0.0211 & $\pm$0.005871 \\
		\multicolumn{6}{c}{rms= 0.0396} \\
		\multicolumn{6}{c}{valid in the range --0.6 $ \leq (G_{BP}-G_{RP})_0 \leq $ 1.7 mag.} \\
		\hline
	\end{tabular}
\end{table*}


\clearpage 

\newpage

\begin{references}

\refitem{Andrae R., et al.}{2018}{A\&A}{616}{A8}

\refitem{Arenou F., et al.}{2018}{A\&A}{616}{A17}

\refitem{Bessell M. S.}{1990}{PASP}{102}{1181}

\refitem{Bessell M. S., Castelli F., Plez B.}{1998}{A\&A}{333}{231}

\refitem{Bressan A., Marigo P., Girardi L., Salasnich B., Dal Cero C., Rubele S., Nanni A.}{2012}{MNRAS}{427}{127}

\refitem{Casagrande L., VandenBerg D. A.}{2018}{MNRAS}{475}{5023}

\refitem{Cayrel R., Castelli F., Katz D., van't Veer C., Gomez A., Perrin M. N.}{1997}{in Bonnet R. M., et al., eds}{ESA Special Publication Vol. 402, Hipparcos - Venice 97.}{433-436}

\refitem{Chen Y., et al.}{2019}{A\&A}{632}{A105}

\refitem{Clayton D. D.}{1968}{Principles of stellar evolution and nucleosynthesis}{}{}

\refitem{Code A. D., Davis J., Bless R. C., Brown R. H.}{1976}{ApJ}{203}{417}

\refitem{Cox A. N.}{2000}{Allen's astrophysical quantities}{}{}

\refitem{Eker Z. et al.}{2018}{MNRAS}{479}{5491}

\refitem{Eker Z., et al.}{2020}{MNRAS}{496}{3887}

\refitem{Eker Z., Bak{\i}\c{s} V., Soydugan F., Bilir S.}{2021a}{MNRAS}{503}{4231}

\refitem{Eker Z., Soydugan F., Bilir S., Bak{\i}\c{s} V.}{2021b}{MNRAS}{507}{3583}

\refitem{Evans D. W., et al.}{2018}{A\&A}{616}{A4}

\refitem{Fitzpatrick E. L.}{1999}{PASP}{111}{63}

\refitem{Flower P. J.}{1977}{A\&A}{54}{31}

\refitem{Flower P. J.}{1996}{ApJ}{469}{355}

\refitem{Gaia Collaboration et al.}{2018}{A\&A}{616}{A1}

\refitem{Girardi L., Bertelli G., Bressan A., Chiosi C., Groenewegen M.
A. T., Marigo P., Salasnich B., Weiss A.}{2002}{MNRAS}{391}{195}

\refitem{Girardi L., et al.}{2008}{PASP}{120}{583}

\refitem{Graczyk D., et al.}{2019}{ApJ}{872}{85}

\refitem{Heintze J. R. W.}{1973}{in Hauck B., Westerlund B. E. eds.}{Problems of Calibration of Absolute Magnitudes and Temperature of Stars.}{54}{231}

\refitem{Johnson H. L.}{1964}{Boletin de los Observatorios Tonantzintla y Tacubaya}{3}{305}

\refitem{Johnson H. L.}{1966}{ARA\&A}{4}{193}

\refitem{Jordi C., Gebran M., Carrasco J. M., de Bruijne J., Voss H.,
Fabricius C., Knude J., Vallenari A., Kohley R., Mora A.}{2010}{A\&A}{532}{48}

\refitem{Kuiper G. P.}{1938}{ApJ}{88}{429}

\refitem{Lacy C. H. S., Torres G., Wolf M., Burks C. L.}{2014}{AJ}{147}{1}

\refitem{Mamajek E.}{2021}{private communication}{}{}

\refitem{Martins F. and Plez B.}{2006}{A\&A}{457}{637}

\refitem{McDonald J. K. and Underhill A. B.}{1952}{ApJ}{115}{577}

\refitem{Pecaut M. J., Mamajek E. E.}{2013}{ApJS}{208}{9}

\refitem{Pedersen M.G., Escorza A., P{\'a}pics P.I., and Aerts C.}{2020}{MNRAS}{495}{2738}


\refitem{Popper D. M.}{1959}{ApJ}{129}{647}

\refitem{Smak J.}{1966}{Acta Astron.}{16}{1}

\refitem{Sung H., Lim B., Bessell M. S., Kim J. S., Hur H., Chun M.-Y., Park B.-G.}{2013}{Journal of Korean Astronomical Society}{46}{103}

\refitem{Tomasella L., Munari U., Siviero A., Cassisi S., Dallaporta S., Zwitter T., Sordo R.}{2008}{A\&A}{480}{465}

\refitem{Torres G.}{2010}{AJ}{140}{1158}

\refitem{Weidemann V., Bues I.}{1967}{Z. Astrophys.}{67}{415}

\refitem{Wenger M., et al.}{2000}{A\&AS}{143}{9}

\refitem{Wildey R. L.}{1963}{Nature}{199}{988}


\refitem{Wilmor Christopher N. A. }{2018}{ApJS}{236}{47}

\end{references}
\end{document}